\documentclass[twocolumn, aps, pra, reprint, floatfix, superscriptaddress,nofootinbib]{revtex4-1}
\usepackage[utf8]{inputenc}
\usepackage{comment}
\usepackage{algorithm}
\usepackage{algpseudocode}
\usepackage{algorithmicx}

\usepackage{amsmath}
\usepackage{amssymb}
\usepackage{appendix}

\usepackage{qcircuit}
\usepackage{bbm}
\usepackage{bm}
\usepackage{dsfont}

\usepackage[caption=false]{subfig}

\usepackage[svgnames]{xcolor}
\usepackage{tikz}

\usepackage{hyperref}
\hypersetup{colorlinks,linkcolor={DarkBlue},citecolor={DarkBlue},urlcolor={DarkBlue}}
\usepackage[capitalize]{cleveref}

\usepackage[color = LightBlue]{todonotes}
\usepackage[normalem]{ulem}

\DeclareMathOperator{\csch}{csch}
\newcommand{\proj}[1]{| #1 \rangle \langle #1|}
\usepackage{braket}
\newcommand{\pr}[1]{\ket{#1}\bra{#1}}
\usepackage{soul}

\newcommand{\id}{\mathds{1}}
\newcommand{\CX}{\text{CX}}
\newcommand{\CZ}{\text{CZ}}
\newcommand{\X}{\text{X}}
\newcommand{\Z}{\text{Z}}

\definecolor{DO}{RGB}{204,102,0}

\begin{document}
\setstcolor{DO}

\title{Fast simulation of bosonic qubits via Gaussian functions in phase space}
\author{J. Eli Bourassa}
\affiliation{Xanadu, 777 Bay Street, Toronto ON,  M5G 2C8, Canada}
\email{eli@xanadu.ai, nicolas@xanadu.ai, ilan@xanadu.ai, \\antal@xanadu.ai, theodor@xanadu.ai, josh@xanadu.ai, \\krishna@xanadu.ai, guillaume@xanadu.ai, ish@xanadu.ai}
\affiliation{Department of Physics, University of Toronto, Toronto ON, M5S 1A7, Canada}
\author{Nicolás Quesada}
\affiliation{Xanadu, 777 Bay Street, Toronto ON,  M5G 2C8, Canada}
\author{Ilan Tzitrin}
\affiliation{Xanadu, 777 Bay Street, Toronto ON,  M5G 2C8, Canada}
\affiliation{Department of Physics, University of Toronto, Toronto ON, M5S 1A7, Canada}
\author{Antal Száva}
\affiliation{Xanadu, 777 Bay Street, Toronto ON,  M5G 2C8, Canada}
\author{Theodor Isacsson}
\affiliation{Xanadu, 777 Bay Street, Toronto ON,  M5G 2C8, Canada}
\author{Josh Izaac}
\affiliation{Xanadu, 777 Bay Street, Toronto ON,  M5G 2C8, Canada}
\author{Krishna Kumar Sabapathy}
\affiliation{Xanadu, 777 Bay Street, Toronto ON,  M5G 2C8, Canada}
\author{Guillaume Dauphinais}
\affiliation{Xanadu, 777 Bay Street, Toronto ON,  M5G 2C8, Canada}
\author{Ish Dhand}
\affiliation{Xanadu, 777 Bay Street, Toronto ON,  M5G 2C8, Canada}

\date{\today}

\begin{abstract}
Bosonic qubits are a promising route to building fault-tolerant quantum computers on a variety of physical platforms. 
Studying the performance of bosonic qubits under realistic gates and measurements is challenging with existing analytical and numerical tools. 
We present a novel formalism for simulating classes of states that can be represented as linear combinations of Gaussian functions in phase space. 
This formalism allows us to analyze and simulate a wide class of non-Gaussian states, transformations and measurements. 
We demonstrate how useful classes of bosonic qubits---Gottesman-Kitaev-Preskill (GKP), cat, and Fock states---can be simulated using this formalism, opening the door to investigating the behaviour of bosonic qubits under Gaussian channels and measurements, non-Gaussian transformations such as those achieved via gate teleportation, and important non-Gaussian measurements such as threshold and photon-number detection. 
Our formalism enables simulating these situations with levels of accuracy that are not feasible with existing methods.
Finally, we use a method informed by our formalism to simulate circuits critical to the study of fault-tolerant quantum computing with bosonic qubits but beyond the reach of existing techniques. Specifically, we examine how finite-energy GKP states transform under realistic qubit phase gates; interface with a CV cluster state; and transform under non-Clifford T gate teleportation using magic states.
We implement our simulation method as a part of the open-source \texttt{Strawberry Fields} Python library.
\end{abstract}

\maketitle

\section{Introduction}

Photonics and superconducting cavities are the leading platforms for building a scalable fault-tolerant quantum computer~\cite{bourassa2021blueprint,chamberland2020building,bartolucci2021fusion,larsen2021fault,CampagneIbarcq2020,Puri2020,Grimm2020,Myers2011,Fukui2020temporal,Fukui2018}.
As continuous-variable (CV) quantum systems, these platforms rely on encoding qubits (two-level quantum systems) into the state of the CV system via so-called bosonic qubits, among which Gottesman-Kitaev-Preskill (GKP) states~\cite{GKP_2001}, cat states~\cite{ralph2003}, and Fock states~\cite{chuang1995simple,knill2001scheme} are the primary. 
Bosonic qubits are especially favourable for quantum computing because of their ability to correct physical errors within the CV space due to loss~\cite{albert2018}, random displacements~\cite{GKP_2001}, and rotations~\cite{grimsmo2020quantum}.

While concrete quantum computing architectures based on bosonic qubits have been proposed, the analysis and simulation of these qubits is challenging because of the infinite-dimensional Hilbert space that they occupy.
This impedes the development and implementation of these architectures since determining fault-tolerance thresholds and overheads is limited by our ability to simulate these physical systems in realistic situations. 
The current most flexible method for simulating bosonic qubits relies on the Fock basis.
Simulations in the Fock basis can be cumbersome, especially for CV states with large energy; in particular, high-quality and therefore high-energy cat and GKP states require a high photon-number cutoff, incurring large memory loads and processing times. Moreover, determining how states change under CV channels and measurements is computationally expensive in the Fock basis representation~\cite{miatto2020fast} as the energy of a given state can increase significantly under paradigmatic CV transformations such as squeezing and displacements.

Here, we overcome the challenge of studying bosonic qubits by introducing a novel formalism that enables their analysis and simulation.
Specifically, we present a mathematical framework for simulating a class of CV states, transformations, and measurements using linear combinations of Gaussian functions in phase space. 
This framework allows us to simulate the transformation of useful bosonic qubits such as GKP, cat, and Fock states under Gaussian channels and measurements, as well as under a class of valuable non-Gaussian channels effected through gate teleportation.
Motivated as a tool to facilitate the design of quantum computing architectures, our framework can model important sources of decoherence (such as optical loss in photonics or dissipation in superconducting cavities) as well as transformations and measurements that are readily-implementable in photonics, such as linear optics, squeezing operations, homodyne and photon-counting detection.
We accomplish this by leveraging the most convenient aspects of the Gaussian CV formalism---namely, the ability to regard the transformation of a state as a transformation of means vectors and covariance matrices---while providing the capability to simulate non-Gaussian systems, which is necessary to the construction of a quantum computer~\cite{Mari2012wigner}. 

Informed by our formalism, we provide a method for the fast and accurate simulation of bosonic qubits.
We find the scaling of the memory and processing time required for our simulator are vastly more favourable than current state-of-the-art Fock basis simulators~\cite{Killoran2019}.
We use this to conduct an in-depth numerical study of bosonic qubits in useful physical circuits under inevitable physical imperfections, including loss in optical components; finite-energy effects in resource states; and finite squeezing in the ancillae of measurement-based squeezing operations, the workhorse of inline squeezing \cite{filip2005}. 
Specifically, we analyze GKP states in three situations that are relevant for quantum computation.
First, we examine GKP states passing through a qubit phase gate (introduced in~\cite{GKP_2001} and studied in Sec. II D of \cite{Tzitrin2020}), a Clifford gate typically used in the universal gate set for GKP qubits and whose CV implementation we simulate being performed with measurement-based squeezing.
Second, we consider the teleportation of a GKP state into a CV cluster state, a scenario present in proposals for measurement-based quantum computation with GKP qubits \cite{Menicucci2014,bourassa2021blueprint, larsen2021fault,pantaleoni2021}.
Third, we study applying a qubit T gate to GKP states via gate teleportation with finite-energy GKP magic states and realistic entangling operations, which is an important scenario because the T gate, in conjunction with experimentally-accessible Gaussian operations, is a standard prerequisite for unlocking universal computation with GKP qubits~\cite{GKP_2001}. 
While gate teleportation is expected to be favourable over other methods~\cite{hastrup2020cubic}, simulation of the technique in the presence of realistic gate noise, to the best of our understanding, has not been performed, likely due to numerical challenges.
Thus, by enabling the study of situations that were hitherto intractable, our work provides a valuable toolkit for the analysis and simulation of quantum computation based on bosonic qubits.

The structure of this paper is as follows. In Section \ref{sec:background} we provide background on the Gaussian CV formalism and on bosonic qubits. In Section \ref{sec:formalism} we introduce our new formalism for simulating a wide class of states which can be expressed as linear combinations of Gaussian functions in phase space. We provide rules for how these states evolve under Gaussian and a class of non-Gaussian transformations and measurements. With the framework in hand, in Section \ref{sec:states} we show how useful classes of bosonic qubits---GKP, cat and Fock states---can be written in our formalism. In Section \ref{sec:ex_transform}, we detail how the formalism can be used for modelling loss, measurement-based squeezing, and useful non-Gaussian GKP qubit operations. Given the formalism, in Section \ref{sec:sim_methods}, we provide our simulation methods, along with a comparison to simulation techniques in the Fock basis. Our formalism and methods allow us to present results from novel simulations of bosonic qubits in Section \ref{sec:simulations}. We discuss additional areas of application for our simulator, and open research problems in Section \ref{sec:conclusion}.

For a more hands-on introduction to the simulations that are enabled by our formalism, we invite the reader to consult the open-source code available in \texttt{Strawberry Fields}~\cite{bosonic_github,Killoran2019} and an accompanying set of tutorials available online~\cite{tutorial1,tutorial2,tutorial3}.

\section{Background}\label{sec:background}

In this section we provide overviews and pointers to the relevant literature of three different threads that are unified in the later parts of the manuscript. 
In Section~\ref{subsec:gauss_bg} we provide a brief survey of the Gaussian formalism for CV quantum systems, including quantum phase-space, the symplectic formalism, and Gaussian states, channels, and measurements.
In Section~\ref{subsec:sf} we provide an overview of \texttt{Strawberry Fields}, the programming library in which we implement the formalism and methods developed in the rest of the paper. 
Finally, in Section~\ref{subsec:bosonic_qubits} we review the GKP and cat qubit encodings, which leverage the large Hilbert space of a CV mode. 

\subsection{Continuous-Variables and the Gaussian Formalism}\label{subsec:gauss_bg}

Multimode continuous-variable systems are best described using the canonical position $\hat{q}_j$ and momentum $\hat{p}_j$ operators acting on the infinite-dimensional Hilbert space associated with the system. It is often convenient to group these operators, representing $N$ modes, into a vector:
\begin{align}
\hat{\bm{\xi}}^T&= (\hat{q}_1,\hat{p}_1,\hat{q}_2,\hat{p}_2,\ldots,\hat{q}_N,\hat{p}_N).
\end{align}
The commutation relations these operators satisfy can be expressed succinctly in terms of the symplectic form 
\begin{equation}
\bm{\Omega}=\bigoplus_{i=1}^N
\begin{pmatrix}
0 & 1 \\
-1 & 0
\end{pmatrix}
\end{equation}
as
\begin{align}
[\hat{\xi}_j, \hat{\xi}_k] \equiv \hat{\xi}_j  \hat{\xi}_k - \hat{\xi}_k \hat{\xi}_j = i \hbar \Omega_{j,k}.
\end{align}
As for any quantum mechanical system, a complete description of its state can be obtained by specifying its density matrix $\hat{\rho}$. For CV systems it is often useful to introduce the characteristic function~\cite{weedbrook2012gaussian}
\begin{align}\label{eq:charac}
\chi(\bm{r};\hat{\rho}) =  \text{tr}\left(\hat{D}(\bm{r}) \hat{\rho} \right),
\end{align}
where $\hat{D}(\bm{r}) = \exp\left( i \hat{\bm{\xi}}^T \bm{\Omega} \bm{r} \right)$ is the Weyl or displacement operator and $\bm{r} \in \mathbb{R}^{2N}$ is  a real vector in phase-space.

\subsubsection{Gaussian States}
\label{sec:gauss_states_background}
We recall Gaussian states~\cite{weedbrook2012gaussian} as the ones whose characteristic function takes the form
\begin{align}
\chi(\bm{r};{\hat{\rho}})={\rm exp}\left(-\tfrac{1}{2}\bm{r}^{T}\bm{\Sigma}
\bm{r}-i\bm{\mu}^T\bm{\Omega} \bm{r}\,\right),
\end{align}
where we introduced the vector of means $\bm{\mu}$ and covariance matrix $\bm{\Sigma}$ of the state $\hat{\rho}$ with elements
\begin{align}
\mu_{i} &= \braket{\hat{\xi}_i},  \\
\Sigma_{i,j} &= \tfrac{1}{2}\braket{\hat{\xi}_i \hat{\xi}_j+\hat{\xi}_j \hat{\xi}_i} - \mu_i \mu_j,
\end{align}
where the expectation value of an operator $\hat{A}$ is defined as
\begin{align}\label{eq:expect-hilbert}
\braket{A} = \text{tr}(\hat{A} \hat{\rho}).
\end{align}
The covariance matrix of a valid quantum state, whether or not it is Gaussian, satisfies the uncertainty relation
\begin{align}\label{eq:UR}
\bm{\Sigma} + i \tfrac{\hbar}{2} \bm{\Omega} \geq 0.
\end{align}

A different characterization of Gaussian pure  states can be obtained by noting that they can be prepared by applying a Gaussian unitary $\hat{U}$ to the multimode vacuum state $\ket{0}^{\otimes N}$\,; that is, $\ket{\psi}_G=\hat{U}\ket{0}^{\otimes N}$, where the unitary is generated by a Hamiltonian that is at most quadratic in the quadrature operators\,\cite{simon94,Simon1987}.
The vacuum state is the unique state that is mapped to zero by its respective destruction operator
\begin{align}
\hat{a}_j \ket{0_j} &= 0, \quad \hat{a}_j = \tfrac{1}{\sqrt{2 \hbar}} (\hat{q}_j + i \hat{p_j}),
\end{align}
and has vector of means and  covariance matrix 
\begin{align}\label{eq:vac}
\bm{\mu}_{\text{vac}} = \bm{0} \text{ and } \bm{\Sigma}_{\text{vac}} = \tfrac{\hbar}{2} \id .
\end{align}

Finally, we introduce the Wigner function, which is the Fourier transform of the characteristic function:
\begin{align}\label{eq:wigner}
W(\bm{\xi};\hat{\rho}) = \int \frac{d^{2N} \bm{r}}{(2 \pi)^{2N}} \exp(- i \bm{\xi}^T \bm{\Omega} \bm{r}) \ \chi(\bm{r}; \hat{\rho}).
\end{align}
The Wigner function of a Gaussian state is a Gaussian function of the phase-space variables $\bm{\xi}$. Later, we introduce a class of states with Wigner functions that can be expressed as a linear combination of Gaussian functions in phase space.

\subsubsection{Gaussian Transformations and Measurements}
A Gaussian unitary transformation $\hat{U}$ is equivalent to a homogeneous linear phase-space transformation $\bm{\xi} \to \bm{S}^T\bm{\xi}$, followed by a phase-space displacement $\bm{\xi} \to \bm{\xi} + \bm{d}$.
Here the symplectic map $\bm{S}$ (satisfying $\bm{S}\bm{\Omega} \bm{S}^T=\bm{\Omega}$)
takes $\chi(\bm{r};\hat{\rho})\to\chi(\bm{S}^T\bm{r}; \hat{\rho})$,
which, for Gaussian states, is equivalent to transforming the covariance matrix as  $\bm{\Sigma} \to \bm{S \Sigma S}^T$ and mean as $\bm{\mu}\to \bm{S}\bm{\mu}$\,\cite{Simon1987}. Finally, the displacement transforms the mean as $\bm{S\mu}\to \bm{S\mu} + \bm{d}$.
As examples, the single-mode displacement and squeezing operators are defined respectively by
\begin{subequations}
\label{eq:sq_and_disp}
\begin{align}
\hat{D}(\alpha) &= \exp(\alpha \hat{a}^\dagger - \alpha^* \hat{a}),\\
\hat{S}(\zeta) &= \exp\left(\tfrac{\zeta}{2} \hat{a}^{2} - \tfrac{\zeta^*}{2} \hat{a}^{\dagger 2}  \right),
\end{align}
\end{subequations}
but in phase-space are represented as
\begin{subequations}
\begin{align}
\bm{S}_{\text{disp.}} &= \id, \quad  \bm{d}_{\text{disp.}}^T = \sqrt{2 \hbar}(\Re(\alpha), \Im(\alpha)), \\
\bm{S}_{\text{sq.}} &= \begin{pmatrix}
e^{-r} & 0 \\
0 & e^r
\end{pmatrix}, \quad \bm{d}_{\text{sq.}}^T = \bm{0},
\end{align}
\end{subequations}
where in the last line we assumed for simplicity that $\zeta= \zeta^* = r$ is real.

A Gaussian channel is a linear completely-positive trace-preserving map from Gaussian states to
Gaussian states.
Gaussian channels can be described by a pair of real matrices $(\bm{X},\bm{Y})$ with $\bm{Y} + i\frac{\hbar}{2}\bm{\Omega} \geq i\frac{\hbar}{2}\bm{X \Omega X}^T$~\cite{book_serafini}.
The action of the Gaussian channel described on the characteristic function is
\begin{equation}
\chi(\bm{r};{\hat{\rho}}) \to \chi'(\bm{r};{\hat{\rho}})  = \chi(\bm{X}\, \bm{r}; {\hat{\rho}})\,
\exp\left({-\frac{1}{2}\bm{r}^{T}\, \bm{Y} \, \bm{r}}\right).\nonumber\\
\end{equation}
It follows that the transformation on the covariance matrix and mean is given by
\begin{equation}
 \bm{\Sigma} \rightarrow \bm{X}^T\bm{\Sigma}\bm{X} +\bm{Y},\ \bm{\mu}\rightarrow \bm{X \mu}.
\label{gact}
\end{equation}
In addition to the mapping provided by $(\bm X,\bm Y)$, a displacement can always be added to a Gaussian channel and the transformation will remain deterministic.

One can also consider the characteristic and Wigner functions of an arbitrary operator $\hat{A}$. This is the so-called Weyl transform of a general Hilbert space operator $\hat{A}$, obtainable by replacing $\hat{\rho}$ with $\hat{A}$ in Eq.~\eqref{eq:charac} and then taking its Fourier transform as in Eq.~\eqref{eq:wigner}. The expectation value of the operator can now be written
\begin{align}\label{eq:weyl_trace}
 \braket{\hat A} = \text{tr}({\hat{\rho}}\hat{A})    = \int d^{2N}   \bm{\xi} \ W(\bm{\xi};\hat{\rho})W(\bm{\xi};\hat{A}).
\end{align}
We allow that $\hat{A}$ is a measurement operator, in which case $\braket{\hat A}$ is the probability of the outcome associated with the operator. In the case the measurement operator describes the detection of a Gaussian state parametrized by $\bm{r}_M$ and $\bm{\Sigma}_M$, this is referred to as a \emph{general-dyne measurement}, and the Weyl transform matches the Wigner function for that state~\cite{book_serafini}. A homodyne measurement is a limiting case of general-dyne measurement corresponding to projection onto an eigenstate of a quadrature operator, which can be treated as an infinitely squeezed state along that quadrature. The probability distribution for the outcome $\bm{r}_M$ of a general-dyne measurement on a Gaussian state, as can be deduced from \cref{eq:weyl_trace}, is itself a Gaussian distribution since the integration is between two Gaussian functions. 

Gaussian measurement motivates a transformation beyond Gaussian channels, namely conditional Gaussian dynamics, i.e., an update to a subset of modes of a multimode Gaussian state conditioned on the outcome of a Gaussian measurement that is performed on the remaining modes. Following~\cite{book_serafini}, the covariance and mean of the multimode state can be written as:
\begin{equation}\label{eq:ABsplit}
        \bm{\Sigma} = \begin{pmatrix}\bm{\Sigma}_{A} & \bm{\Sigma}_{AB}\\
        \bm{\Sigma}_{AB}^T & \bm{\Sigma}_{B}\end{pmatrix} \text{ and } \ \bm{\mu} = \begin{pmatrix}\bm{\mu}_{A} \\ \bm{\mu}_{B}\end{pmatrix},
\end{equation}
where $A$ denotes the modes that will remain active and $B$ those that will be measured.
If the Weyl transform of the measurement operator $\hat{M}$ corresponds to a Gaussian state with mean $\bm{r}_M$ and covariance $\bm{\Sigma}_M$, then the partial trace $\text{tr}_B(\hat{\rho}_{AB}\hat{M})$ corresponds to a Gaussian integral in phase space over the quadrature variables of modes $B$, yielding a Gaussian state in modes $A$ with the following covariances and means:
\begin{equation}\label{eq:cond_dyn}
    \begin{split}
        \bm{\Sigma}_{A} &\rightarrow \bm{\Sigma}_{A} - \bm{\Sigma}_{AB}(\bm{\Sigma}_{B}+\bm{\Sigma}_M)^{-1}\bm{\Sigma}_{AB}^T, \\
        \bm{\mu}_{A} &\rightarrow  \bm{\mu}_{A} + \bm{\Sigma}_{AB}(\bm{\Sigma}_{B}+\bm{\Sigma}_M)^{-1}(\bm{r}_M - \bm{\mu}_{B}).
    \end{split}
\end{equation}

\begin{figure}[!t]
	\begin{minipage}{0.45\textwidth}
	\providecommand{\ket}[1]{\left|#1\right\rangle}
\providecommand{\bra}[1]{\left\langle#1\right|}
\begin{tikzpicture}[scale=1,x=1pt,y=1pt]
\filldraw[color=white] (0, 0) rectangle (60, 30);

\draw[color=black] (0,30) -- (60,30);
\draw[color=black] (0,30) node[left] {$\hat{\rho}$};

\draw[color=black] (0,15) -- (60,15);
\draw[color=black] (0,15) node[left] {$\ket{0}$};

\draw[color=black] (0,0) -- (35,0);
\draw[color=black] (45,0) -- (60,0);
\draw[color=black] (0,0) node[left] {$\ket{0}$};

\draw (10,15) -- (10,0);
\filldraw (10, 15) ellipse(4pt and 1pt);
\filldraw (10, 0) ellipse(4pt and 1pt);

\draw[color=black] (35, -5) rectangle (45, 5);
\draw[color=black] (40,0) node {$\hat{\Phi}$};

\draw (40,30) -- (40,15);
\filldraw (40, 30) ellipse(4pt and 1pt);
\filldraw (40, 15) ellipse(4pt and 1pt);

\draw[color=black] (60,15) node[right] {$\bra{0}$};
\draw[color=black] (60,30) node[right] {$\bra{0}$};
Warning: 1 unused label(s) on wire a1

\draw[color=black] (60,0) node[right] {$\hat{\Phi}(\hat{\rho})$};
\end{tikzpicture}
    \end{minipage}
	\caption{A circuit representation of the Choi-Jamiolkowki method for applying the most general Gaussian CP map to a state. Here, $\ket{0}$ denotes the vacuum state; the vertical line linking two modes with ellipses on the ends indicates the two-mode squeezing operation (in this case, we take the limit as squeezing goes to infinity); and $\hat{\Phi}(\cdot)$ is a general Gaussian CP map applied to one half of the Choi state so that it can be teleported onto the target state $\hat{\rho}$. Since the two-mode squeezing operation is a Gaussian transformation, this circuit can be treated with conditional Gaussian dynamics when $\hat{\rho}$ is a Gaussian state.}
	\label{fig:choi}
\end{figure}
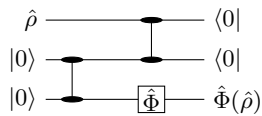

Conditional Gaussian dynamics also opens the door to effecting the most general Gaussian completely positive maps via their Choi state representation~\cite{book_serafini}. A CP map on a state is related by the Choi-Jamiolkowski isomorphism to a Choi state, which is found by acting the CP map on one half of a maximally entangled state. For CV systems, the maximally entangled state is the Gaussian two-mode squeezed state in the limit of infinite squeezing; a two-mode squeezed state can be constructed by interfering $q$- and $p$-squeezed vacuum states at a 50:50 beam-splitter. Consider the two-mode Choi state (covariance matrix $\bm{\Sigma}_{C}$ and mean $\bm{\mu}_C$) associated to the CP map. The CP map can then be teleported onto an arbitrary target Gaussian state $\hat{\rho}$ by projecting $\hat{\rho}$ and the untouched half of the Choi state onto the maximally entangled state, which is again just the two-mode squeezed state. We summarize the circuit in \cref{fig:choi} Since the Choi state and the target are both Gaussian, before the projection they can be expressed in the form of \cref{eq:ABsplit}, with the off-diagonal matrices being $\bm{0}$ since the states have not yet interacted. Because the measurement is also onto a Gaussian state, the final output state has covariances and means of the form \eqref{eq:cond_dyn}. For the exact mapping, one must take care to evaluate the infinite squeezing limit for the two-mode squeezed states in the final expression for the means and covariances. If $\hat{\rho}$ is a multimode state on $n$ modes, the same procedure can be applied using $n$ copies of the maximally entangled state on $2n$ modes~\cite{book_serafini}.

As we show in Section \ref{sec:formalism}, we can take inspiration from the Gaussian phase space mathematical framework we have reviewed to introduce a class of states and measurements that can be expressed as a linear combination of Gaussian functions in phase space, along with how such states transform. Importantly, we find in Section \ref{sec:states} that common bosonic qubit encodings fall within this formalism. Next, we review the definitions and properties of those bosonic qubits.

\subsection{Continuous-Variable Simulation with Strawberry Fields}\label{subsec:sf}

\texttt{Strawberry Fields (SF)} is a full-stack Python library for programming, designing, simulating, and optimizing continuous-variable quantum optical circuits~\cite{Killoran2019,bromley2020applications}. The library has a unified frontend that allows to write CV quantum circuits and programs at a high-level. Moreover, it allows users with basic knowledge about CV and quantum photonics to access an application layer that can be used to solve practical problems in graph theory, point processes and chemistry. The frontend also provides functionality for gate decomposition and program compilation and verification. Once programs are verified and compiled they are passed to a software backend or directly to cloud-available hardware~\cite{arrazola2021quantum}.
The software or quantum photonic hardware can return a number of useful results to the frontend, including batches of samples, cost functions for further numerical optimization or representations of the quantum state. The frontend provides further functionality for exploration such as sample processing and plotting. The three software backends handle the actual simulation using different internal numerical representations of quantum states, each with their own unique advantages and weaknesses.
The \texttt{gaussian} backend simulates Gaussian states undergoing Gaussian and non-Gaussian (threshold and photon-number-resolving) measurements~\cite{gupt2019walrus,quesada2020exact}. The \texttt{fock} and \texttt{tf} backends use a Fock basis truncation to represent CV quantum states and operations as high-dimensional tensors~\cite{miatto2020fast}. They differ in the tools they rely on for the numerical implementation of the tensor operations: the former uses the \texttt{NumPy} package~\cite{harris2020array}, while the latter employs \texttt{TensorFlow}~\cite{abadi2016tensorflow}.

We implement the results of this manuscript as a new, fourth backend of \texttt{SF} \cite{bosonic_github}. This \texttt{bosonic} backend can be regarded as a generalization of the \texttt{gaussian} backend, and integrates directly into the rest of the \texttt{SF} stack. For a series of beginner to advanced tutorials implemented by the authors on using the new backed, see \cite{tutorial1,tutorial2,tutorial3}. The \texttt{bosonic} backend implements much of the formalism and methods we discuss in Sections \ref{sec:formalism} through \ref{sec:sim_methods}, enabling new simulation capabilities while benefiting from the unified high-level frontend functionality of the rest of the library.

\subsection{Bosonic Qubits}\label{subsec:bosonic_qubits}
Residing in a two-dimensional subspace of the infinite-dimensional Hilbert space of a CV mode, the bosonic qubit is robust unit for quantum computation in platforms such as photonics. Several classes of bosonic qubits can moreover correct errors within the CV space, adding an additional level of protection against physical noise. Here, we review the definitions and main properties of two promising encodings: GKP and cat states. Understanding how these states behave in realistic settings is especially valuable as their use becomes more widespread in quantum technologies. As we show in Section \ref{sec:states}, we are able to express these qubits as linear combinations of Gaussian functions in phase space, allowing us to model them under realistic conditions such as loss, finite-energy, and noisy gate teleportations.

\subsubsection{GKP States}
The ideal square-lattice GKP logical states are defined as infinite combs of Dirac delta functions spaced by $2\sqrt{\pi\hbar}$ in the position quadrature:
\begin{equation}\label{eq:idealGKP}
    \ket{k}_{\rm gkp} = \sum_{s = -\infty}^{\infty} \ket{\sqrt{\pi\hbar}(2s+k)}_q, \ k = \{0,1\},
\end{equation}
where $\ket{\cdot}_q$ denotes an eigenstate of the position quadrature and $k$ denotes the logical value. Rectangular and hexagonal lattice encodings are related to the square lattice via symplectic transformations~\cite{GKP_2001}, a mapping that we show falls neatly within our formalism. One advantage of the GKP encoding is that Clifford gates and measurements correspond to Gaussian transformations~\cite{GKP_2001}, which are experimentally accessible in the photonics context, as we review in Appendix \ref{app:mbgates}. Pauli $\X$ and $\Z$ gates correspond to displacements by $\sqrt{\pi\hbar}$ along the $q$ and $p$ quadratures, respectively. The Hadamard gate is a rotation by $\pi/2$ in phase space. The qubit phase, $\CX$, and $\CZ$ gates correspond to a CV quadratic phase, $\CX$, and $\CZ$ gates, which are active Gaussian transformations, in the sense of requiring a squeezing component. In practice, (measurement-based) inline squeezers are challenging to implement but are nonetheless feasible and deterministic, as we discuss in Section \ref{subsec:mbgates}. Pauli $\X$ and $\Z$ measurements correspond to homodyne measurements along $q$ and $p$ quadratures. A universal gate set can be completed with the qubit T gate; in the GKP encoding, this gate can be implemented through gate teleportation with a magic state, a process we review and align with our formalism in Section \ref{subsec:Tgate}.

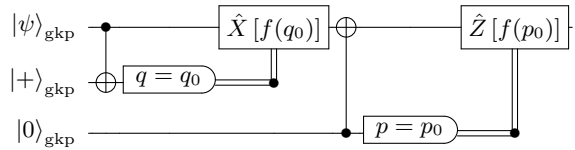
\begin{figure}
\centering
$$
\Qcircuit @C=.15em @R=.7em { 
\lstick{\Ket{\psi}_{\rm gkp}} & \qw & \ctrl{1} & \qw & \qw & \gate{\hat{X}\left[f(q_0)\right]} & \qw & \targ & \qw & \qw & \gate{\hat{Z}\left[f(p_0)\right]} & \qw \\
\lstick{\Ket{+}_{\rm gkp}} & \qw  & \targ & \qw & \measureD{q=q_0} & \cctrl{-1}  \\
\lstick{\Ket{0}_{\rm gkp}} & \qw & \qw & \qw & \qw & \qw & \qw & \ctrl{-2} & \qw & \measureD{p=p_0} & \cctrl{-2}
}
$$
\caption{Error syndrome measurement with normalizable GKP states following the approach from~\cite{GKP_2001}. First, shifts
in $q$ are corrected: an encoded data qubit $\Ket{\psi}_{\rm gkp}$ and
an ancilla $\Ket{+}_{\rm gkp}$ are sent through a SUM gate, and $\Ket{\psi}_{\rm gkp}$
is displaced according to the result of a homodyne $q$ measurement
on the ancilla, mod $\sqrt{\pi\hbar}$. A similar procedure follows for shifts in $p$.  \label{fig:syndrome_meas}}
\end{figure}

GKP states can correct small displacement errors in phase space, and and can reduce larger displacement errors to qubit-level Pauli errors. A qubit error-correction code concatenated with GKP states can then be used to correct these discrete errors~\cite{GKP_2001}. In \cref{fig:syndrome_meas} we review the GKP error-correction circuit from~\cite{GKP_2001}, noting that various other decompositions of the circuit exist~\cite{glancy,Wan2020}. Briefly, two ancillary GKP states are entangled with the data mode to be corrected and measured with homodyne detectors; the outcomes of these measurements determine the displacement that is then applied to the data mode to correct for the error (up to a logical Pauli error). We discuss in Section \ref{subsec:Tgate} how our formalism can treat this circuit.

To the chagrin of experimentalists, ideal GKP qubits have infinite energy; to the chagrin of theorists, we must consider their finite-energy, normalizable forms. One such form is obtained by replacing each Dirac delta with a Gaussian peak corresponding to a squeezed state of variance $\Delta^2/2$, and then applying an overall Gaussian envelope of width $1/\Delta^2$ so that peaks further from the origin are suppressed~\cite{GKP_2001}: 
\begin{align}\label{eq:finiteE_GKP}
\begin{split}
    \ket{k^\Delta}_{\rm gkp} \equiv& \frac{1}{\mathcal{N}_k}\int_{-\infty}^{+\infty}dx\sum_{s}e^{-\Delta^2[(2s+k)\sqrt{\pi}]^2/2} \\&\hspace{1.5cm}\times e^{-[x-(2s+k)\sqrt{\pi\hbar}]^2/2\hbar\Delta^2}\ket{x}_q.
\end{split}
\end{align}
This normalization process is not symmetric in phase space because the peaks are constrained to remain centred at the initial positions of the ideal state~\cite{Matsuura2020}. A related normalization process, which has the Fock damping operator $E(\epsilon) = e^{-\epsilon\hat{n}}$ applied to the ideal state, is symmetric since the number operator $\hat{n}$ acts symmetrically in phase space; we denote such states as $\ket{k^\epsilon}_{\rm gkp}$. In~\cite{Matsuura2020} the authors provide a thorough review of the connections and mappings between these finite energy forms of GKP states, noting that they can be related to each other by a simple squeezing operation; in~\cite{Mensen2020} the authors explore alternative normalization envelopes to Gaussians, demonstrating sufficient conditions for the normalization process to yield physical states. Yet another option for finite energy GKP states are comb states~\cite{shukla2020squeezed}; these correspond to taking only a finite superposition of evenly-weighted $q$-squeezed states centred at the location of the peaks in the ideal state.

\subsubsection{Cat States}\label{subsubsec:cat_background}
To define cat states, one starts with coherent states $\ket{\alpha} = \hat{D}(\alpha)\ket{0} =  e^{-|\alpha|^2/2} \sum_{n=0}^\infty \tfrac{\alpha^n}{\sqrt{n!}}\ket{n}$, $\alpha \in \mathbb{C}$, which are Gaussian states with covariance matrix equal to the vacuum covariance matrix $\bm{\Sigma}_{\text{vac}} = \tfrac{\hbar}{2} \id $ and displacement vector $\bm{\mu} = [\sqrt{2 \hbar} \Re(\alpha), \sqrt{2 \hbar} \Im(\alpha)]$. Ideal (two-lobe) cat states are superpositions of coherent states~\cite{ralph2003}:
\begin{equation}
    \ket{k^\alpha}_{\text{cat}} = \sqrt{\mathcal{N}}(\ket{\alpha} + e^{i \pi k} \ket{-\alpha}),\ k = \{0,1\},
\end{equation}
with normalization
\begin{equation}
    \mathcal{N} = \frac{1}{2(1+e^{-2 |\alpha|^2} \cos(\pi k))}.
\end{equation}
There is some freedom in the choice of logical basis states. For example, for resource-efficient preparation of Pauli $\X$ eigenstates, one can identify them with coherent states $\ket{\pm\alpha}$, which are approximately orthogonal as $\alpha$ increases in magnitude. Then, the Pauli $\Z$ gate is given by a rotation in phase space by $\pi$. Pauli $\Z$ measurements become photon number parity measurements, since the wavefunction for $k = 0 \ (1)$ is symmetric (antisymmetric) and therefore only contains even (odd) photon numbers. A cat-qubit Bell state can be prepared by splitting a higher-energy cat state $\ket{k^{\sqrt{2}\alpha}}_{\text{cat}}$ at a 50:50 beam-splitter. Bell state measurements on cat qubits can be performed by interacting two states at a beam-splitter, then measuring photon number patterns at the output. The Pauli $\X$ gate, small single qubit rotations, and two-qubit entangling gates can all be applied deterministically via gate teleportation, conditional on the availability of cat Bell states~\cite{ralph2003}. In Section \ref{subsec:cats}, we show how cat states can be written as a linear combination of Gaussian functions in phase space, and in Section \ref{subsec:Fock} we show the same for Fock states. This means that our formalism enables simulation of the states, teleportation-based gates and measurements required for quantum computation with cat states.

\section{Linear combinations of Gaussians in phase space}\label{sec:formalism}

\begin{figure*}
	\centering
	\includegraphics[width = 0.97\textwidth]{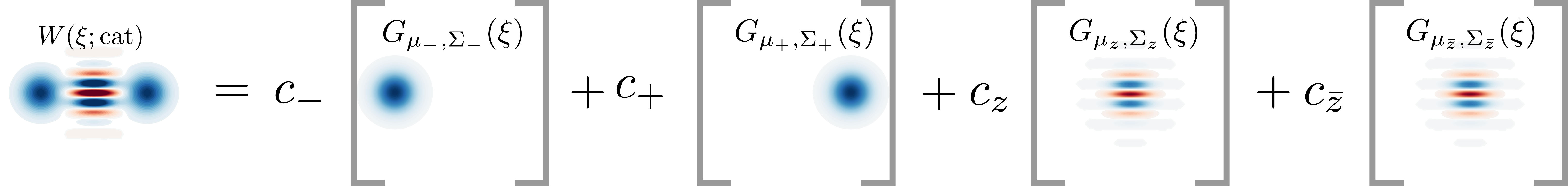}
	\caption{Gaussian decomposition of the Wigner function of the (non-Gaussian) cat state. We use the labels $\mathcal{M} = \{+,-,z,\bar{z} \}$ for the four Gaussians needed. Note that the last two terms have complex coefficients and means and are also complex conjugates of each other; thus the imaginary part of the Gaussians cancel out and we only plot the real part. The details of this particular decomposition are provided in Section~\ref{subsec:cats}. For an introductory tutorial implemented by the authors on using the \texttt{bosonic} backend of \texttt{Strawberry Fields} to obtain this figure, see \cite{tutorial1}.}
	\label{fig:wignerdecomp}
\end{figure*}

Having reviewed the relevant background material on CV Gaussian formalism in Section \ref{subsec:gauss_bg}, we present a new formalism for simulating a wide class of CV states, transformations and measurements. Specifically, in Section \ref{subsec:wigner}, we introduce states with Wigner functions that can be expressed as a linear combination of Gaussian functions in phase space. Next, in Section \ref{subsec:meas}, we provide the framework for describing Gaussian and a class of non-Gaussian measurements on the aforementioned states. Finally, in Section \ref{subsec:gauss_transf}, we detail how the states in our formalism transform under deterministic and conditional Gaussian maps, as well as under a class of non-Gaussian transformations.

\subsection{States in the Wigner Representation}\label{subsec:wigner}
In this work, we consider $n$-mode states whose Wigner function can be written as a linear combination of Gaussian functions in phase space:
\begin{equation} \label{eq:lin_comb_gauss}
        W \left(  \bm{\xi};{\hat{\rho}} \right) = \sum_{m \in {\mathcal M} } c_m G_{ \bm{\mu}_m,\bm{\Sigma}_m} \left( \bm{\xi} \right),
\end{equation}
where ${\cal M}$ is the set of indices which can in general be multiparametered, $\bm{\xi} \in \mathbb{R}^{2n}$ is the phase space variable for an $n$-mode CV quantum system, and $\hat{\rho}$ is the corresponding density matrix operator in Hilbert space. Each Gaussian in the linear combination is associated with a weight $c_m$, a $2n$-dimensional mean $\bm{\mu}_m$ and a $2n\times 2n$ covariance matrix $\bm{\Sigma}_m$, all complex-valued in general.
The normalized multivariate Gaussian distribution $G$ is defined as
\begin{align}
    G_{ \bm{\mu},\bm{\Sigma}} (\bm{\xi}) \equiv \frac{\exp\left[-\tfrac12(\bm{\xi}-\bm{\mu})^T \bm{\Sigma}^{-1} (\bm{\xi}-\bm{\mu})\right]}{\sqrt{\text{det}(2\pi \bm{\Sigma})}}.
\end{align}
If $\hat{\rho}$ is a physical density matrix---in particular, if it has unit trace---then its corresponding Wigner function is normalized through
\begin{align}
    \sum_{m \in \cal{M}} c_m = 1.
\end{align}
At this point, we do not restrict our means, covariances and weights to be real, as imaginary components from these quantities can be required to produce interference fringes and negativity in the Wigner function; however, Hermiticity of the density matrix implies, at least, that the total Wigner function is real. Furthermore, we require the real part of the covariance matrices to be positive-definite so that the distribution remains bounded; however, the covariance matrices need not always respect the uncertainty relation from \cref{eq:UR}. 

As an example of a wide class of states that fall into this framework, consider a pure state that consists of a superposition of Gaussian pure states (i.e. displaced squeezed vacuum states):
\begin{equation}\label{eq:gauss_superposition}
    \ket{\psi} = \sum_{n} \kappa_n \ket{\gamma_n,\zeta_n} = \sum_n \kappa_n \hat{D}(\gamma_n)\hat{S}(\zeta_n)\ket{0}.
\end{equation}
Here, $\gamma_n$ and $\zeta_n$ are the complex displacement and squeezing parameters of the $n^{\text{th}}$ term in the superposition; in this context $\hat{D}(\gamma)$ is the displacement operator and $\hat{S}(\zeta)$ is the squeezing operator (cf. Eq.~\eqref{eq:sq_and_disp}). The density matrix for this state is a sum of terms of the form $\kappa_m \kappa_n^*\ket{\gamma_m,\zeta_m}\bra{\gamma_n,\zeta_n}$. Since the Weyl transform is linear, the Wigner function for the state is a linear combination of functions in phase space, each associated with an operator $\ket{\gamma_m,\zeta_m}\bra{\gamma_n,\zeta_n}$. For $m=n$, the phase space functions are simply products of $|\kappa_n|^2$ and the Wigner function for $\ket{\gamma_n,\zeta_n}$, which is a Gaussian state and hence has a Gaussian Wigner function. In Appendix \ref{app:dyad}, we prove that, for $m\neq n$, the phase space function is still Gaussian, albeit with complex weight, means and covariances. Thus, states of the form \eqref{eq:gauss_superposition} can be expressed in the form \eqref{eq:lin_comb_gauss}.

As we discuss in Section \ref{sec:states}, the representation in \cref{eq:lin_comb_gauss} is useful for describing salient families of continuous variable states that can act as bosonic qubits. Using the derivation from the previous paragraph, as well as additional, tailored derivations, we show how to write GKP and cat qubits as linear combinations of Gaussian functions in phase space. Moreover, we show how Fock states, as well as superpositions of Fock states created by performing photon-number-resolving measurements on some modes of a multimode Gaussian state can be expressed in the form of \cref{eq:lin_comb_gauss}. In addition to giving us the ability to write down states useful for quantum computing, our formalism is well-suited to subjecting such states to general Gaussian transformations and measurements, as well as certain non-Gaussian transformations via gate teleportation, as we explore in the next few sections.

\subsection{Gaussian and a Class of Non-Gaussian Measurements}\label{subsec:meas} 
Given a state whose Wigner can be written as a linear combination of Gaussians in phase space, we now describe the formalism for Gaussian measurements on such states. A general-dyne Gaussian measurement on $n$ modes is characterized by the $2n\times2n$ covariance matrix $\bm{\Sigma}_M$ of the Gaussian state onto which one projects.
The outcome of a general-dyne measurement is a point in phase space, $\bm{r}_M$~\cite{book_serafini}.
Given a state that can be described by \cref{eq:lin_comb_gauss}, the probability of outcome $\bm{r}_M$ is
\begin{equation}\label{eq:outcome_prob}
    \begin{split}
        p \left( \bm{r}_M; {\hat{\rho}}, \bm{\Sigma}_M\right) &= \int d\bm{\xi} \,W \left(  \bm{\xi};{\hat{\rho}} \right) \left(  \bm{\xi} \right) G_{ \bm{r}_M,\bm{\Sigma}_M}\left(  \bm{\xi} \right)\\
        &= \sum_{m \in {\mathcal M} } c_m G_{ \bm{\mu}_m,\bm{\Sigma}_M+\bm{\Sigma}_m}\left(  \bm{r}_M \right).
    \end{split}
\end{equation}

A special case of general-dyne measurement is homodyne measurement. Without loss of generality, we consider a measurement of the $\bm{q}$ quadrature, for which
\begin{equation}
\bm{\Sigma}_M = \lim_{\epsilon\rightarrow 0} \frac{\hbar}{2}\begin{pmatrix}\epsilon\id & \bm{0} \\ \bm{0} & \epsilon^{-1}\id\end{pmatrix},
\end{equation}
and the measurement outcome becomes $\bm{r}_M \rightarrow \bm{q}_M$.
Since the $\bm{q}$-homodyne distribution can be retrieved by integrating out the $\bm{p}$ quadrature, and since the Wigner function is a linear combination of Gaussians, we have
\begin{equation}\label{eq:homodyne}
    p(\bm{q}_M;\hat{\rho}_M) = \sum_{m \in {\mathcal M} } c_m G_{ \bm{\mu}^{(\bm{q})}_m,\bm{\Sigma}^{(\bm{qq})}_m}\left(  \bm{q}_M \right),
\end{equation}
where $\bm{\mu}^{(\bm{q})}_m$ and $\bm{\Sigma}^{(\bm{qq})}_m$ denote the $\bm{q}$-quadrature components of the means and covariances. Importantly, the distribution is yet another linear combination of Gaussians, this time of a variable in $n$ rather than $2n$ dimensions. Later, in Section \ref{sec:sim_methods}, we provide a tailored simulation method for sampling outcomes of Gaussian measurements from states in our formalism.

The mathematical method for calculating the probability distribution of a Gaussian measurement can be straightforwardly generalized to a class of non-Gaussian measurements for which the measurement operator can itself be represented in phase space as a linear combination of Gaussian functions. 
Assuming the Weyl transform of a measurement operator $\hat{M}$ associated with outcome $M$ is of the form
\begin{equation}\label{eq:lin_comb_meas}
    \Phi(\bm{\xi};\hat{M}) = \sum_{j\in\mathcal{J}} d_j G_{\bm{\mu}_j,\bm{\Sigma}_j}(\bm{\xi}),
\end{equation}
the probability of obtaining $M$ is simply given by:

\begin{equation}\label{eq:nG_out_prob}
\begin{split}
       p(M;\hat{\rho}) &= \int d\bm{\xi} \Phi(\bm{\xi};\hat{M})W \left(  \bm{\xi};{\hat{\rho}} \right)\\
       &= \sum_{m\in\mathcal{M}}\sum_{j\in\mathcal{J}} c_m d_j G_{\bm{\mu}_m,\bm{\Sigma}_m+\bm{\Sigma}_j}[\bm{\mu}_j].
\end{split}
\end{equation}
Note that, for measurement operators, we do not necessarily have the unit trace condition, so $\sum_{j\in\mathcal{J}} d_j \neq 1$ in general. In Section \ref{subsec:Fock}, we show, as a pertinent example, how Fock states can be expressed as a linear combination of Gaussians in phase space, which means photon-number-resolving measurements can be described with this formalism.

\subsection{Gaussian and a Class of Non-Gaussian Transformations in a Gaussian-Inspired Framework}\label{subsec:gauss_transf}
As we saw in Section \ref{subsec:gauss_bg}, Gaussian transformations in phase space map Gaussian states to Gaussian states.
As we show next, these maps motivate a wider class of Gaussian and non-Gaussian transformations for states with Wigner functions that can be represented as linear combinations of Gaussian functions, as in \cref{eq:lin_comb_gauss}.

The first class of transformations we consider are deterministic Gaussian completely positive and trace preserving (CPTP) maps, that is, transformations that are not conditioned on any probabilistic measurement outcomes. As we reviewed in Section \ref{subsec:gauss_bg}, deterministic Gaussian transformations acting on an $n$-mode Gaussian state can be parametrized by two $2n\times2n$ matrices, $\bm{X}$ and $\bm{Y}$, and a length-$2n$ vector $\bm{d}$ that together transform the covariance matrix and mean of the Wigner function~\cite{book_serafini}. 
Since the mapping is linear over phase space variables, it generalizes straightforwardly for a linear combination of Gaussian functions in phase space:
\begin{equation}\label{eq:gauss_cptp}
    \begin{split}
        \bm{\Sigma}_m &\rightarrow \bm{X\Sigma}_m \bm{X}^T + \bm{Y},\\
        \bm{\mu}_m &\rightarrow \bm{X\mu}_m + \bm{d}.
    \end{split}
\end{equation}
When $\bm{X}$ is a symplectic matrix and $\bm{Y} =\bm{0}$, this corresponds to a Gaussian unitary transformation. 

Deterministic Gaussian transformations modify neither the number nor the weighting of the peaks in the linear combination, which makes them easy to apply to states of the form \eqref{eq:lin_comb_gauss}. A wide class of CV operations---displacement, squeezing, rotation, and beam-splitters---as well as common noise models---loss and Gaussian random displacements---fall under this umbrella. By contrast, simple transformations such as displacements and squeezing can quickly push Fock distributions beyond the energy cutoff, an important limitation of the Fock representation.

The second class of transformations we consider are conditional dynamics: how a measurement of some modes updates the remaining active modes. In this case, our formalism opens the door to a class of non-Gaussian transformations of the Wigner function; if we take a set of target modes of the form \eqref{eq:lin_comb_gauss} and interact them with a set of non-Gaussian ancillae, also of the form \eqref{eq:lin_comb_gauss}, and then perform a non-Gaussian measurement on the ancillary modes of the form \eqref{eq:lin_comb_meas}, then the effective transformation on the target modes is non-Gaussian in general. This conclusion applies even if the ancillary modes or the measurement---but not both---are Gaussian. Effecting a non-Gaussian transformation on a state through an interaction with non-Gaussian ancillary modes or through a non-Gaussian measurement has been studied extensively in the context of CV gate teleportation and state preparation~\cite{GKP_2001,Ghose2007,Sefi2013,Marshall2015,Miyata2016,Arzani2017,Sabapathy2019,Su2019,Tzitrin2020}. However, we show next that we can represent such non-Gaussian transformations in the spirit of conditional Gaussian dynamics reviewed in Section \ref{subsec:gauss_bg}.

To understand these Gaussian-inspired but nonetheless non-Gaussian transformations, consider two sets of modes: the set $A$ of active modes, described initially by a Wigner function of the form \eqref{eq:lin_comb_gauss} with weights $a_\ell$, means $\bm{\mu}_\ell$ and covariances $\bm{\Sigma}_\ell$, for $\ell\in \mathcal{L}$; and the set $B$ of modes that will eventually be measured, with corresponding parameters $b_k$, $\bm{\mu}_k$ and $\bm{\Sigma}_k$, for $k\in\mathcal{K}$. If the two sets of modes are entangled via a deterministic Gaussian transformation parametrized by $(\bm{X},\bm{Y},\bm{d})$, the weights, means and covariances become:
\begin{equation}
    \begin{split}
        c_m &= a_\ell b_k,\ m = (\ell,k)\in \mathcal{M} = (\mathcal{L},\mathcal{K}),\\
        \bm{\mu}_m &= \bm{X}(\bm{\mu}_\ell\oplus\bm{\mu}_k) + \bm{d},\\
        \bm{\Sigma}_m &= \bm{X}(\bm{\Sigma}_\ell\oplus\bm{\Sigma}_k)\bm{X}^T + \bm{Y}.
    \end{split}
\end{equation}
In turn, we can express the means and covariances as
\begin{equation}\label{eq:AB_mean_cov}
        \bm{\Sigma}_m = \begin{pmatrix}\bm{\Sigma}_{m,A} & \bm{\Sigma}_{m,AB}\\
        \bm{\Sigma}_{m,AB}^T & \bm{\Sigma}_{m,B}\end{pmatrix},\ \bm{\mu}_m = \begin{pmatrix}\bm{\mu}_{m,A} \\ \bm{\mu}_{m,B}\end{pmatrix},
\end{equation}
where $A$ and $B$ indicate the active and measured modes, respectively. 

Consider now a measurement $\hat{M}$ with outcome $M$ on modes $B$ with a phase space representation of the form from \cref{eq:lin_comb_meas}, parametrized by weights $d_j$, means $\bm{\mu}_j$ and covariances $\bm{\Sigma}_j$, with $j\in\mathcal{J}$. Since the partial trace is linear, the corresponding phase space integral is a sum of many partial Gaussian integrals. Thus the covariances and means of modes $A$ update as~\cite{book_serafini}:
\begin{equation}\label{eq:update_covs_means}
    \begin{split}
        \bm{\Sigma}_{m,A} &\rightarrow \bm{\Sigma}_{m,A} - \bm{\Sigma}_{m,AB}(\bm{\Sigma}_{m,B}+\bm{\Sigma}_j)^{-1}\bm{\Sigma}_{m,AB}^T, \\
        \bm{\mu}_{m,A} &\rightarrow  \bm{\mu}_{m,A} + \bm{\Sigma}_{m,AB}(\bm{\Sigma}_{m,B}+\bm{\Sigma}_j)^{-1}(\bm{\mu}_j - \bm{\mu}_{m,B}).
    \end{split}
\end{equation}
Until this point, the update rules have been inspired by the conditional dynamics for Gaussian states. However, our consideration of multiple Gaussians instead of just one necessitates a novel rule: an expansion and re-weighting of the peaks in phase space. While outcome $M$ occurs with probability $p(M;\hat{\rho}_{AB})$, as in \cref{eq:nG_out_prob}, each peak from modes $B$ and from the measurement operator contribute differently to this probability, with a weight given by:
\begin{equation}
    w(M|\bm{\mu}_{m,B},\bm{\Sigma}_{m,B},\bm{\mu}_j,\bm{\Sigma}_j) = c_m d_j G_{ \bm{\mu}_{m,B},\bm{\bm{\Sigma}_{m,B}+\Sigma}_j}\left(  \bm{\mu}_{j} \right).
\end{equation}
Therefore, given the result $M$, the weights update as:
\begin{equation}\label{eq:update_weights}
    c_m \rightarrow \gamma_{m,j} =  \frac{w(M|\bm{\mu}_{m,B},\bm{\Sigma}_{m,B},\bm{\mu}_j,\bm{\Sigma}_j)}{p(M;\hat{\rho}_{AB})},
\end{equation}
where we include a normalization by all the new weights. As a result of this process, the total number of Gaussians in the Wigner function for modes $A$, as well as their associated weights, means and covariances, is been grown both by modes $B$ and by the measurement $M$:
\begin{equation}
    \ell \xrightarrow[]{B} m = (\ell,k)\xrightarrow[]{M} (m,j) = (\ell,k,j) \in(\mathcal{L},\mathcal{K},\mathcal{J}),
\end{equation}
such that the final number of Gaussian functions required to describe mode $A$ is the product of the initial number of functions in modes $A$ and $B$, and in the Weyl transform of $\hat{M}$. 

While we have only been tracking transformations of Gaussian functions in phase space, it is worth emphasizing that these conditional dynamics increase the number of Gaussian functions initially in modes $A$, thereby necessarily describing non-Gaussian transformations. The cost of modelling non-Gaussian transformations with our formalism is that the number of peaks one needs to consider can grow; still, we show that this is a worthwhile trade-off compared to alternative methods for simulating certain classes of bosonic qubits.

In the case that modes $B$ and the measurement are truly Gaussian states, as opposed to linear combinations of Gaussian functions, then $b_k = d_j = 1$, and $(\mathcal{L},\mathcal{K},\mathcal{J}) \rightarrow \mathcal{L}$. This means the initial number of weights does \textit{not} increase, and the initial means $\bm{\mu}_\ell$ and covariances $\bm{\Sigma}_\ell$ follow the traditional Gaussian conditional update rule reviewed in Section \ref{subsec:gauss_bg}:
\begin{equation}\label{eq:update_covs_means_gauss}
    \begin{split}
        \bm{\Sigma}_{m,A} &\rightarrow \bm{\Sigma}_{m,A} - \bm{\Sigma}_{m,AB}(\bm{\Sigma}_{m,B}+\bm{\Sigma}_M)^{-1}\bm{\Sigma}_{m,AB}^T, \\
        \bm{\mu}_{m,A} &\rightarrow  \bm{\mu}_{m,A} + \bm{\Sigma}_{m,AB}(\bm{\Sigma}_{m,B}+\bm{\Sigma}_M)^{-1}(\bm{r}_M - \bm{\mu}_{m,B}),
    \end{split}
\end{equation}
where $\bm{\Sigma}_M$ and $\bm{r}_M$ parametrize the Gaussian measurement as in \cref{eq:outcome_prob}. The peak reweighting from \cref{eq:update_weights} is still required; however, it is simpler because the total number of weights does not increase and $d_j = 1$. Finally, we point out that since we can implement conditional Gaussian dynamics in this framework, we can apply the most general Gaussian CP maps via the Choi-Jamiolkowski isomorphism using the circuit in \cref{fig:choi}.

In Section \ref{sec:ex_transform}, we explore how various transformations such as loss, Fock damping, measurement-based squeezing, and gate teleportation onto bosonic qubits can be described in this formalism. Before this, in the following section, we present the description common bosonic states as linear combinations of Gaussians in phase space.

\section{Useful Families of States Expressed as Linear Combinations of Gaussians}\label{sec:states}
Having provided a general formalism in Section \ref{sec:formalism}, we now demonstrate how bosonic qubits, namely GKP, cat and Fock states, can be written in the form of Eq. \ref{eq:lin_comb_gauss}.

\subsection{GKP States}\label{subsec:GKP}
Care must be taken when dealing with the finite-energy forms of GKP states. Before we tackle those, let us recall the Wigner representation of the ideal states.

\subsubsection{Ideal GKP States}
Any pure GKP state can be expressed in the Bloch sphere representation as
\begin{equation}\label{eq:gkp_ideal_psi}
    \vert \psi \rangle = \cos{\tfrac{\theta}{2}} \vert 0 \rangle_{\rm gkp} + e^{-i \phi} \sin{\tfrac{\theta}{2}} \vert 1 \rangle_{\rm gkp}.
\end{equation}
The Wigner function of an ideal single-mode square-lattice GKP state can be expressed in terms of the parameters of \cref{eq:lin_comb_gauss} in the following way~\cite{GKP_2001}:
\begin{align}\label{eq:gkp_ideal}
    &\mathcal{M} =\{ m \equiv (k \sqrt{\pi\hbar}/2, \ell \sqrt{\pi\hbar}/2)\,|\, k, \ell \in \mathbb{Z}\}, \nonumber\\
    & \bm{\Sigma}_m = \lim_{\delta \to 0+} \delta \id, \ \bm{\mu}_m = m.
\end{align}
The weights $c_m(\theta,\phi)$---which encode the logical content of the state---are presented in  Appendix\,\ref{anc:coeffs} as given in Ref.~\cite{Garcia_2021}. Recall each Gaussian peak in the Wigner function is a function of $\bm{\Sigma}_m^{-1}$, so we cannot directly evaluate the limit as $\delta \rightarrow 0+$. The Gaussian distributions for the ideal GKP Wigner function are Dirac delta functions located at the lattice points enumerated in $\cal{M}$. This means that the ideal GKP state has infinite energy and cannot be normalized, rendering it unphysical. However, the covariances matrices do not vary with the index and are proportional to the identity, a feature that is put to use in the following section. Note that GKP states corresponding to alternative lattice spacings, such as rectangular or hexagonal GKP states, can be related to square-lattice states by symplectic transformations in phase space~\cite{GKP_2001}. As we showed in Section \ref{subsec:gauss_transf}, symplectic transformations are straightforward to apply to states that can be expressed as linear combinations of Gaussians in phase space, so it is easy to transform between lattices in our framework. Additionally, GKP qudits simply correspond to a different linear combination of $\delta$-functions in phase space, so our formalism can be employed to treat those states as well.

\subsubsection{Finite-Energy GKP States}
There are different ways to obtain finite-energy versions of the GKP states. A summary of these  is presented in Fig. 1 of Ref.~\cite{Tzitrin2020}. Here, we focus on a commonly-used model of finite-energy GKP states: a Fock damping operator 
\begin{equation}\label{eq:fock_damping}
    \hat{E}(\epsilon ) = e^{-\epsilon \hat n},\;\epsilon > 0,
\end{equation}
applied to the ideal GKP state in \cref{eq:idealGKP}. Because $\hat{E}(\epsilon )$ is a single non-unitary Kraus operator, it is non-trace-preserving, meaning one needs to carefully account for the normalization of the resulting states.

A straightforward but somewhat lengthy calculation (expounded in Appendix~\ref{subann:GKP_Moyal}) shows that the  Wigner function corresponding to the $\hat{E}(\epsilon)$ operator applied to an ideal state can be represented in the same form as Eq.\,\eqref{eq:lin_comb_gauss}. With the set ${\cal M}$, ideal coefficients $c_m(\theta,\phi)$, and ideal means $\bm{\mu}_m$ given in Eq.\,\eqref{eq:gkp_ideal}, we have that:
\begin{align} \label{eq:gkp_wigner_explicit}
    &c_{m}(\epsilon;\theta,\phi) = \frac{c_m(\theta,\phi)}{\mathcal{N}_\epsilon} \exp\left[- \frac{1-e^{-2\epsilon}}{\hbar(1+e^{-2\epsilon})} \bm{\mu}_m^T\bm{\mu}_m\right]\nonumber,\\
    & \bm{\mu}_m(\epsilon) = \frac{2e^{-\epsilon}}{1+e^{-2\epsilon}} \bm{\mu}_m,\ \bm{\Sigma}_m(\epsilon) =  \frac{\hbar}{2}\frac{1-e^{-2\epsilon}}{1+e^{-2\epsilon}}\id.
\end{align}
Here $\mathcal{N}_\epsilon$ is chosen such that $\sum_{m\in\mathcal{M}}c_{m}(\epsilon;\theta,\phi) = 1$. In this representation, the weights and means are real. In practice, one can take a finite number of terms from $\mathcal{M}$---a numerical cutoff---depending on the desired precision. The value of the cutoff would depend on the strengths of the weights and the normalization of the Gaussian distributions in the Wigner expansion. A simpler derivation motivated by the dilation of the operator $\hat{E}(\epsilon)$ is provided in Appendix \ref{subann:GKP_optical}.

We compare the model for finite-energy GKP states we have derived in Eq. \ref{eq:gkp_wigner_explicit} to the approach from \cite{Menicucci2014}. There, the Dirac delta spikes of the ideal GKP Wigner function are replaced with Gaussians of non-zero variance. While that approach can capture the broadening of peaks due to finite-energy and other noise effects, and while the covariance matrix of each peak can be updated as Gaussian channels are applied, the state does not have an envelope, so it still has infinite energy and remains unphysical. Moreover, in that model, the locations of peaks are not tracked, so their contraction towards the origin due to finite energy effects and any shifts under Gaussian channels are ignored. This can have an impact on the estimation of qubit-level errors. By contrast, the mathematical form for finite-energy GKP states provided here offers a rich noise model that captures these effects.

For an alternative phase-space representation of finite-energy GKP states, one can apply $\hat{E}(\epsilon)$ to the wavefunction to obtain a superposition of squeezed states, and then use this form to calculate the Wigner function directly. This representation differs from the one previously presented in two ways. First, the weights and means of the resulting Wigner function are complex rather than real numbers. Second, the covariances of the individual Gaussian functions now respect the uncertainty relation. Since in this representation, each peak is directly mappable to squeezed states in the wave function, it can be more useful for converting effects observed at the wavefunction level to phase space transformations. The derivation of this alternative representation is provided in Appendix\,\ref{subann:GKP_alt}.

Let us denote a logical GKP qubit by $\ket{\psi(\bm{a})} = a_0 |0\rangle_{\rm gkp} + a_1 |1\rangle_{\rm gkp}$. We find that the parameters from  \cref{eq:lin_comb_gauss} become

\begin{equation}
    \begin{split}
        &\mathcal{M} = \{m\equiv (k,\ell,s,t) \, | \, s,t \in \{ 0,1\} \,\& \, k, \ell \in \mathbb{Z} \} ,\\
        &\bm{\mu}_{m}(\epsilon) = -\frac{\beta \sqrt{\pi\hbar}}{2}\begin{pmatrix}\frac{ (t+s+2\ell+2k)}{\alpha}\\  i(s-t+2\ell-2k)\end{pmatrix},\\
        &\bm{\Sigma}_m(\epsilon) = \frac{\hbar}{2}\begin{pmatrix}
                                       1/\alpha & 0\\
                                       0 & \alpha
                                        \end{pmatrix},\ (\alpha,\beta) = (\coth(\epsilon), -\csch(\epsilon)),\\
        &c_m= \frac{a_s a^*_t}{\mathcal{N}(\bm{a},\epsilon)}\exp\left\{-\frac{\alpha\pi}{2} \left[(s+2\ell)^2 + (t+2k)^2\right]\right\}\\
        &\hspace{1.5cm}\times\exp\left[ \frac{\beta^2\pi(t+s+2\ell+2k)^2}{4\alpha}\right].
    \end{split}
\end{equation}
Here $\mathcal{N}{(\bm{a},\epsilon)}$ is an overall normalization constant such that $\sum_{m\in\mathcal{M}}c_m = 1$; it depends on the choice of the GKP state and the strength of the Fock-damping operator applied to it. This time, the Gaussians all have the same covariance matrix, and this matrix respects the uncertainty relation\,\cite{simon94}. While the coefficients can be complex, this only stems from the initial qubit coefficients $a_0, a_1$. Finally, we can see that the means $\bm{\mu}_m{(\epsilon)}$ are complex; the resulting sinusoidal oscillations in phase space are what generate the interference fringes and negative regions of the Wigner function in this representation.

\subsubsection{Squeezed Comb States}
Squeezed comb states form another class of finite-energy
GKP states~\cite{shukla2020squeezed}. These states can be obtained from the ideal GKP states by applying a symmetric step function in the position basis to pick out a certain number of peaks and then replacing the infinitely squeezed position kets with their finitely squeezed analogues.

Mathematically, squeezed comb states can be written in the following way~\cite{shukla2020squeezed}:
\begin{align}
\ket{0}_{\text{Comb}} &= \frac{1}{\sqrt{\mathcal{N}}} \sum_{n=1}^N \ket{\psi_n}, \\
\ket{1}_{\text{Comb}} &= \hat{D}\left(\frac{d}{2\sqrt{2 \hbar}}\right) \ket{0}_{\text{Comb}},
\end{align}
where
\begin{equation}
\ket{\psi_n} = \hat{D}\left(\frac{\bar{q}_n}{\sqrt{2 \hbar}}\right) \hat{S}(r) \ket{0}
\end{equation}
and $\bar{q}_n = -(N+1)\tfrac{d}{2} + n d$. The normalizaton constant is given by
\begin{align}
\mathcal{N} = \sum_{k,l=1}^N \exp\left(-\tfrac{1}{4 \hbar } e^{2r}\left[\bar{q}_k - \bar{q}_l \right]^2 \right).
\end{align}
We can then write
\begin{align}
\pr{0}_{\text{Comb}}  = \frac{1}{\mathcal{N}} \sum_{n,m=1}^N \ket{\psi_n} \bra{\psi_m}.
\end{align}
Using the linearity of the mapping between density operators and Wigner functions in Eq.~\eqref{eq:wigner} and the results from Appendix~\ref{app:dyad}, we find the Wigner function has the canonical form \eqref{eq:lin_comb_gauss} with
\begin{align}
\mathcal{M} &= \{m\equiv (k,l)| k,l \in 1\ldots,d \}, \\
\bm{\mu}_m &= \tfrac{1}{2}\left(\bar{q}_k+\bar{q}_l, i e^{2r}[\bar{q}_k-\bar{q}_l] \right), \\
\bm{\Sigma}_m &= \frac{\hbar}{2} \begin{bmatrix}
e^{-2r} & 0\\
0 & e^{2 r}
\end{bmatrix}\\
c_m &= \frac{1}{\mathcal{N}} \exp\left( -\tfrac{1}{4\hbar} e^{2r} (\bar{q}_k-\bar{q}_l)^2 \right).
\end{align}
For a representation of comb states in terms of real-valued Gaussian functions in phase space, see Appendix~\ref{app:comb}.

\subsection{Cat States}\label{subsec:cats}

As is the case with squeezed comb states, cat states are linear superpositions of pure Gaussian states. We can write the density matrix of these states as
\begin{align}
\pr{k^\alpha}_{\text{cat}}=& \mathcal{N} \Big( \proj{\alpha}+\proj{-\alpha} \\ \Big.
& \quad \Big.+  \ket{\alpha} \bra{-\alpha} e^{-i \pi k } + \ket{-\alpha} \bra{\alpha} e^{i \pi k } \Big). \nonumber
\end{align}
Again by linearity of the mapping between density operators and Wigner functions, we see that the Wigner functions of the first two terms $\proj{\pm \alpha}$ are Gaussian functions with the covariance matrix of the vacuum and displacement $\bm{\mu}_\pm = \pm \sqrt{2 \hbar} ( \Re(\alpha), \Im(\alpha) )$.
Using the results from Appendix \ref{app:dyad} we find the Wigner function of  $\ket{\alpha}\bra{-\alpha}$  and $\ket{-\alpha}\bra{\alpha}$ to be complex-valued Gaussians with prefactor $e^{-2|\alpha|^2}$, vacuum covariance matrix, and complex-valued vectors of means $\bm{\mu}_z = \sqrt{2 \hbar} ( i \Im(\alpha),-i \Re(\alpha))$ with $\bm{\mu}_{\bar{z}} = \bm{\mu}_{{z}}^*$. From these results, we have that
\begin{align} \label{eqn:Wigner_Cat}
\mathcal M &=  \{+,-,z,\bar{z} \}, \nonumber \\
c_{\pm} &= \mathcal{N}, \quad c_{z} = \left( c_{\bar{z}} \right)^* = e^{-i \pi k -2|\alpha|^2  }\mathcal{N}  \nonumber \\
\bm{\mu}_{\pm} &= \pm \sqrt{2 \hbar} ( \Re(\alpha), \Im(\alpha) ), \nonumber \\
\bm{\mu}_{z} &= (\bm{\mu}_{\bar{z}} )^* = \sqrt{2 \hbar} ( i \Im(\alpha),-i \Re(\alpha)),
\end{align}
and where all the Gaussian functions have the same vacuum covariance matrix given in~\eqref{eq:vac}.

Cat states can also be approximated, to arbitrary precision, by a linear combination of Gaussian functions where the weights, means, and covariance matrices are all real. The trade-off for this real-valued representation is an increase in the number of terms in the linear combination. The derivation, shown in Appendix~\ref{app:realCat}, leads to 

\begin{align} 
\begin{split}
\label{eqn:Wigner_Cat_real}
\mathcal M &= \{+,-\} \cup \{ m \;|\; m \in \mathbb Z \}, \\
c_m &= \mathcal N \frac{\sqrt{\pi \hbar} e^{\frac{ \pi^2 D}{4}}}{4 \alpha \left(\Sigma_{\alpha} + v \right)^{\frac{1}{2}}} e^{- \frac{ {b}^2_{m/2}}{2 (\Sigma_{\alpha} + v)}} \\
& \times \begin{cases} (-1)^{m/2} \cos \phi \quad \text{ if } m/2 \in \mathbb Z, \\ (-1)^{\frac{m-1}{2}+1} \sin \phi \quad \text{ otherwise,}  \end{cases} \\
\bm \mu_m &= \left( 0, \cfrac{4 \alpha \sqrt{\hbar} m}{\sqrt{2} \pi D \left(16 \alpha^2 + 2 D \right)} \right), \\
\bm \Sigma_m &= \begin{pmatrix} v & 0 \\ 0 & \frac{\Sigma_{\alpha} + v}{v \Sigma_{\alpha}} \end{pmatrix}, \quad v = \frac{\hbar}{2}, \\
\Sigma_{\alpha} &= \frac{\pi^2 \hbar D}{16 \alpha^2}, \quad b_m = \frac{\pi \sqrt{\hbar} m}{2 \sqrt{2} \alpha},
\end{split}
\end{align}
where $\alpha \in \mathbb R_{\neq 0}$, $c_{\pm}, \bm{\mu}_{\pm}$ and $\bm{\Sigma}_{\pm}$ are the same as in Eq.~\eqref{eqn:Wigner_Cat}, and $D$ is a positive real parameter controlling the precision of this approximation through $\mathcal O (e^{-2 \pi D^2})$.
The general case of a cat state with $\alpha = |\alpha| e^{i \theta}$ can be understood as a $\theta$-rotation in phase space of a cat state with parameter $|\alpha|$; as rotations correspond to symplectic transformation, we know from Section~\ref{subsec:gauss_transf} that this is easily implementable. Alternatively, a generalization of the description presented in Eq.~\eqref{eqn:Wigner_Cat_real} can be found in Appendix~\ref{app:realCat}.

\subsection{Fock States}\label{subsec:Fock}

\begin{figure}[!t]
	\begin{minipage}{0.15\textwidth}
		(a)
	\providecommand{\ket}[1]{\left|#1\right\rangle}
\providecommand{\bra}[1]{\left\langle#1\right|}
\begin{tikzpicture}[scale=1.000000,x=1pt,y=1pt]
\filldraw[color=white] (0.000000, -7.500000) rectangle (18.000000, 22.500000);
\draw[color=black] (0.000000,15.000000) -- (18.000000,15.000000);
\draw[color=black] (0.000000,15.000000) node[left] {$\ket{\psi}$};
\draw[color=black] (0.000000,0.000000) -- (18.000000,0.000000);
\draw[color=black] (0.000000,0.000000) node[left] {$\ket{0}$};
\draw (9.000000,15.000000) -- (9.000000,0.000000);
\filldraw (9.000000, 15.000000) ellipse(4pt and 1pt);
\filldraw (9.000000, 0.000000) ellipse(4pt and 1pt);
\draw[color=black] (18.000000,15.000000) node[right] {$\hat{a}_0^\dagger \ket{\psi}$};
Warning: 1 unused label(s) on wire a1

\draw[color=black] (18.000000,0.000000) node[right] {$\bra{1}$};
\end{tikzpicture}
\end{minipage}
\begin{minipage}{0.25\textwidth}
	(b)
	\providecommand{\ket}[1]{\left|#1\right\rangle}
\providecommand{\bra}[1]{\left\langle#1\right|}
\begin{tikzpicture}[scale=1.000000,x=1pt,y=1pt]
\filldraw[color=white] (0.000000, -7.500000) rectangle (72.000000, 82.500000);
\draw[color=black] (0.000000,75.000000) -- (72.000000,75.000000);
\draw[color=black] (0.000000,75.000000) node[left] {$\ket{\psi}$};
\draw[color=black] (0.000000,60.000000) -- (72.000000,60.000000);
\draw[color=black] (0.000000,60.000000) node[left] {$\ket{0_1}$};
\draw[color=black] (0.000000,45.000000) -- (72.000000,45.000000);
\draw[color=black] (0.000000,45.000000) node[left] {$\ket{0_2}$};
\draw[color=black] (0.000000,30.000000) -- (72.000000,30.000000);
\draw[color=black] (0.000000,30.000000) node[left] {$\ket{0_3}$};
\draw[color=white] (0.000000,15.000000) -- (72.000000,15.000000);
\draw[color=black] (-7.000000,17.000000) node[left] {$\vdots$};
\draw[color=black] (0.000000,0.000000) -- (72.000000,0.000000);
\draw[color=black] (0.000000,0.000000) node[left] {$\ket{0_n}$};
\draw (9.000000,75.000000) -- (9.000000,60.000000);
\filldraw (9.000000, 75.000000) ellipse(4pt and 1pt);
\filldraw (9.000000, 60.000000) ellipse(4pt and 1pt);
\draw (27.000000,75.000000) -- (27.000000,45.000000);
\filldraw (27.000000, 75.000000) ellipse(4pt and 1pt);
\filldraw (27.000000, 45.000000) ellipse(4pt and 1pt);
\draw (45.000000,75.000000) -- (45.000000,30.000000);
\filldraw (45.000000, 75.000000) ellipse(4pt and 1pt);
\filldraw (45.000000, 30.000000) ellipse(4pt and 1pt);
\draw (63.000000,75.000000) -- (63.000000,0.000000);
\filldraw (63.000000, 75.000000) ellipse(4pt and 1pt);
\filldraw (63.000000, 0.000000) ellipse(4pt and 1pt);
\draw[color=black] (72.000000,75.000000) node[right] {$\hat{a}_0^{\dagger n}\ket{\psi}$};
Warning: 1 unused label(s) on wire a1

\draw[color=black] (72.000000,60.000000) node[right] {$\bra{1_1}$};
\draw[color=black] (72.000000,45.000000) node[right] {$\bra{1_2}$};
\draw[color=black] (72.000000,30.000000) node[right] {$\bra{1_3}$};
\draw[color=black] (79.000000,17.000000) node[right] {$\vdots$};
\draw[color=black] (72.000000,0.000000) node[right] {$\bra{1_n}$};
\end{tikzpicture}
\end{minipage}
	\caption{\label{additioncirc} Heralding scheme to do (a) single- and (b) many-photon addition. The vertical line with two ellipses at the end is used to denote a two-mode squeezing operation. Note that we always postselect on a single click for all but the first mode. }
\end{figure}
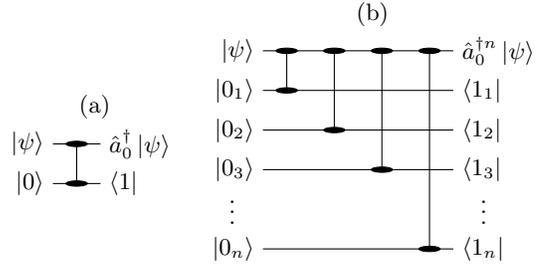

In this section we consider the representation of Fock states as linear combinations of Gaussians. Compared with GKP and cat states, it is less clear whether Fock states can be can be written this way, since they cannot be represented as discrete coherent superpositions of Gaussian states. To obtain a representation for Fock states we use the idea of photon addition; that is, we study a quantum-optical circuit in which postselected heralded outcomes in certain ancillary modes allow us to apply the creation operation of a given mode to an arbitrary input state. 

We first consider photon addition applied to the vacuum state for the generation of a single photon, as considered in Ref.~\cite{fiuravsek2005conditional,Quesada_2019,horoshko2019thermal,thomas2020general}. The circuit for the addition of a single photon is shown in \cref{additioncirc} (a), where the vertical line with ellipses on the ends corresponds to a two-mode squeezing operation $\hat{S}_{0,1}(r) = \exp\left( r \left[\hat{ a}_0^\dagger \hat{a}_1^\dagger - \hat{ a}_0 \hat{a}_1  \right] \right)$ with squeezing parameter $r\ll 1$. To see how it works, we simply need to calculate
\begin{align}
\Big(\mathbb{\hat{I}}_0 & \otimes \pr{1_1} \Big) \hat{S}^{(2)}_{0,1}(r) \left( \ket{\psi_0} \otimes \ket{0_1} \right) \nonumber \\
\approx& \left(\mathbb{\hat{I}}_0 \otimes \pr{1_1} \right) (\mathbb{\hat{I}}+r \hat{a}_0^\dagger \hat{a}_1^\dagger) \left( \ket{\psi_0} \otimes \ket{0_1} \right) \\
=&r \left(\hat{a}_0^\dagger \ket{\psi_0} \right) \otimes \ket{1_1},
\end{align}
where $\mathbb{\hat{I}}_j$ is the identity operation in the Hilbert space of mode $j$.
If $\ket{\psi} = \ket{0_0}$ is the vacuum state of mode 0, then the state at the output is a single photon in this mode.
Since we are working in the regime where $r \ll 1$, to make progress we replace the photon-number-resolving detection $\pr{1}$ by its poor-man's version, $\mathbb{\hat{I}} - \pr{0}$, the threshold detection. 

We can now write the probability of successful and failed heralding more formally as
\begin{gather}
p_{1} = \text{tr}\left( \left[\mathbb{\hat{I}}_1 - \pr{0_1}\right] \pr{\Psi} \right) = 1-
p_{0},\\
p_0  = \frac{1}{1+\bar{n} },
\end{gather}
where $\ket{\Psi} = \hat{S}_{0,1}(r) \ket{0_0 0_1}$ and $\bar{n}=\sinh^2 r$ is the mean photon number in either of the two modes, 0 or 1. The state conditioned on successful heralding with threshold detectors in mode 0 is
\begin{align}\label{eq:fock1_approx}
\pr{1} \approx \hat{\rho}_{1} &= \frac{ \text{tr}_1 \left(  \left[\mathbb{\hat{I}}_1 - \pr{0_1}\right] \pr{\Psi} \right) }{p_1}\\
&= \frac{\hat{\rho}^{\text{th}} - p_0 \pr{0}}{p_1},
\end{align}
which is a linear combination of density matrices for two Gaussian states, namely, a thermal state with mean photon number $\bar{n}$ and the single-mode vacuum. 
A thermal state with mean photon number $\braket{\hat{a}^\dagger \hat{a}} = \braket{\hat{n}} = \bar{n}$ is a mixed Gaussian state with zero mean and covariance matrix $\bm{\Sigma} = \hbar ({\bar n} + \tfrac12) \id$ and can be expressed in the Fock basis as~\cite{weedbrook2012gaussian}
\begin{align}\label{eq:thermalstate}
\hat{\rho}^{\text{th}} = \frac{1}{1+ \bar{n}}  e^{-\epsilon \hat{n}} = \frac{1}{1+ \bar{n}}  \sum_{m=0}^\infty e^{-\epsilon m} \pr{m},
\end{align}
where $\bar{n}=[e^{\epsilon} -1]^{-1}$ or, equivalently,  $e^{\epsilon}=1+1/\bar{n}$.
With this expression one can easily confirm that $\hat{\rho}_{1}$ in Eq.~\eqref{eq:fock1_approx} approaches a single photon in the limit $r \to 0$.

\begin{figure}
	\centering
	\includegraphics[width = \columnwidth]{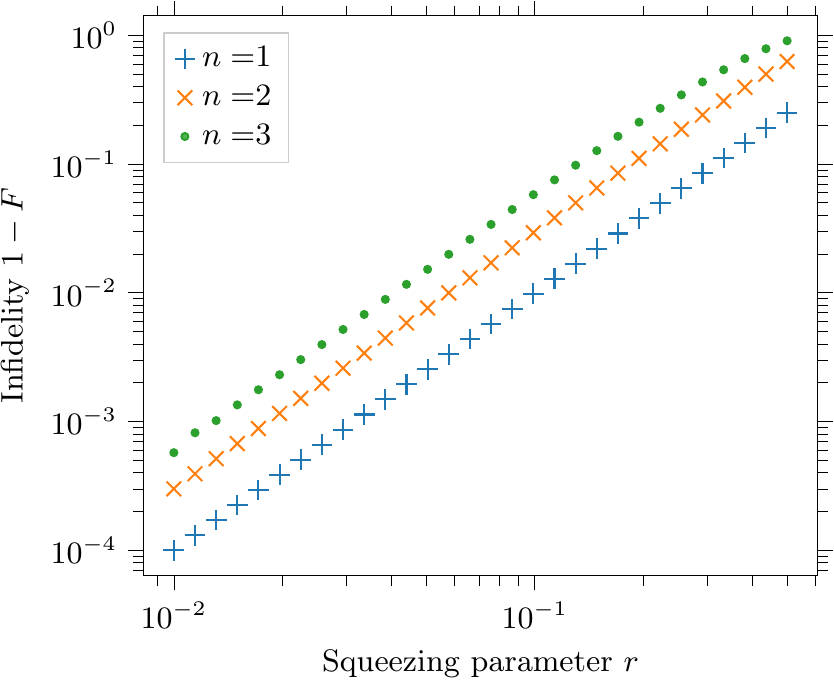}
	\caption{Infidelity between the Gaussian approximation of a Fock state $\hat{\rho}_n$ with Wigner function described in Eq.~\eqref{eq:Fockn} and an exact Fock state $\ket{n}$ as a function of the squeezing parameter $r$. The fidelity is defined as $F =  \braket{n|{\hat{\rho}}_n|n}$.}
	\label{fig:fockinfidelities}
\end{figure}

We generalize this scheme to an $n$-photon addition that, as schematically shown in Fig.~\ref{additioncirc} (b), gives a Fock state with $n$ photons  when applied to the vacuum. The mathematical details of the derivation are provided in Appendix \ref{app:Fock}; the final result is that we can approximate any $n$-particle Fock state using the general notation of Eq.~\eqref{eq:lin_comb_gauss} with
\begin{equation}\label{eq:Fockn}
\begin{split} 
\mathcal{M} &= \{0,\ldots, n\}, \\
c_m &=   \frac{(-1)^{n-m}}{\mathcal{N}_n} {\binom{n}{m}}  \left[
\frac{1-n r^2}{1-(n-m)r^2} \right], \\
\bm{\Sigma}_m &= \frac{\hbar}{2} \frac{1+(n-m)r^2}{1-(n-m)r^2} \id, \\
\bm{\mu}_m &= \bm{0}, \\
\mathcal{N}_n &= \frac{ n! \left(n
	\left( \frac{-r^2-1}{r^2} \right)!+\left(-\frac{1}{r^2
	}\right)!\right)}{ \left( \frac{n r^2-1}{r^2} \right)!} .
\end{split}
\end{equation}
Although the equations above were derived with the assumption that the squeezing parameter $r \ll 1$, and they depend explicitly on this parameter, as long as $r<1/\sqrt{n}$, the expressions correspond to a physical state. Formally, it is only in the limit $r \to 0$ that one recovers with perfect fidelity a Fock state. In Fig.~\ref{fig:fockinfidelities} we study the behaviour of the (in)fidelity between an ideal Fock state and our approximation. For $r\sim 10^{-2}$ one gets a fidelity of at least 99.9\% for $n=1,2,3$.

It is worth adding that a path to approximating superpositions of Fock states as linear combinations of Gaussian states to high fidelity is accessible through our formalism. We recall Gaussian Boson Sampling-type state preparation devices, wherein Fock measurements on all but one mode of a multimode Gaussian state herald a superposition of Fock states in the remaining mode. The exact form of the superposition can be tuned using the means and covariance matrix of the multimode Gaussian, as well as the choice of which Fock measurements to postselect upon~\cite{Sabapathy2019,Su2019,Quesada2019,Tzitrin2020}. As the input state is Gaussian, and we can approximate Fock measurements to arbitrarily high accuracy as linear combinations of Gaussians, the output state can be calculated using the partial trace form of \cref{eq:weyl_trace}; this materializes as a linear combination of Gaussian integrals over all but the two phase space quadratures of the output mode, yielding a linear combination of Gaussian functions in phase space for that mode. Notably, this state preparation device can be used as a means to prepare bosonic qubits, including GKP and cat states.

\section{Useful Transformations and Measurements in a Gaussian-Inspired Framework}\label{sec:ex_transform}

Given our formalism from Section \ref{sec:formalism}, and examples from Section \ref{sec:states} for writing bosonic qubits as linear combinations of Gaussians, we now examine how our formalism can be used to treat useful transformations for studying quantum computing with bosonic qubits, namely loss, measurement-based squeezing and non-Gaussian gate teleportation.

\subsection{Loss Channel and Fock Damping}
Loss is a physical process modeled as an interaction with a thermal environment through a beam-splitter transformation, resulting in a bosonic Gaussian channel. The loss channel is closely related to the additive random noise channel by composition with an amplifier channel\,\cite{sabapathy11,raul2012,Caruso2006,solomon2010}. The strength of the loss parameter of the channel is set by the beam-splitter angle. The pure loss channel is easily described in the form \eqref{eq:gauss_cptp} with the matrix pair 
\begin{equation}\label{eq:loss}
(\bm{X}_\eta,\bm{Y}_\eta) = (\sqrt{\eta} \id, (1-\eta)\hbar \id/2),
\end{equation}
where $\eta$ is sometimes referred to as the transmission or transmissivity parameter, and where the environment is assumed to have zero temperature. Thermal loss---interaction with a thermal environment with covariance matrix $\hbar (\bar{n}+\tfrac12) \id$---can also be incorporated into this description by multiplying $\bm{Y}_\eta$ by $(2\bar{n}+1)$. The Kraus operators and other details of the pure loss channel are provided in great detail in\,\cite{solomon2010}. Loss is the dominant imperfection in the photonics context.

The Fock damping or finite-energy operator is given by \eqref{eq:fock_damping}.
It is valuable for converting infinite-energy states, such as ideal position and momentum eigenstates and GKP states, into their normalizable, finite-energy forms. The operator is directly linked to the loss channel as it forms a leading-order Kraus operator of the channel (see Eq. 4.6 of~\cite{solomon2010}), while maintaining the purity of the state. In~\cite{Noh2019capacity}, in the context of GKP states, it was shown that $\hat{E}(\epsilon)$ can be derived by passing a state through a beam-splitter of transmissivity $\cos\theta = e^{-\epsilon}$ with an ancillary vacuum state, and postselecting the ancillary mode on vacuum (see also Appendix B1 of~\cite{Tzitrin2020} for a simple derivation of this result).

As we show in Appendix \ref{app:fock_damping}, the use of only Gaussian states and operations in deriving $\hat{E}(\epsilon)$ makes the operator fit neatly into the formalism for Gaussian transformations developed in Section \ref{subsec:gauss_transf}. If the initial means and covariances of the mode are $(\bm{\mu}_{m,0},\bm{\Sigma}_{m,0})$, then we can simply use \cref{eq:AB_mean_cov} with:
\begin{equation}\label{eq:fock_damping_update}
    \bm{\Sigma}_m = \bm{S}_\theta \begin{pmatrix}\bm{\Sigma}_{m,0} & \textbf{0} \\ \textbf{0} & \hbar\id/2\end{pmatrix}\bm{S}_\theta^T,\ \bm{\mu}_m = \bm{S}_\theta \begin{pmatrix}\bm{\mu}_{m,0}\\ \textbf{0}\end{pmatrix},
\end{equation}
where $\bm{S}_\theta = \begin{pmatrix}
\cos\theta\id & \sin\theta\id \\ -\sin\theta\id & \cos\theta\id
\end{pmatrix}$ is the symplectic matrix for a beam-splitter assuming a mode-wise ordering $(q_1,p_1,q_2,p_2)$. From there one can proceed with the update rule in \cref{eq:update_covs_means_gauss}, as well as the per-peak reweighting in \cref{eq:update_weights} with $d_j = 1$, $\bm{r}_M = \bm{0},$ and $\bm{\Sigma}_M = \hbar\id/2,$ since the projection is onto vacuum.

\subsection{Squeezed Ancilla-Assisted Gates}
\label{subsec:mbgates}

While passive Gaussian transformations can be effected in optics using phase shifters and beam-splitters, and while squeezing applied to vacuum can be achieved with nonlinear cavity resonators~\cite{vernon2019scalable} or waveguides~\cite{helt2020degenerate}, inline squeezing---squeezing applied directly to an arbitrary input state---poses a greater challenge. A feasible approach to inline squeezing is possible by consuming an ancillary squeezed vacuum state~\cite{filip2005}. To effect squeezing in $q$ ($p$), one first interferes the target state with a highly $q$-squeezed ($p$-squeezed) vacuum on a beam-splitter parametrized by angle $\theta$. Next, one performs a $p$-homodyne ($q$-homodyne) measurement on the ancillary mode with a detector of efficiency $\eta$, producing result $p_M$. Finally, one applies a feedforward $p$-displacement ($q$-displacement) by $\sqrt{\eta^{-1}}\tan\theta p_M$ to the target mode. As we discuss next, the resulting average map effects a transformation equivalent to squeezing the target mode by $\cos\theta$ in $q$, along with some noise dependent on the level of squeezing of the ancillary state and the efficiency of the homodyne detection. Squeezing along any quadrature can be achieved by preceding and following inline squeezing with rotations.

Since the ancillary state, the transformation between the target and ancilla, the measurement on the ancilla, and the feedforward are all Gaussian, inline squeezing using a squeezed ancillary state---commonly referred to as measurement-based squeezing---falls within the formalism developed in Section \ref{subsec:gauss_transf}. In Appendix \ref{app:mbgates}, we derive the resulting transformation on the covariance matrices, means and modes of the linear combination of Gaussians in phase space that form the target state. On average (integrating over $p_M$), the map due to this type of squeezing is a deterministic Gaussian CPTP map parametrized by:
\begin{equation}\label{eq:mbsq_avg}
\begin{split}
    \bm{X}^{(q)}_{s} &= \begin{pmatrix}
    \cos\theta & 0 \\
    0 & \frac{1}{\cos\theta}
    \end{pmatrix},\, \cos\theta = e^{-s},\\
    \bm{Y}^{(q)}_{s,r,\eta} &= \frac{\hbar}{2} \begin{pmatrix}
   \sin^2\theta e^{-2r} & 0 \\
    0& \tan^2{\theta}\frac{1-\eta}{\eta}
    \end{pmatrix},
\end{split}
\end{equation}
where $r$ is the squeezing parameter for the ancilla. The superscript $q$ denotes squeezing in the $q$ quadrature. This result almost matches the result from~\cite{filip2005}; however, we avoid any extra displacement on the target state by using knowledge of the homodyne detection efficiency in the feedforward displacement.

Importantly, while measurement-based squeezing produces noisy squeezed states on average, in the single-shot case (not averaging over $p_M$), the covariances, means and weights of the Gaussian functions of the target state transform in a non-linear fashion that precludes description with a deterministic CPTP map. We show this fact in Appendix \ref{app:mbgates}. For initial means and covariances $(\bm{\mu}_{m,0},\bm{\Sigma}_{m,0})$, we can set \cref{eq:AB_mean_cov} to be:
\begin{equation}
\begin{split}
        \bm{\Sigma}_m &= \bm{X}_{\eta,2}\bm{S}_\theta \bm{S}_{r,2} \begin{pmatrix}\bm{\Sigma}_{m,0} & \textbf{0} \\ \textbf{0} & \hbar\id/2\end{pmatrix}\bm{S}_{r,2}^T\bm{S}_\theta^T\bm{X}_{\eta,2}^T + \bm{Y}_{\eta,2},\\
        \bm{\mu}_m &= \bm{X}_{\eta,2}\bm{S}_\theta \bm{S}_{r,2}\begin{pmatrix}\bm{\mu}_{m,0}\\ \textbf{0}\end{pmatrix},\\
        \bm{S}_{r,2} &= \id\oplus\bm{S}_r,\bm{S}_r = \begin{pmatrix}e^{-r} & 0 \\ 0 & e^r\end{pmatrix} \\
        \bm{X}_{\eta,2} &= \id\oplus \bm{X}_{\eta},\ \bm{Y}_{\eta,2} = \bm{0}\oplus \bm{Y}_{\eta},
\end{split}
\end{equation}
where $\bm{S}_{r,2}$ models squeezing of the ancillary vacuum, $\bm{S}_\theta$ is the beam-splitter symplectic, and $(\bm{X}_{\eta,2},\bm{Y}_{\eta,2})$ model inefficiency in the homodyne detector for the second mode. Next, we use the update rule in \cref{eq:update_covs_means_gauss}, as well as the per-peak reweighting in \cref{eq:update_weights} with $d_j = 1$, $\bm{r}^T_M = (0,p_M)$ and $\bm{\Sigma}_M = \lim_{\epsilon\rightarrow 0}\frac{\hbar}{2}\begin{pmatrix}\epsilon^{-1} &0 \\ 0& \epsilon\end{pmatrix}$, as we are modelling $p$-homodyne measurements. Finally, to that result we apply a displacement in $p$-quadrature to the means by $\sqrt{\eta^{-1}}\tan\theta p_M$.

Inline squeezing is an especially valuable operation to be able to model, as it is required to perform the most general Gaussian unitary operations~\cite{dutta1995}. Moreover, for GKP states specifically, inline squeezing  is required to perform certain Clifford qubit gates such as the phase gate and the controlled-NOT and controlled-Z gates, as we review in \cref{fig:mbsq_circuits} of Appendix \ref{app:mbgates}. Using our formalism, we now can model realistic noise effects from inline squeezing applied to bosonic qubit states in both the average and single-shot case. This is useful, for example, when quantifying the build-up of noise in stitching together cluster states of bosonic qubits using active entangling operations~\cite{Menicucci2014, bourassa2021blueprint}, although there exist methods for tailored clusters that do-away with inline squeezing~\cite{walshe2020continuous}.

\subsection{GKP T gate and Error Correction}\label{subsec:Tgate}
Just as inline squeezing operations can be teleported onto a mode via the consumption of a squeezed vacuum ancilla, non-Gaussian gates can be effected via gate teleportation using non-Gaussian ancillary states or measurements~\cite{GKP_2001,Ghose2007,Sefi2013,Marshall2015,Miyata2016,Arzani2017,Sabapathy2019,Su2019,Tzitrin2020}. As a pertinent example, we consider the implementation of the T gate for GKP states. A single-qubit T gate is a non-Clifford gate that, in conjunction with Clifford gates, completes the set of universal operations. Physically, in the GKP encoding, it is implemented through a non-Gaussian transformation, the only one strictly required in the universal set. To effect a T gate via gate teleportation, an ancillary GKP magic state $\ket{M}_{\rm gkp}=\frac{1}{\sqrt{2}}(e^{-i\pi/8}\ket{0}_{\rm gkp}+e^{i\pi/8}\ket{1}_{\rm gkp})$ is first rotated in phase space by $\pi/2$ and entangled with the target mode at a $\CZ$ gate. Next, a $p$-homodyne measurement is performed on the ancilla; the measurement outcome is rounded to the nearest $n\sqrt{\pi\hbar}$, and the parity of $n$ determines whether a feedforward phase gate is applied on the target mode~\cite{GKP_2001}. The non-Gaussianity of the gate enters via the ancillary GKP resource state, which has significant Wigner negativity~\cite{Garcia_2021}. See \cref{fig:Tgate} for a summary, and \cref{fig:mbsq_circuits} of Appendix \ref{app:mbgates} for decompositions of the $\CZ$ and phase gate. 

\begin{figure}
$$
\Qcircuit @C=1.0em @R=.7em { 
\lstick{\Ket{\psi}_{\rm gkp}} & \qw & \multigate{1}{\hat{CZ}} & \qw & \gate{\hat{P}\left[s(p_0)\right]} & \qw & &  {\hat{T}\Ket{\psi}_{\rm gkp}}\\
\lstick{\Ket{M}_{\rm gkp}} & \gate{\hat{H}}  &  \ghost{\hat{CZ}} & \measureD{p=p_0} & \cctrl{-1}
}
$$
\caption{Optical implementation of the GKP qubit T gate up to global phase, following the method from~\cite{GKP_2001}. Here, in the ideal limit, $\ket{M}_{\rm gkp}=\frac{1}{\sqrt{2}}(e^{-i\pi/8}\ket{0}_{\rm gkp}+e^{i\pi/8}\ket{1}_{\rm gkp})$. The GKP Hadamard gate (with operator $\hat{H}$) is a $\pi/2$ phase shift, and the GKP $\CZ$ gate is discussed in Appendix \ref{app:mbgates}. The output $p_0$ of the $p$-homodyne measurement is processed by the function $s(p_0)$, which rounds the value to the nearest $n\sqrt{\pi\hbar}$ and returns the parity of $n$. If $n$ is even (odd), no phase gate (a phase gate $\hat{P}(s) = e^{is\hbar\hat{q}^2/2}$ with $s=1$) is applied.  
}
\label{fig:Tgate}
\end{figure}
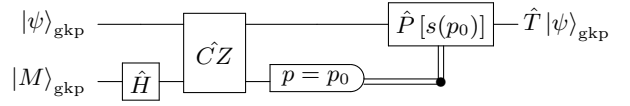

We emphasize that the ancillary GKP magic state and the target GKP state can both be described as Wigner functions of the form \eqref{eq:lin_comb_gauss}. The $\CZ$ and phase gates---implemented directly or via ancilla-assisted squeezing---can be described through an average map or as a single-shot transformation, as we explored generally in Section \ref{subsec:gauss_transf} and specifically to these transformations in Appendix \ref{app:mbgates}. Finally, a $p$-homodyne measurement of the ancillary mode containing the magic state falls neatly into the formalism from Sections \ref{subsec:meas} and \ref{subsec:gauss_transf}. Unlike measurement-based squeezing, we cannot write the average transformation effected by this circuit as a Gaussian CPTP map on the original state, but we can simulate the single-shot transformation of the gate teleportation by selecting or sampling an outcome for the last homodyne measurement and using Eqs. \eqref{eq:update_covs_means} and \eqref{eq:update_weights}.

While we can now implement single-shot non-Gaussian operations via gate teleportation, a simple average map for T gate teleportation is untenable for several reasons. First, the ancillary state is itself described by a linear combination of Gaussian functions in phase space, in contrast to the single Gaussian peak of the ancillary squeezed state used for inline squeezing. In the latter case, when the ancillary mode is added to the description, the weights $\{c_m\}$ in the Wigner function of \cref{eq:lin_comb_gauss} do not change, since they are each multiplied by $1$, while new rows and columns for the additional mode are added to the covariances and means of the Gaussian functions. When the ancillary magic state is added, however, in addition to new rows and columns being added to the covariances and means, the weights and the set of indices $\mathcal{M}$ of the Wigner function \textit{also} change, since we must consider all the cross-terms of different Gaussian peaks in the target and ancillary state. This means that the final resulting state after the teleportation operation has a different set of weights than what it started with, which cannot be captured by a simple average map on each Gaussian function. Second, when applying the feedforward phase gate, the homodyne data is processed by a highly non-linear function because of the binning and parity check, so there would again be no opportunity for the integration over homodyne outcomes to yield a simple Gaussian CPTP map as in \cref{eq:gauss_cptp}.

Notice that the GKP error-correction circuit in \cref{fig:syndrome_meas} is just another example of a gate teleportation circuit that can be treated by our formalism. The initial ancillary modes are GKP states, and hence can be expressed as a linear combination of Gaussian functions; the entangling operations are Gaussian $\CX$ gates; and the homodyne measurements and feedforward displacements are also all Gaussian. Thus, single-shot error correction, with appropriate noise sources such as loss and Fock damping incorporated into the circuit, can be studied using our formalism.

\section{Simulation methods}\label{sec:sim_methods}
Having provided our general formalism in Section \ref{sec:formalism}, and useful states and transformations in Sections \ref{sec:states} and \ref{sec:ex_transform}, we now describe our simulation methods. In Section \ref{subsec:tracking}, we summarize how to track the weights, means and covariance matrices associated with the linear combination of Gaussian functions in phase space that describes the state of the modes in a photonic circuit. In Section \ref{subsec:gauss_sampling}, we detail a technique for sampling outcomes of Gaussian measurements on states in our formalism. Finally in Section \ref{Sec:Comparison}, we compare our simulation method to a state-of-the-art method the employs the Fock basis, demonstrating the advantage of our formalism and method as compared to this alternative leading technique.

\subsection{Tracking Weights, Means and Covariances}\label{subsec:tracking}

Our method for simulating states and transformations from our formalism is relatively straightforward. It can be summarized as follows:
\begin{enumerate}
    \item Initialize the weights, vectors of means, and covariance matrices for each mode at the start of the circuit. For example, Eq. \eqref{eqn:Wigner_Cat} provides these parameters for the cat state.
    
    \item Combine the initial weights, means and covariances for each mode into weights, means and covariances for the multimode state of the whole circuit as follows:
    \begin{equation}
        \begin{array}{c c c}
         &\text{Mode 1}&\text{Mode 2}  \\
         \text{Weights}& a_\ell & b_k \\
         \text{Means}& \bm\mu_\ell & \bm\mu_k \\
         \text{Covariances}& \bm\Sigma_\ell & \bm\Sigma_k \\
        \end{array} 
        \rightarrow\begin{array}{c}
         \text{Multimode}  \\
         a_\ell b_k \\
         \bm\mu_\ell\oplus\bm\mu_k \\
         \bm\Sigma_\ell\oplus\bm\Sigma_k \\
        \end{array}
    \end{equation}
    This procedure can be applied recursively to build the weights, means and covariances for the full multimode initial state.
    
    \item For each gate or measurement in the circuit, apply the associated transformation to the weights, means and covariances of the state. For example, to apply the loss channel, the matrices that parametrize the channel from Eq. \eqref{eq:loss} are used in Eq. \eqref{eq:gauss_cptp}. Alternatively, for a given measurement outcome on a subset of modes, the other modes are updated using Eqs. \eqref{eq:update_covs_means} and \eqref{eq:update_weights}.
    
    \item The output of the simulation consists of the weights, means and covariances of all the modes after the application of transformations and measurements in the circuit, along with any of the measurement outcomes collected along the way. Since these weights, means and covariances can be used to construct the Wigner function of the state as in Eq. \eqref{eq:lin_comb_gauss}, we have access to full state information.
\end{enumerate}
For some states, such as GKP and cat states, all the Gaussian peaks in the linear combination representing that mode have the same covariance matrix. In those cases, to save memory, we need not initialize the same copy many times, and can simply track a single covariance matrix. 

In the next part, we detail our method for sampling outcomes of Gaussian measurements for states in our formalism.

\subsection{Sampling Outcomes of a Gaussian Measurement}\label{subsec:gauss_sampling}

To further the utility of our formalism and method for performing simulations, we now provide an algorithm for sampling outcomes of Gaussian measurements on states that can be expressed as linear combinations of Gaussian functions in phase space. In Section \ref{subsec:meas}, we detailed in Eq. \eqref{eq:outcome_prob} how to calculate probability distributions for Gaussian measurements within our formalism. A general approach to sampling from non-trivial Wigner functions can be found in Appendix D of~\cite{Quesada_2019}; however, we find that a more tailored approach one can use is a rejection-sampling technique~\cite{mcbook}. In~\cite{rabusseau_2014}, the authors consider a rejection-sampling method for a distribution composed of real-valued Gaussian functions with (positive or negative) real weights. Here, we extend the method to include Gaussian functions with complex weights and means. We summarize the technique in Algorithm \ref{alg:rejection_sampling}. First, we separate the peaks into those which have negative weights and real-valued means $\mathcal{M}^- = \{m | c_m < 0 \}\cap\{m|\Im(\bm{\mu}_m) = 0\}$, and everything else $\{m \not \in \mathcal{M}^-\}$:
\begin{align}
p \left( \bm{r}_M; {\hat{\rho}}, \bm{\Sigma}_M\right) =& \sum_{m \in \mathcal{M}^{-} } c_m G_{ \bm{\mu}_m,\bm{\Sigma}_M+\bm{\Sigma}_m}\left(  \bm{r}_M \right) \nonumber\\
&+ \sum_{m' \not \in \mathcal{M}^{-}} c_{m'} G_{ \bm{\mu}_{m'},\bm{\Sigma}_M+\bm{\Sigma}_{m'}}\left(  \bm{r}_M \right).
\end{align}
By writing $\bm{\mu}_{m'} = \Re(\bm{\mu}_{m'}) + i \Im(\bm{\mu}_{m'})$ we can obtain
\begin{align}
 &G_{ \bm{\mu}_{m'},\bm{\Sigma}_M+\bm{\Sigma}_{m'}}\left(   \bm{r}_M \right)\\
&=  e^{\frac{1}{2}\Im(\bm{\mu}_{m'})^T (\bm{\Sigma}_M + \bm{\Sigma}_{m'})^{-1} \Im(\bm{\mu}_{m'})} \nonumber \\
& \quad\times 
G_{ \Re(\bm{\mu}_{m'}),\bm{\Sigma}_M+\bm{\Sigma}_{m'}}\left(   \bm{r}_M \right) \nonumber \\
& \quad \times 
 e^{i [\bm{r}_M -\Re(\bm{\mu}_{m'})]^T (\bm{\Sigma}_M + \bm{\Sigma}_{m'})^{-1} \Im(\bm{\mu}_{m'})}. \nonumber
\end{align}
This allows us to define an upper-bounding function to the distribution of interest:\footnote{Note that we do not include any weights from $\mathcal{M}^-$ in this bound since bounding a real-valued Gaussian with a negative weight by a Gaussian with the absolute value of the negative weight is a looser upper bound than simply bounding it by 0.}
\begin{gather}
    p \left( \bm{r}_M; {\hat{\rho}}, \bm{\Sigma}_M\right) \leq g(\bm{r}_M)= \sum_{m \not\in \mathcal{M}^{-}} \tilde{c}_{m}  G_{ \Re(\bm{\mu}_{m}),\bm{\Sigma}_M+\bm{\Sigma}_{m}}\left( \bm{r}_M \right), \nonumber \\
    \tilde{c}_{m} = |c_{m}| e^{\frac{1}{2}\Im(\bm{\mu}_{m})^T (\bm{\Sigma}_M + \bm{\Sigma}_m)^{-1} \Im(\bm{\mu}_{m})}.
\end{gather}

Next, we note that $g(\bm{r}_M)$ is a scalar multiple of a probability distribution $q\left( \bm{r}_M\right)$:
\begin{equation}
    g(\bm{r}_M) = \mathcal{N} q \left( \bm{r}_M\right), \ \mathcal{N} = \sum_{m \not\in \mathcal{M}^{-}} \tilde{c}_{m}.
\end{equation}
Importantly, one can sample from $q \left( \bm{r}_M\right)$ more straightforwardly: first, one samples a value $m_0 \not\in \mathcal{M}^-$ according to the distribution $p_m = \tilde{c}_m/\mathcal{N}$; since the weights $\tilde{c}_m$ from $g(\bm{r}_M)$ are now all positive, we can associate them with probabilities after proper normalization by $\mathcal{N}$. Next, one samples a phase space value $\bm{r}_0$ according to $G_{ \Re(\bm{\mu}_{m_0}),\bm{\Sigma}_M+\bm{\Sigma}_{m_0}}\left(  \bm{r} \right)$. Given $\bm{r}_0$, one then samples $y_0$ uniformly form $[0,g(\bm{r}_0)]$. If $y_0 \leq p \left(\bm{r}_0; {\hat{\rho}}, \bm{\Sigma}_M\right)$, then $\bm{r}_0$ is kept as the sampled value of $p \left(\bm{r}_M; {\hat{\rho}}, \bm{\Sigma}_M\right)$; otherwise, it is rejected and the process is restarted until a sample is drawn. 

This sampling technique leverages the form of the Wigner function of our states: we use the fact that our distribution can be bounded above by a mixture of Gaussian functions with all-positive weights, and that it is computationally easy to sample a value from a single Gaussian distribution.

\begin{algorithm}[H]
\begin{algorithmic}
\State \textbf{Input:} Weights $c_m$, means $\bm{\mu}_m$, and covariances $\bm{\Sigma}_m$ of the initial state. Covariance $\bm{\Sigma}_M$ of the measurement.
\State \textbf{Output:} Sampled phase-space position outcome $\bm{r}_0$
\newline
\State $\mathcal{M}^- = \{m | c_m < 0 \}\cap\{m|\Im(\bm{\mu}_m) = 0\}$
\State $\tilde{c}_{m} = |c_{m}| e^{\frac{1}{2}\Im(\bm{\mu}_{m})^T (\bm{\Sigma}_M + \bm{\Sigma}_m)^{-1} \Im(\bm{\mu}_{m})}$
\State drawn $\leftarrow$ False
\While{not drawn}

    Sample $m_0 \not\in \mathcal{M}^-$ according to $p_m = \tilde{c}_{m}/\mathcal{N}$ \;
    
    Sample $\bm{r}_0$ according to $G_{ \Re(\bm{\mu}_{m_0}),\bm{\Sigma}_M+\bm{\Sigma}_{m_0}}\left(  \bm{r} \right)$\;
    
    Sample $y_0$ uniformly from $[ 0, g(\bm{r}_0)]$\;
    
    \If{ $y_0 \leq  p \left(\bm{r}_0; {\hat{\rho}}, \bm{\Sigma}_M\right)$}
    
        drawn $\leftarrow$ true 
        
    \EndIf
\EndWhile

\State \textbf{return} $\bm{r}_0$
\newline
\end{algorithmic}
\caption{Simulating Gaussian measurement outcomes using rejection sampling}
\label{alg:rejection_sampling}
\end{algorithm}

With the simulation methods in hand, we now compare our simulation method to simulation methods in the Fock basis.

\subsection{Comparison to Fock Basis Simulations}
\label{Sec:Comparison}

\begin{figure*}
	\centering
        \includegraphics[width = \textwidth]{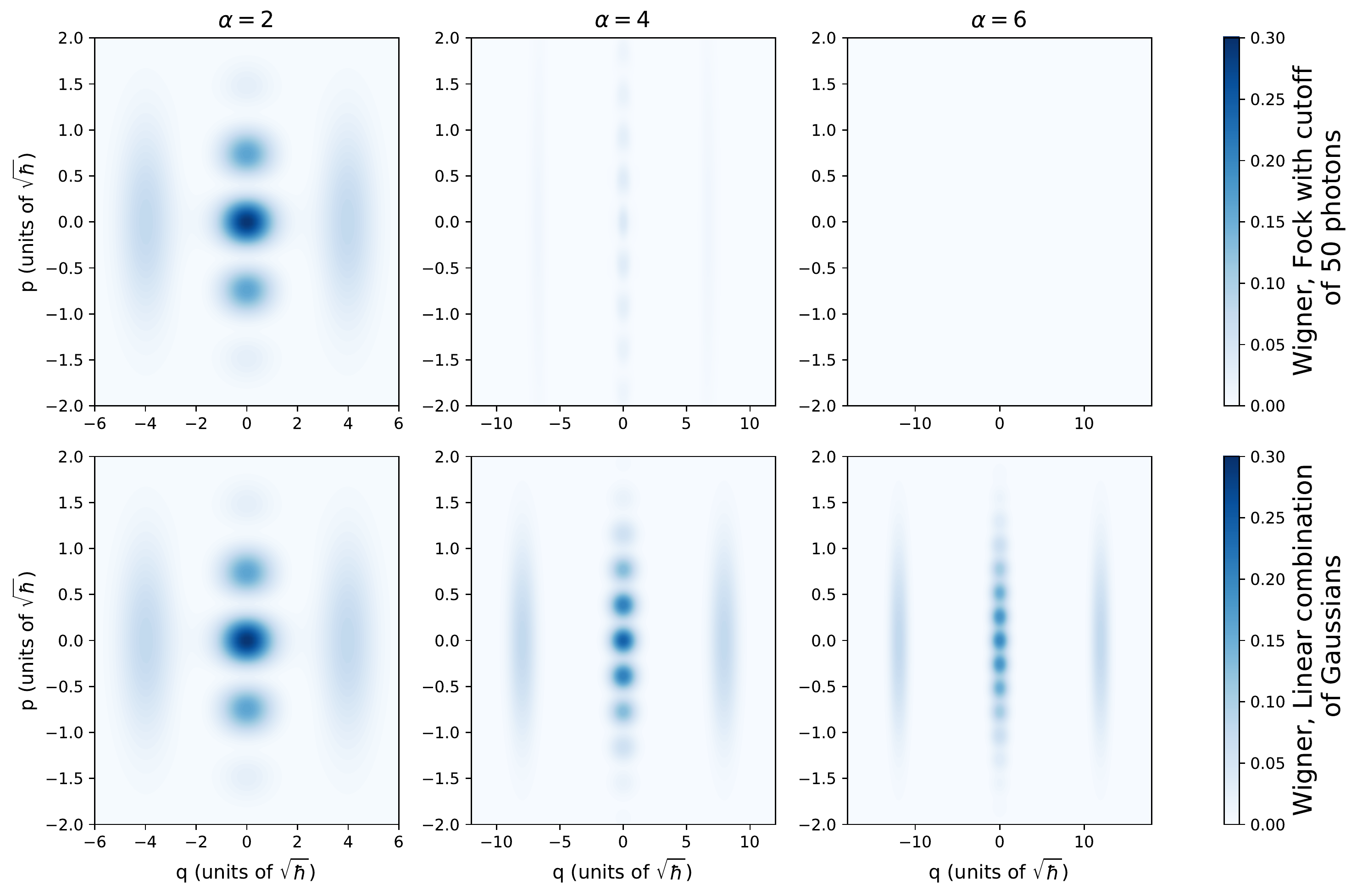}
	\caption{Simulation of two cat states passing through a lossy beam-splitter, with the Wigner function of the first mode displayed for increasing values of $\alpha$. Note the difference in scale between $q$ and $p$. We compare the output of the simulation using the \texttt{fock} backend of \texttt{Strawberry Fields} with a cutoff of 50 photons (top row), and using a linear combination of Gaussians in the \texttt{bosonic} backend (bottom row). While the number of photons in the state increases with $\alpha$, saturating the cutoff for the Fock basis simulation and leading to an incorrect output, the number of Gaussian peaks in phase space that require tracking remains constant. Note additionally that the mixing introduced by loss necessitates the full density matrix being tracked in the Fock basis, while loss does not increase the complexity of simulation for the linear combination of Gaussians.}
	\label{fig:fock_vs_bosonic_cat}
\end{figure*}

\begin{figure}
	\centering
        \includegraphics[width = \columnwidth]{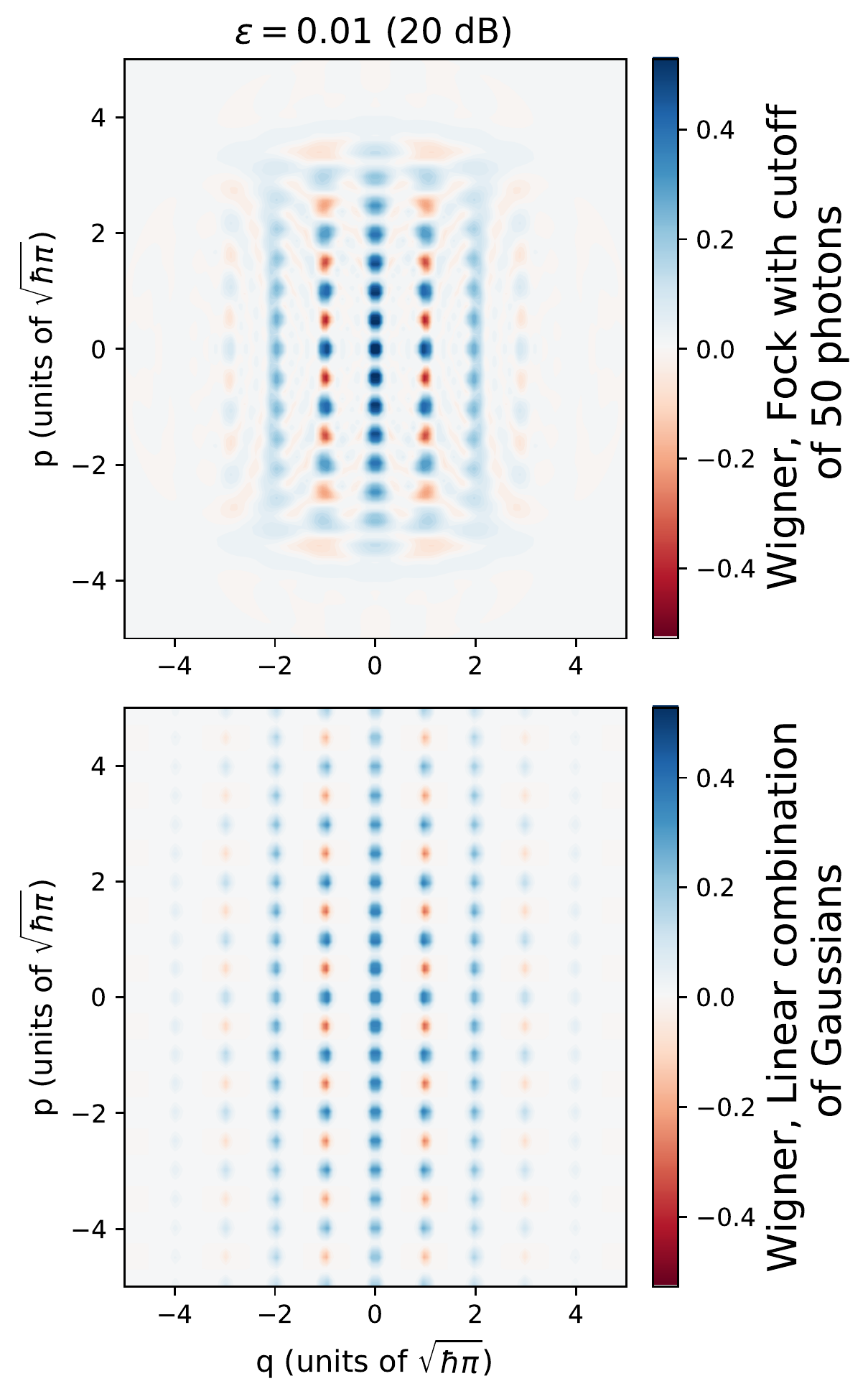}
	\caption{Wigner functions for a $\ket{+^\epsilon}_{\rm gkp}$ state with $\epsilon=0.01$ (20 dB of per-peak squeezing) teleported onto a 15 dB squeezed state. We entangle the states via a CV $\CZ$ gate and and subject them 1\% loss in each mode; next, we measure the mode originally containing the GKP state in the $p$-basis, postselecting on $p=0$. In the top panel, we perform the simulation using the \texttt{fock} backend in \texttt{Strawberry Fields} with a cutoff of 50 photons. In the bottom row, we employ the \texttt{bosonic} backend. We see that the \texttt{bosonic} backend can maintain a correct description of the state even as the number of photons becomes very high due to the large per-peak squeezing.}
	\label{fig:fock_vs_bosonic_gkp}
\end{figure}

It is well-known that representing a Gaussian state in terms of its vector of means and its covariance matrix is a more efficient representation than writing the state in the Fock basis, especially considering the energy of the state increases under displacements and squeezing. Here, we draw a similar conclusion for the states we have studied under our formalism. In this and upcoming sections, we implement our simulation technique in the new \texttt{bosonic} backend of \texttt{Strawberry Fields}.

Consider that the Hamiltonian for a single mode is given by $\hat{H} = \frac{1}{2}(\hat{q}^2 + \hat{p}^2) = \frac{1}{2}\hat{r}^2 = \hbar(\hat{n}+\frac{1}{2})$. This means that the Wigner function of a Fock state $\ket{n}$ has an associated radius in phase space of roughly $|\bm{r}_n| = \sqrt{\hbar(2n+1)}$, beyond which the function decays monotonically. This fact can also be demonstrated rigorously using an analytic representation of Fock Wigner functions and the properties of Laguerre polynomials~\cite{Kenfack2004,Temme1990}. Thus, to determine the Fock representation for a state which has a phase-space Gaussian peak at a point $\bm{r}_0$, a conservative estimate has that we would need a photon number of at least
\begin{equation}\label{eq:fock_radius}
    n(\bm{r}_0) \sim \frac{|\bm{r}_0|^2}{2\hbar}
\end{equation}
to reach the required radius in phase space. Furthermore, Fock states beyond this value may be necessary, for example, to shape the phase space peak for a desired level of squeezing. In the presence of realistic noise sources like loss, the state additionally becomes mixed, requiring a density matrix rather than a state vector representation in the Fock basis, squaring the number of elements which one needs to track. From these considerations, we see that for a state with a phase space peak at $\bm{r}_0$, the number of elements one needs to track in the Fock basis to have a faithful representation of the state scales like $\frac{|\bm{r}_0|^4}{\hbar^2}$.

Let us compare this Fock-basis scaling to what we would obtain by expressing our states of interest as linear combinations of Gaussian functions in phase space. The trivial case is a Gaussian state; regardless of its position, orientation, and level of squeezing in phase space, one need only track a $2$-component vector of means and a $2\times2$ covariance matrix. For an $N$ mode Gaussian state, the number of elements to track scales like $4N^2 + 2N$---no exponential scaling since all the modes are represented by a single Gaussian function. Adding a mode with a Gaussian state to a series of other modes with states represented by a linear combination of Gaussian functions only increases the dimension of the covariance matrices and means by $2$, but does not change the number of weights, means or covariances one needs to track. This is especially useful for the GKP encoding which, when combined with Gaussian states, enables universal quantum computation~\cite{Baragiola2019,Yamasaki2020}.

We have shown in Section \ref{subsec:cats} that single-mode two-lobe cat states can be represented using one $2\times2$ covariance matrix, along with four weights and $2$-component vectors of means---two real-valued and two complex-valued. Crucially, the size of the mathematical objects required for this representation is independent of the energy of the cat state associated with $\alpha$, and is invariant under Gaussian transformations. While the covariance for $N$ modes encoded as cat qubits scales like $4N^2$, the number of weights still scales exponentially, like $4^N$, and the number of elements in the means grows like $2N(4^N)$. This is still preferable to the Fock representation, for which the number of density matrix elements scales like $|\alpha|^{4N}$; notably, the scaling for the linear combination of Gaussians representation does not depend on the energy of the state or modifications under Gaussian transformations. This can be valuable since, for example, the error rates for some qubit gates on cat states can scale like $1/|\alpha|$~\cite{chamberland2020building}, so to examine regimes of low error, one must increase the energy of the cat state. In \cref{fig:fock_vs_bosonic_cat}, we examine the case of two cat states sent through a lossy beam-splitter. We trace out one of the modes and plot the Wigner function of the remaining mode for increasing values of energy, as parametrized by $\alpha$. We compare the results using the \texttt{fock} backend of \texttt{Strawberry Fields} using a photon number cutoff of 50 photons per mode--- going beyond this value saturates the memory of the standard desktop terminal used for simulation---and using the \texttt{bosonic} backend, representing states as linear combinations of Gaussian functions in phase space. We find that, for $\alpha = 2$, the Fock representation still works well, albeit requiring more memory and running more slowly when simulating the lossy beam-splitter. For $\alpha = 4$ and 6, the Fock representation with 50 photons quickly becomes insufficient. This is unsurprising: for these values of $\alpha$, the action of the beam-splitter leads to Gaussian peaks in phase space at distances, respectively, of $8\sqrt{\hbar}$ and $12\sqrt{\hbar}$ from the origin, requiring, by the conservative estimate in \cref{eq:fock_radius}, greater than 32 and 72 photons for each case. In fact, although $n=32$ is comfortably within the cutoff of 50 photons, the Fock representation cannot accurately construct the Wigner function for $\alpha=4$, confirming that \cref{eq:fock_radius} is an underestimation of the required photon number cutoff.

As seen in Section \ref{subsec:Fock}, even Fock states of $n$ photons can be approximated by $n$ real-valued weights, $n$ $2\times2$ covariance matrices, and one $2$-component vector of means. This is perhaps more memory-intensive than representing a pure number state in the Fock basis; however, under Gaussian transformations such as displacements or squeezing, or the common loss channel, representing the state by a linear combination of Gaussians becomes advantageous.

Finally, while GKP states have an infinite number of peaks, we can consider only a finite number since the weights in \cref{eq:gkp_wigner_explicit} decay exponentially under the Fock damping operator. Since the Fock damping operator decays exponentially in Fock space, an analogous procedure can be executed to establish a reasonable photon number cutoff. If a single-mode GKP state is approximated by a finite number of peaks, with the furthest peak located near $\frac{\sqrt{\pi}}{2}(k,\ell)$, then the number of peaks one needs to track scales like $k^2 + \ell^2$, and consequently so does the number of elements for the vectors of means; by contrast, only a single $2\times 2$ covariance matrix is required. Compare this with the Fock representation, where, by \cref{eq:fock_radius}, the number of density matrix elements scales like $(k^2 + \ell^2)^2$. For $N$ modes encoded as GKP states, we track $(k^2 + \ell^2)^N$ weights and means, and a single $2N\times 2N$ covariance matrix, an improvement over the $(k^2 + \ell^2)^{2N}$ scaling for the number of density matrix elements in the Fock basis. As a demonstration of the advantage of the \texttt{bosonic} backend, in \cref{fig:fock_vs_bosonic_gkp} we plot the Wigner function for the output mode of a simulated CV teleportation of a high-energy GKP state  ($\epsilon=0.01$, corresponding to 20 dB of per-peak squeezing) onto a $p$-squeezed state with 15 dB of squeezing using a lossy $\CZ$ gate, homodyne measurement, and feedforward displacement. We see that the \texttt{fock} backend is limited in the radius of phase space that it can accurately capture, while the \texttt{bosonic} backend produces all peaks correctly. Moreover, the all-Gaussian nature of the teleportation circuit allows for more efficient computation when representing the states as linear combinations of Gaussians over the Fock representation, which requires many tensor contractions for large density matrices.

\section{Numerical Simulations}\label{sec:simulations}
Having established our simulation method and its advantage over competing techniques in Section \ref{sec:sim_methods}, we now put it to use in simulations of bosonic qubits, leveraging a numerical implementation of our method available through the \texttt{bosonic} backend of \texttt{Strawberry Fields} \cite{bosonic_github}. In Section \ref{subsec:basic_eg}, we provide a basic test-run of our simulator, plotting Wigner functions of and sampling from bosonic qubits. In Section \ref{Sec:NewSimulations}, we provide novel simulations of bosonic qubits undergoing realistic gate implementations, with a focus on GKP states. For an advanced tutorial implemented by the authors on using the \texttt{bosonic} backend of \texttt{Strawberry Fields} to simulate realistic GKP qubit gates, see \cite{tutorial3}.

\subsection{Basic Examples: Wigner Plots and Homodyne Sampling}\label{subsec:basic_eg}

\begin{figure*}
	\centering
	\includegraphics[width = \textwidth]{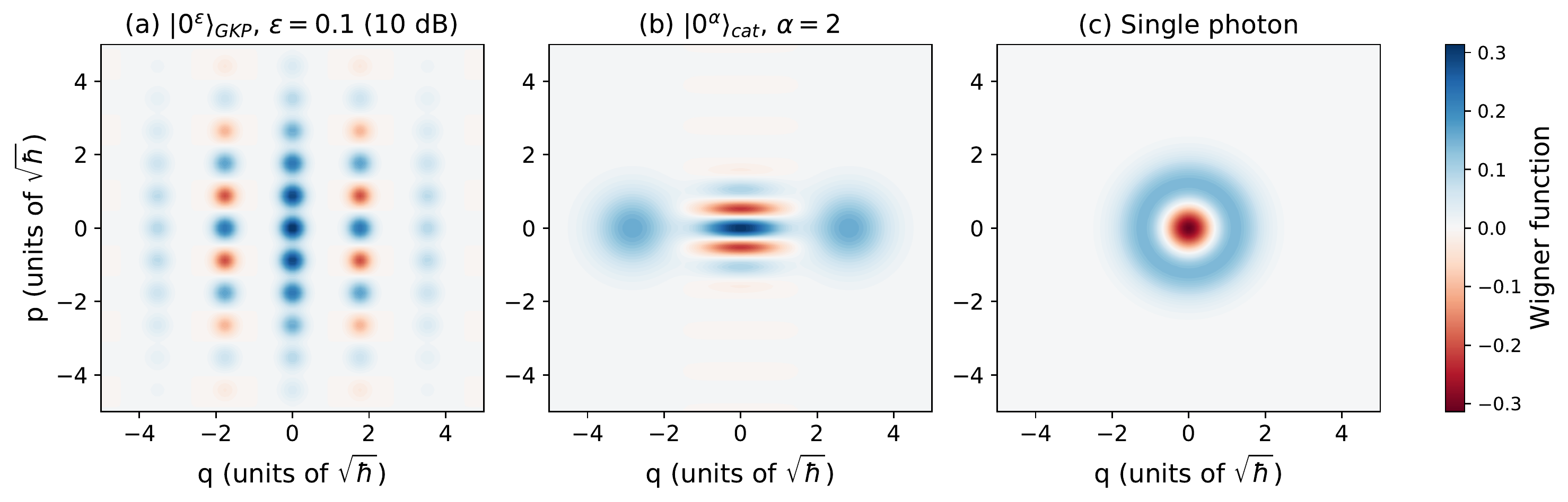}
	\caption{Wigner functions for GKP, cat, and Fock states. All figures are produced using linear combinations of Gaussians in phase space, as presented in Section \ref{sec:states}. For the GKP state, negative regions correspond directly to Gaussian peaks with negative weights, while for the cat state negativity is produced by the sinusoidal oscillations of complex-valued Gaussian peaks. In the single photon case, negativity is produced by the difference of two zero-mean Gaussians with slightly different covariance matrices.}
	\label{fig:wigners}
\end{figure*}

As a first simple demonstration of simulations with our technique, in \cref{fig:wigners} we plot the Wigner functions for $\ket{0^\epsilon}_{\rm gkp}$ with $\epsilon=0.1$ (10 dB of squeezing per peak of the wavefunction), for $\ket{0^\alpha}_{\text{cat}}$ with $\alpha=2$, and for the single-photon Fock state. For the GKP state, we can identify the following features: the Gaussian functions centred near integer and half-integer multiples of $\sqrt{\pi}$; the positive and negative peaks determined by the weighting function; and the per-peak variance, which is smaller than that of the vacuum. For the cat state, we see a similar distribution to \cref{fig:wignerdecomp}, where the interference fringes are produced by the Gaussians with complex weights and means. Finally, for the Fock state, we see how two Gaussians, both with zero means but with slightly different covariance matrices, can combine to form a rotationally symmetric Wigner function with a region of negativity in the centre.

\begin{figure*}
	\centering
	\subfloat{\includegraphics[width = 0.75\textwidth]{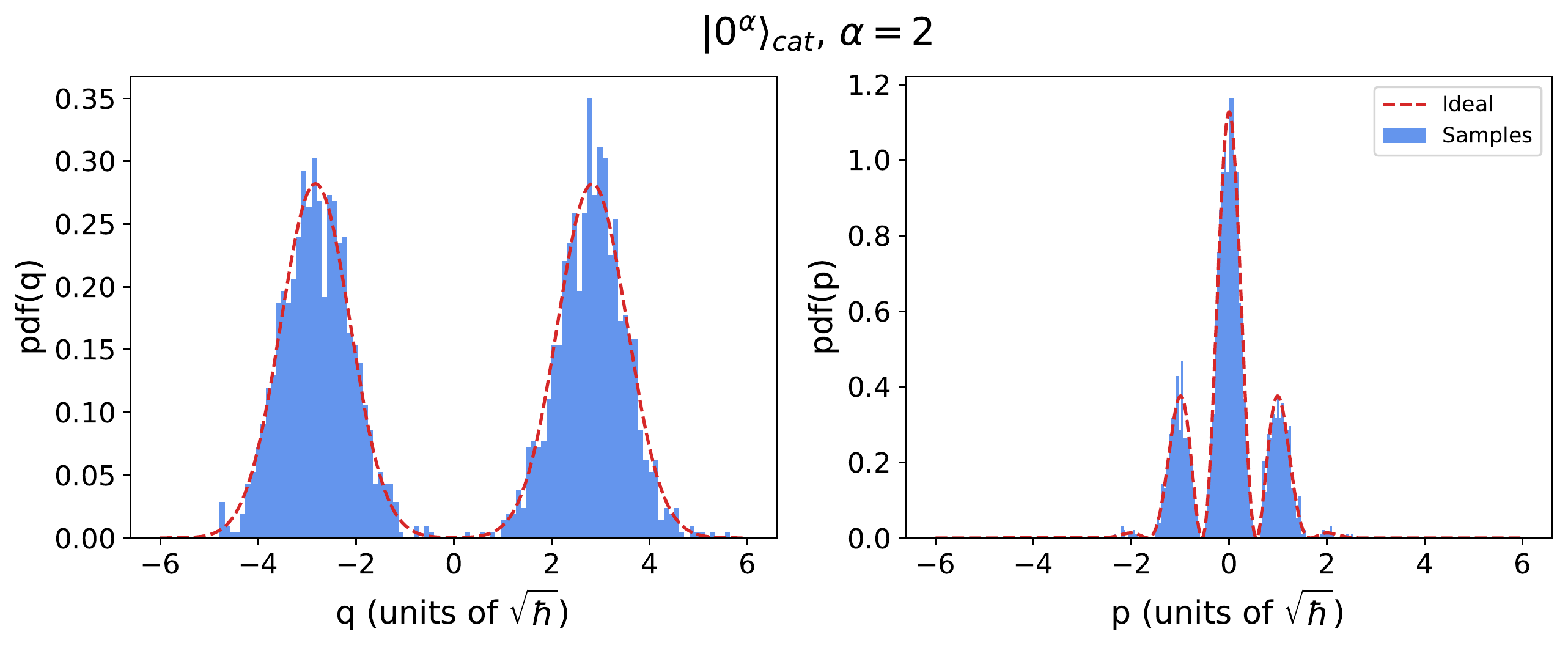}}
	\hfill
	\subfloat{\includegraphics[width =0.75 \textwidth]{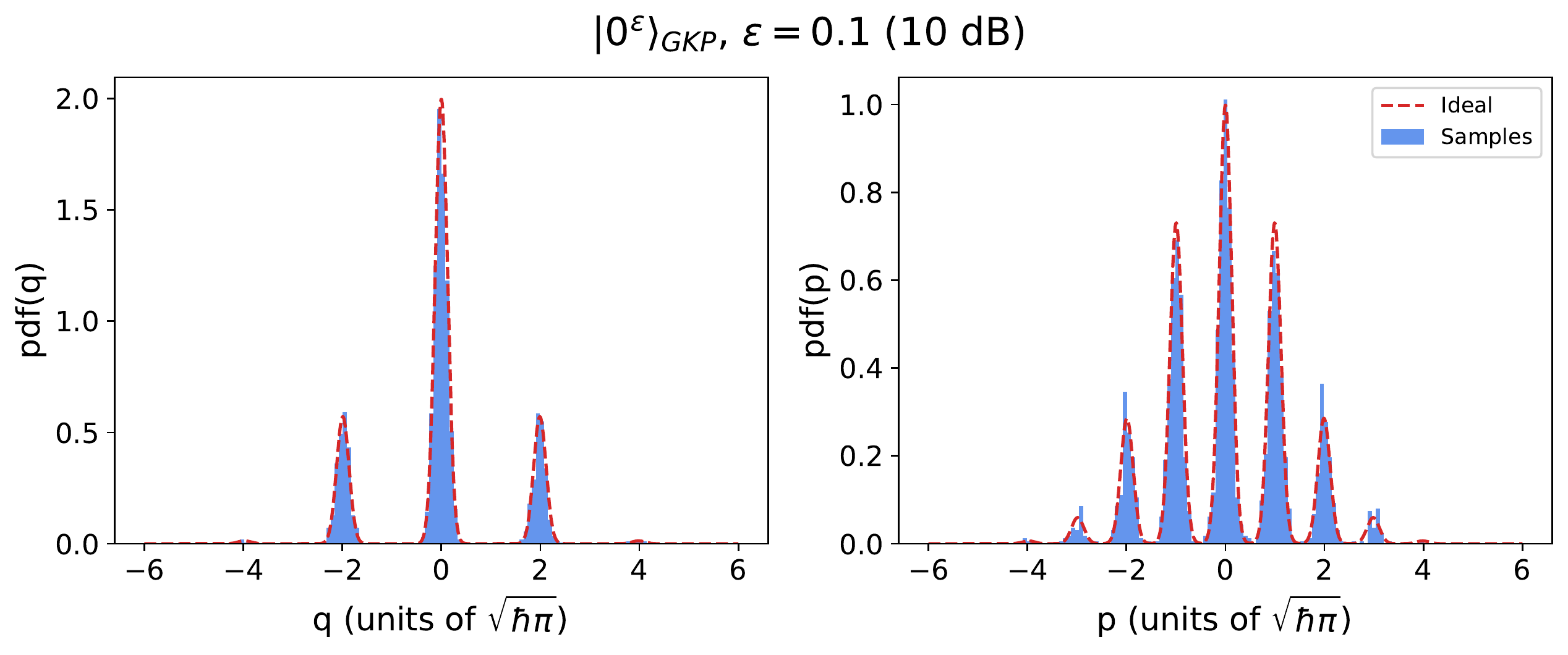}}
	\caption{2000 homodyne measurement samples in both quadratures for a cat (top row) and a GKP (bottom row) state obtained using Algorithm \ref{alg:rejection_sampling}. We compare histograms obtained from finite sampling to the asymptotic distributions, obtained easily from the linear combination of the marginals of each Gaussian function in the Wigner function, as in \cref{eq:homodyne}. For an intermediate tutorial implemented by the authors on how these results can be generated with the \texttt{bosonic} backend of \texttt{Strawberry Fields}, see \cite{tutorial2}.} 
	\label{fig:homodyne}
\end{figure*}

Algorithm \ref{alg:rejection_sampling} offers a straightforward method for sampling the results of general-dyne measurements of states with Wigner functions that are expressed as linear combinations of Gaussian functions. We use the algorithm to simulate 2000 samples of $q$ and $p$ quadrature homodyne measurements of a $\ket{0^{\alpha}}_{\text{cat}}$ with $\alpha=2$, and of a $\ket{0^\epsilon}_{\rm gkp}$ with $\epsilon = 0.1$ (10 dB per peak), and plot the output in \cref{fig:homodyne}. We see that the results align with the asymptotic marginal distributions for these quadratures. The marginals can also be easily attained in our formalism: integrating out one quadrature for a linear combination of Gaussian functions amounts to dropping the entries associated with that quadrature from each mean and covariance matrix, as we saw in \cref{eq:homodyne}.

For the cat state, we see the $q$ quadrature distribution is the same as for mixture of coherent states centred at $\pm\sqrt{2\hbar}\alpha$, since the interference fringes cancel out, while they are prominent in the $p$ quadrature. For the GKP state, we can clearly identify the peaks at even (all) integer multiples of $\sqrt{\pi\hbar}$ in $q$ ($p$) quadrature.

\subsection{Novel Simulations of Useful CV Circuits}
\label{Sec:NewSimulations}

\subsubsection{Measurement-Based Squeezing}
\begin{figure*}
	\centering
        \includegraphics[width = \textwidth]{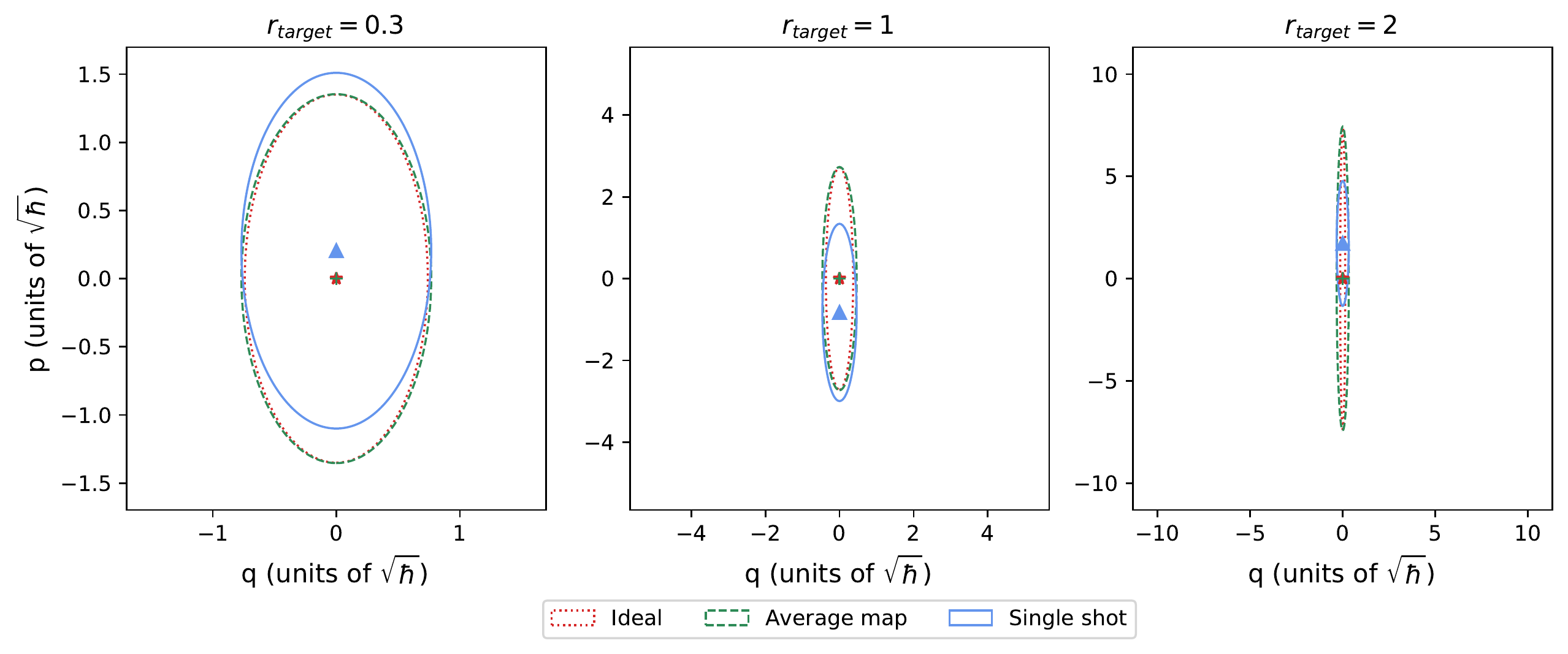}
	\caption{For different levels of target squeezing on vacuum $r_{\text{target}}$, a comparison of the states produced from ideal squeezing (red), the average map from measurement-based squeezing (green), and a single-shot occurrence of measurement-based squeezing (blue). The centre and axis lengths are set by the mean and variances of the output state. We simulate using an ancillary squeezed state of $r=1.2$ ($\sim$10.5 dB), and a homodyne detection efficiency of 0.99. A key takeaway is that the single-shot instances have less antisqueezing and a distribution of locations for where they can be centred in phase space, indicating the level of antisqueezing and the correct mean value observed in the average map are a result of the averaging procedure itself.}
	\label{fig:mbsqueeze}
\end{figure*}

In Section \ref{subsec:mbgates} and in Appendix \ref{app:mbgates}, we examine how to use our formalism to describe measurement-based squeezing, a feasible method for applying in-line squeezing operations. Later subsections examine gates that employ measurement-based squeezing in more complicated circuits with bosonic qubits, but we restrict ourselves to a simpler case to develop a clear a picture of the action of the transformation. In \cref{fig:mbsqueeze} we plot the  the Wigner function for measurement-based squeezing applied to vacuum.  We assume that the ancillary state has a fixed level of $r = 1.2$ ($\sim 10.5$ dB) of squeezing, and that the efficiency of the homodyne detection is 0.99. We consider three different target levels of squeezing, $r_{\text{target}} = 0.3, 1, \text{ and }2$, to apply to the vacuum state, and calculate the Wigner functions in the case of ideal, direct inline squeezing; the average map of measurement-based squeezing; and a single-shot occurrence of measurement-based squeezing.

There are a few key observations we can make from the simulation. First, if $r_{\text{target}}$ is comparable to or greater than the level of resource squeezing $r$, then the variance in the $q$ quadrature exceeds the desired value. This is to be expected: \cref{eq:mbsq_avg} tell us that, while the level of noise in the $q$ quadrature is modulated by $e^{-2r}$ (constant across the simulations), it grows with $r_{\text{target}}$. Next, we see that, in the average case, the level of anti-squeezing essentially matches the ideal case. This is due to the high efficiency of the homodyne measurement, which forces the level of $p$ quadrature noise---proportional to $\frac{1-\eta}{\eta}$, per Eq.~\eqref{eq:mbsq_avg}---to be small. Additionally, the mean in the average case matches the ideal case, as guaranteed by \cref{eq:mbsq_avg}. Finally, in the single-shot case, the location at which the state is centred along the $p$ quadrature can vary on the order of the ideal anti-squeezing; moreover, the level of anti-squeezing in the single-shot case becomes significantly lower than ideal as $r_{\text{target}}$ becomes comparable to or greater than the level of resource squeezing $r$. This indicates that the accurate level of anti-squeezing and the correct mean observed on average is an artifact of the averaging process itself: the states that factor into the average have less anti-squeezing but a broad distribution of where they are centred along the $p$ quadrature, resulting in a broadened average output state. The differences between ideal, average, and single-shot instances of measurement-based squeezing are crucial to understand for realistic implementations of bosonic codes; our formalism and simulations are valuable for such analysis.

\subsubsection{GKP Phase Gate}\label{subsubsec:Pgate}
\begin{figure*}
	\centering
        \includegraphics[width = \textwidth]{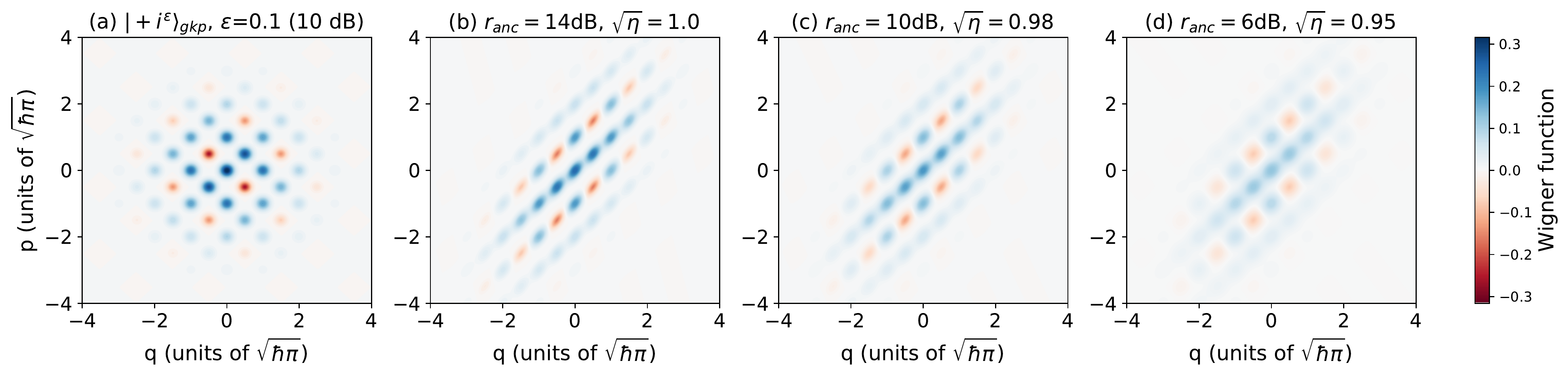}
	\caption{(a) Wigner function for $\ket{+i^\epsilon}_{\rm gkp}$ simulated directly with $\epsilon=0.1$ (10 dB). We compare this to a method for ideally preparing the same state by generating $\ket{+^\epsilon}_{\rm gkp}$ with $\epsilon=0.1$, then applying a GKP phase gate $\hat{P} = e^{i\hbar\hat{q}^2/2}$. Since the phase gate requires inline squeezing (see Appendix \ref{app:mbgates}), we simulate measurement-based squeezing with various levels of ancillary squeezing and ancilla detector efficiency, which we label in (b)-(d). We see that phase gates apply a shearing effect to the envelope and peaks of the finite energy GKP state, with lower quality measurement-based squeezing adding additional broadening to the peaks. This simulation was performed with the \texttt{bosonic} backend of \texttt{Strawberry Fields}.}
	\label{fig:Pgate_wigner}
\end{figure*}   
The phase gate is a pertinent example of a single-mode GKP gate that employs inline squeezing, an operation that may be performed, in practice, through measurement-based methods. In Appendix \ref{subsec:mbsq_gates}, we provide details for such an optical implementation. Here, we present results of a simulation of a realistic phase gate applied to $\ket{+^\epsilon}_{\rm gkp}$ with $\epsilon=0.1$ (10 dB). In \cref{fig:Pgate_wigner}, we plot the Wigner functions of the average output states, varying the level of squeezing of the ancillary squeezed state and the efficiency of the homodyne detection. We compare the graphs to those for $\ket{+i^\epsilon}_{\rm gkp}$ with $\epsilon=0.1$ (10 dB), which is the finite energy version of the ideal application $\hat{P}\ket{+}_{\rm gkp}$. We find several differences. First, while the finite-energy $\ket{+i^\epsilon}_{\rm gkp}$ has a symmetric envelope in phase space, and each Gaussian peak has an isotropic spread about its mean, the application of a phase gate to a finite-energy $\ket{+^\epsilon}_{\rm gkp}$ results in an asymmetric envelope and anisotropic Gaussian peaks about each mean in phase space. This would be the case even if the phase gate was applied perfectly to a finite energy state: the finite-energy $\ket{+^\epsilon}_{\rm gkp}$ is non-periodic due to the envelope, and each peak has finite width; even a perfect phase gate would cause shearing effects in both the envelope and the individual peaks, an effect not seen in ideal states with perfect periodicity and infinitely narrow peaks. Second, as the quality of the squeezed ancilla resource and homodyne efficiency worsen, we find that the peaks in phase space broaden, albeit asymmetrically, decreasing the height of each peak.
  
\begin{figure}
	\centering
        \subfloat[]{\includegraphics[width = \columnwidth]{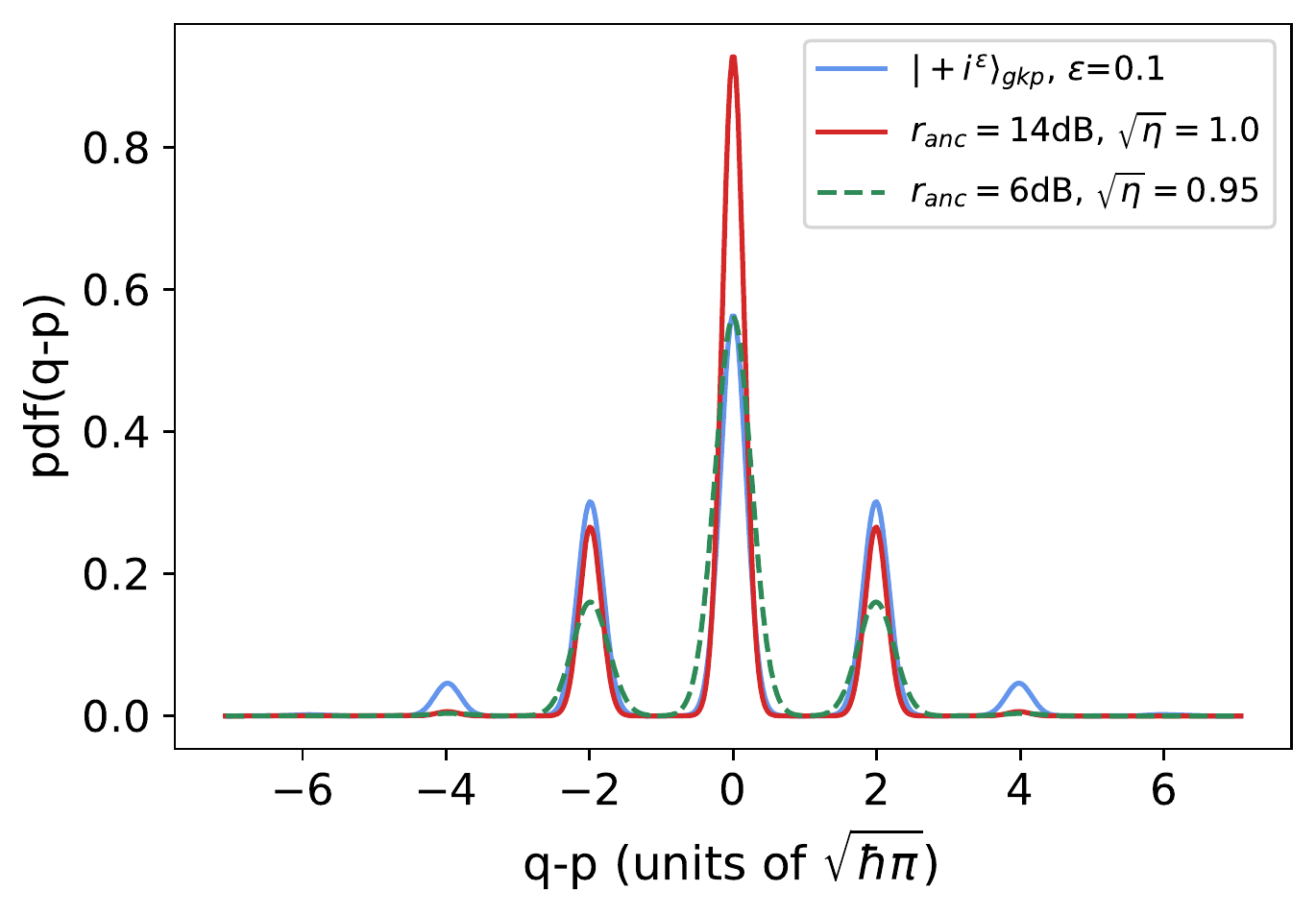}}
        \newline
        \subfloat[]{\includegraphics[width = \columnwidth]{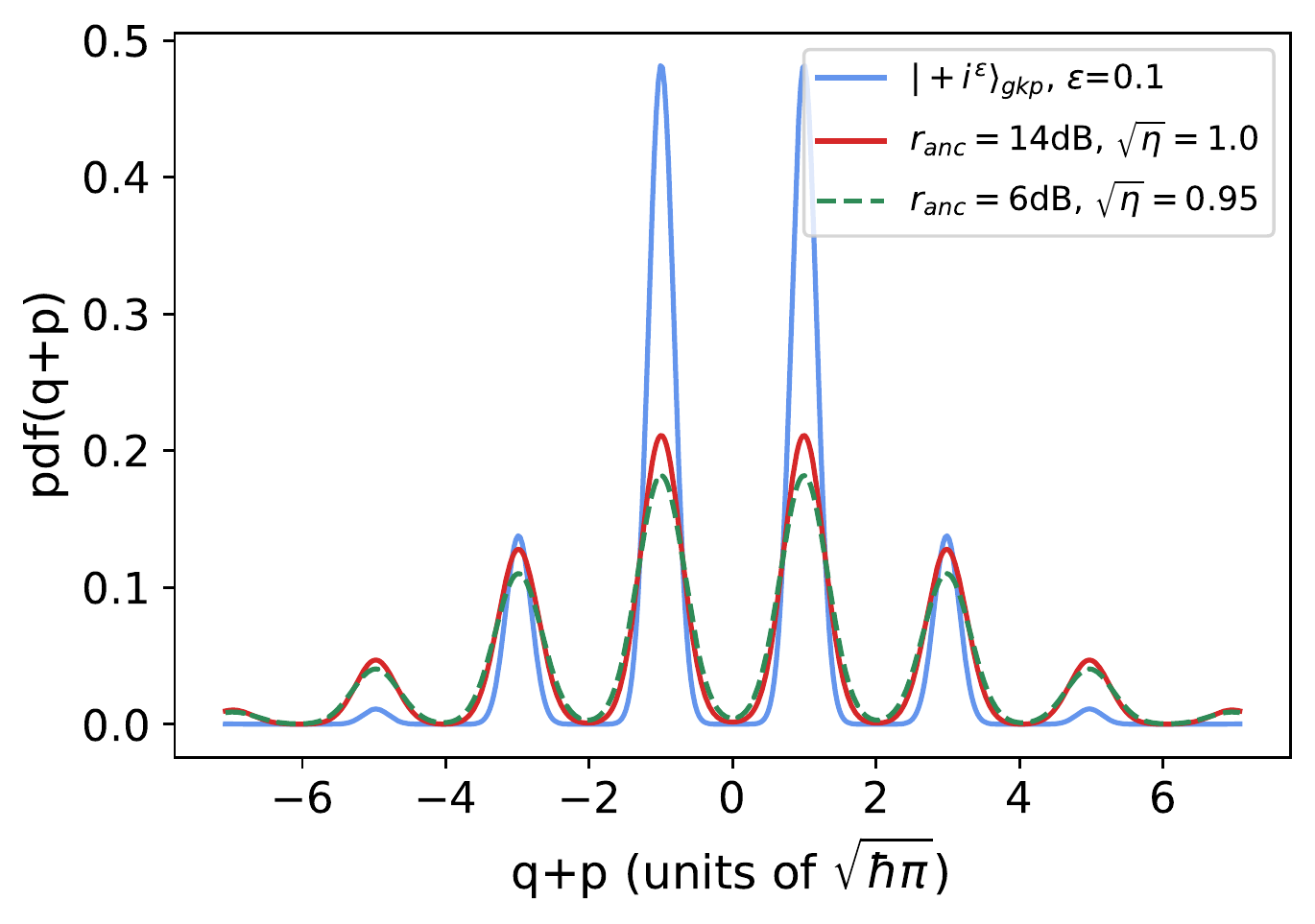}}
	\caption{The marginal distribution for a homodyne measurement along (a) $\frac{q-p}{\sqrt{2}}$ and (b) $\frac{q+p}{\sqrt{2}}$, each followed by a rescaling of the results by $\sqrt{2}$. Binning the result to the nearest $n\sqrt{\pi}$ and taking the parity of (a) $n$ or (b) $n+1$ effects a GKP qubit $Y$ measurement. We present the marginals for Wigner functions (a), (b), and (d) from Fig.~\ref{fig:Pgate_wigner}. While marginal distributions along either $q-p$ or $q+p$ could be sampled for a Pauli Y measurement, the narrower peaks along $q-p$ yield higher likelihood of falling within the correct bin, emphasizing the value of tracking shearing effects to optimize readout fidelity.}
	\label{fig:Pgate_marginal}
\end{figure} 

The differences in the Wigner function do not tell the full story, however; we are also interested in how they affect readout of the state. In \cref{fig:Pgate_marginal}, we plot the marginal distribution along $q-p$ and $q+p$ of three of the states in \cref{fig:Pgate_wigner}. Recall that binning the outcome along $q-p$ to the nearest $n\sqrt{\pi}$ and taking the parity of $n$ corresponds, ideally, to the a measurement in the qubit Pauli $Y$ basis; alternatively, one can bin the outcome along $q+p$ to the nearest $n\sqrt{\pi}$ and take the parity of $n+1$. This choice comes from the fact that $\hat{H}\hat{\sigma}_y \hat{H}^\dagger = -\hat{\sigma}_y$ for qubits, where $\hat{\sigma}_y$ is the Pauli $Y$ operator and $\hat{H}$ is a qubit Hadamard gate operator, effected for GKP states by a CV phase space rotation by $\pi/2$. In practice, these two CV operators can be measured by performing homodyne detection along $\frac{q\pm p}{\sqrt{2}}$ and then rescaling the outcome by $\sqrt{2}$. We find that the choice of readout quadrature affects the probability of obtaining the correct qubit result. Because the Wigner function is sheared in phase space, measuring along $q-p$ results in a narrower envelope and narrower peaks. For the purpose of qubit readout, the peak width is what matters since the likelihood of falling outside a 0-bin does not depend on the number of peaks. The results are presented in Table \ref{tab:Pgate_results}. There is a sizable drop in readout quality going from $q-p$ to $q+p$, even though both represent the same ideal qubit measurement. This is further confirmation that the one-to-many mapping of qubit-to-CV operators affords GKP states an advantageous flexibility~\cite{Tzitrin2020}. Using our simulator, we identify how to adapt the readout strategy based on the noise to obtain a more faithful measurement of the binary qubit outcomes.

\begin{table}[]
    \centering
    \begin{tabular}{p{3.7cm}|p{0.5cm} p{1.5cm} p{1cm}}
        &\multicolumn{3}{c}{$p$(reading out a qubit 0)}\\
      Simulation parameters &  &  $q-p$ & $q+p$\\
      \hline
      $\ket{+i^\epsilon}_{\rm gkp}$, $\epsilon=0.1$ &  & 99.5\% & 99.5\%\\
      $r_{\rm anc}=14$ dB, $\sqrt{\eta}=1$& & 99.9\% & 92.2\%\\
      $r_{\rm anc}=6$ dB, $\sqrt{\eta}=0.95$& & 95.2\% & 87.2\%
    \end{tabular}
    \caption{For the marginal distributions in \cref{fig:Pgate_marginal}, we calculate the probability of correctly identifying a qubit 0 readout from a Pauli Y measurement, depending on whether the CV homodyne measurement is performed along $q-p$ or $q+p$. $q-p$ provides better readout, due to narrowed peaks in the marginal distribution.}
    \label{tab:Pgate_results}
\end{table}

\subsubsection{GKP to CV Cluster Teleportation}
CV cluster states---multimode Gaussian states constructed by stitching together momentum-squeezed states with CV $\CZ$ gates---are a valuable resource for measurement-based quantum computing with GKP states. In particular, a GKP state can be added to and teleported along the cluster using the same operations as would be used for a momentum-squeezed state~\cite{Menicucci2014}. The teleportation circuit begins with a GKP state and a momentum-squeezed state interacting via a $\CZ$ gate; next, a $p$-homodyne measurement is applied to the GKP mode, yielding outcome $p_0$; finally, a shift in $q$ by $-p_0$ is applied to the remaining mode. Here, we seek to understand the dynamics of a realistic optical implementation of GKP teleportation into a CV cluster. As discussed in Appendix \ref{subsec:mbsq_gates}, we break apart the $\CZ$ gate into beam-splitters and inline squeezing operations effected using measurement-based squeezing, and then investigate the effects of finite squeezing and loss in the teleportation circuit. The circuit requires three squeezed states: the ancilla mode onto which the GKP is teleported and the resource states for measurement-based squeezing within the decomposition of the $\CZ$ gate. We vary the level of squeezing of these Gaussian modes, while fixing the GKP state to be $\ket{0^\epsilon}_{\rm gkp}$ with $\epsilon=0.1$ (10 dB). Additionally, we vary loss within the circuit; we apply a loss channel of transmissivity $\eta$ to each mode after initialization and at the output of each of the four beam-splitters in the circuit (two for the decomposition of the $\CZ$ gate and one for each inline squeezing operation). For the three homodyne detectors in the circuit, we fix an efficiency of $\eta_{\text{det}}=99\%$.

\begin{figure}
	\centering
	\subfloat[]{\includegraphics[width = \columnwidth]{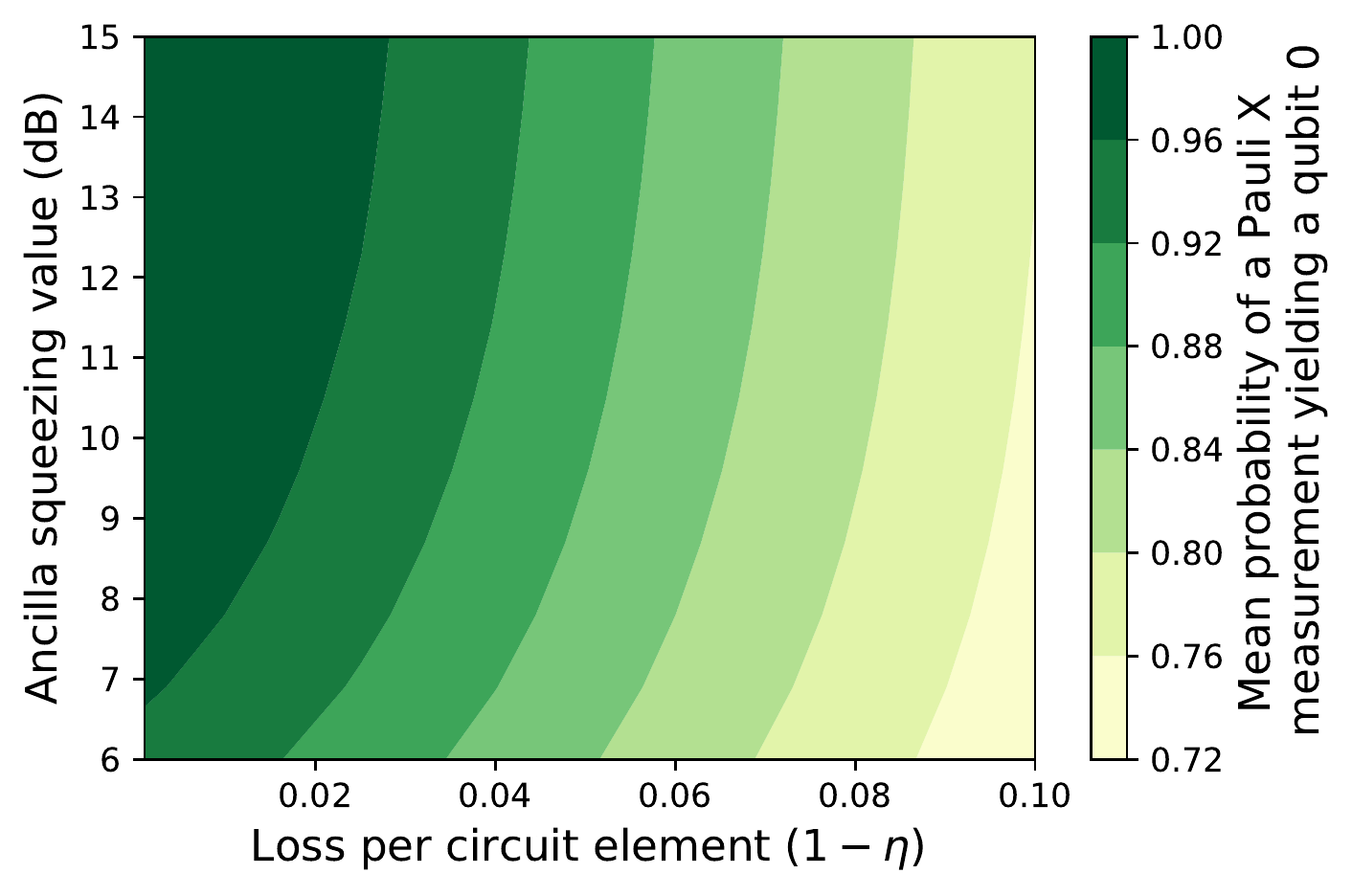}}\\[-0.5ex]
	\subfloat[]{\includegraphics[width = \columnwidth]{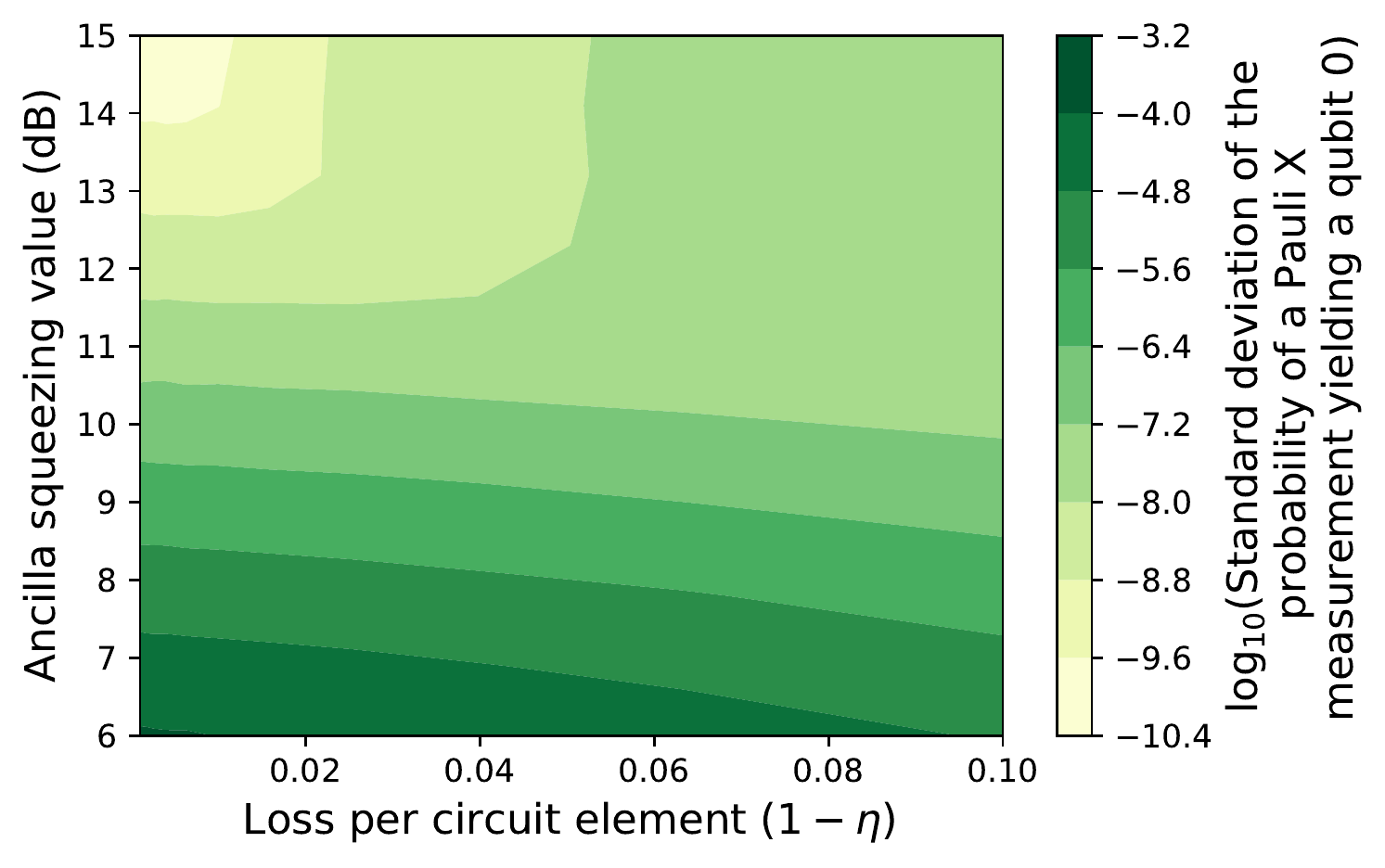}}\\[-0.5ex]
	\subfloat[]{\includegraphics[width = \columnwidth]{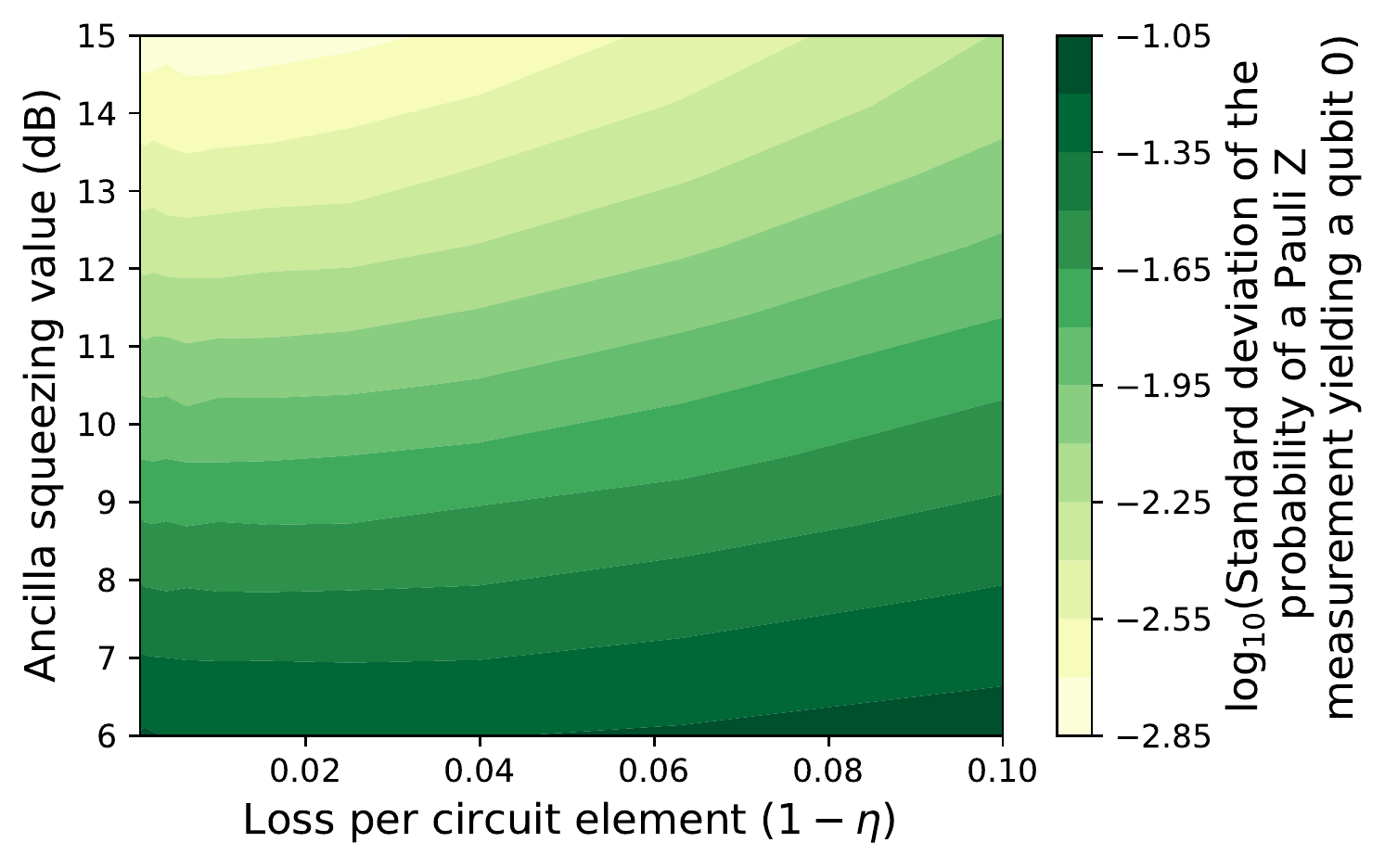}}
	\caption{Teleportation of $\ket{0^\epsilon}_{\rm gkp}$ onto a $p$-squeezed state. Ideally, the teleported state should be $\ket{+}_{\rm gkp}$, with Pauli X (Z) measurement of the state yielding a qubit 0 outcome with 100\% (50\%) probability. However, with noisy teleportation circuits (loss and measurement-based squeezing in the CZ gate), each instance of the circuit, conditioned on the probabilistic value of the teleportation feedforward, yields a slightly different state. We simulate the noisy teleportation circuit 500 times. In (a) and (b), we plot the mean and the (log of the) standard deviation for the probability of a Pauli X measurement yielding 0, as a function of circuit loss and squeezing. Having found the mean probability of a Pauli Z measurement yielding 0 to be roughly $50\%$ across squeezing and loss levels, in (c) we provide the (log of the) standard deviation for the probability a Pauli Z measurement yields 0.} 
	\label{fig:GKP_teleport}
\end{figure}

For each value of squeezing and $\eta$, we run 500 simulations, the teleported state output by each run depending on the probabilistic feedforward value. Then, given the final state in each run, we measure the qubit Pauli $\X$ and $\Z$ operators, achieved by performing homodyne measurements along $p$ and $q$, respectively, binning the outcome to the nearest $n\sqrt{\pi}$, then taking the parity of $n$. Ideally, teleportation of $\ket{0}_{\rm gkp}$ onto a CV cluster yields $\ket{+}_{\rm gkp}$, which would return a qubit 0 outcome $100\%$ of the time when measuring Pauli X, and $50\%$ of the time when measuring Pauli Z. In \cref{fig:GKP_teleport} (a), we plot---as a function of squeezing and loss---the mean probability of obtaining the outcome 0 from a Pauli X measurement. As expected, we see that high squeezing and low loss provide the best readout. Interestingly, we also find that above 10 dB of squeezing (an amount comparable to the per-peak squeezing of the teleported state), loss is the major determinant of readout quality for this Pauli measurement. We are also interested in how far individual runs can deviate from the mean probability of correct readout, so in \cref{fig:GKP_teleport} (b), we plot the spread of the probability from (a) as a function of squeezing and loss. We find that the spread is quite small (only a tenth of a percentage in the worst case), meaning that individual runs yield values close to the correct readout. 

The story differs for the Pauli Z measurement. \cref{fig:GKP_teleport} (c), where we plot the spread for the probability of reading out a qubit 0 outcome in the Pauli Z measurement, shows that there can be significant deviation from the $50\%$ mean value in the presence of finite squeezing and loss. While in the ideal case, every instance of the teleportation circuit yields a teleported state that has a 50-50 chance of generating outcomes 0 or 1 in a Pauli Z measurement, finite squeezing and loss can distort the readout odds: certain instances of the teleportation circuit (heralded by homodyne measurements of the ancillae) can cause one outcome to be favoured by several extra percentage points. Interestingly, in contrast to the Pauli X measurement, the spread of results in the Pauli Z measurement are more sensitive to the level of squeezing than to the level of loss.

While it is intuitive that low loss and high squeezing should yield qubit outcomes closer to ideal, it is interesting---for this choice of initial GKP states---that that loss (squeezing) is the greater noise source for Pauli X (Z) measurements.  
It is worth investigating, however, whether a different feedforward function or postselection of outcomes can improve the faithfulness of qubit readout, and whether the homodyne outcomes can be used to assign a confidence rating to the qubit readout of the teleported mode. Such strategies have been explored in the context of fault-tolerant quantum computing architectures with GKP qubits~\cite{Fukui2018,Vuillot2019,bourassa2021blueprint}.

\subsubsection{GKP T Gate Teleportation}
While Clifford gates can be performed using Gaussian resources in the GKP encoding, one requires a non-Clifford---for GKP states, non-Gaussian---gate to achieve universality. As we discussed in Section \ref{subsec:Tgate}, the qubit T gate can be applied via gate teleportation using a GKP magic state $\ket{M}_{\rm gkp}$. A realistic T gate teleportation is no small feat: an ancillary GKP state is required in addition to the three ancillary squeezed states for the $\CZ$ gates and the feedforward phase gate, and the multiple beam-splitters each incur loss.  As far as we are aware, no investigation of realistic T gate application has been carried out, likely due to the difficulty of simulation. Let us close this gap.

\begin{figure*}
	\centering
        \includegraphics[width = \textwidth]{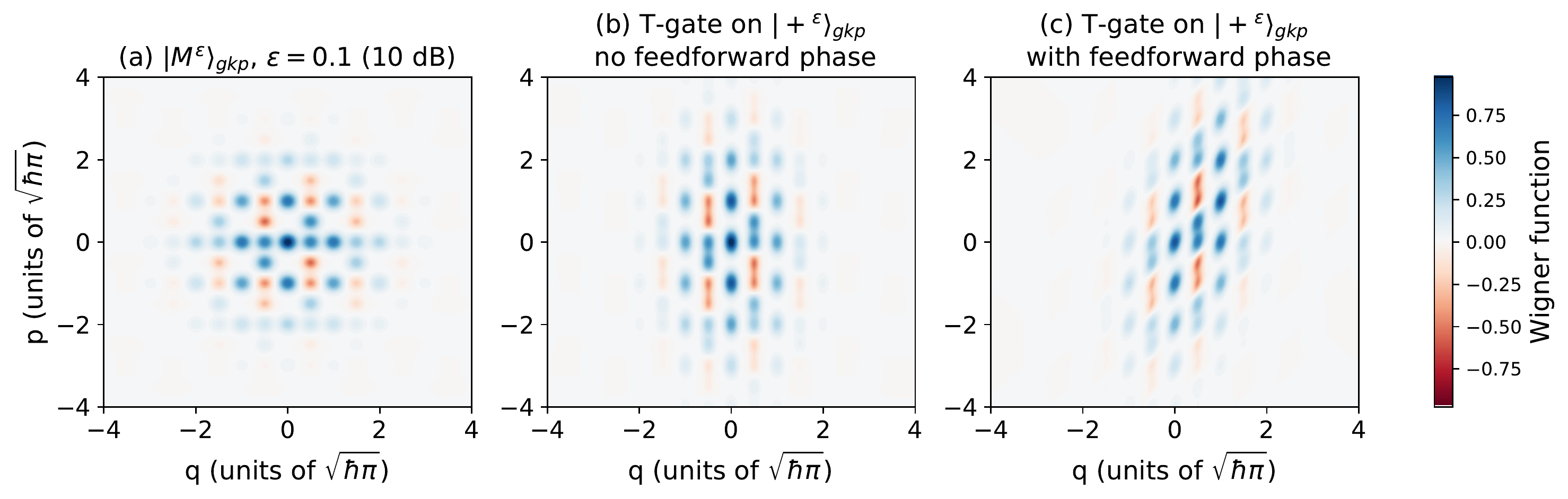}
	\caption{(a) Wigner function for a GKP magic state $\ket{M^\epsilon}_{\rm gkp}=\hat{E}(\epsilon)[\frac{1}{\sqrt{2}}(e^{-i\pi/8}\ket{0}_{\rm gkp}+e^{i\pi/8}\ket{1}_{\rm gkp})]$ with $\epsilon=0.1$ (10 dB). (b) Wigner function for T gate teleportation applied to $\ket{+^\epsilon}_{\rm gkp}$ with $\epsilon=0.1$ (10 dB) using the circuit from \cref{fig:Tgate} and postselecting on an outcome that does not require the feedforward phase gate. (c) Output from the same circuit, but postselecting on an outcome that does require the feedforward phase gate. The gate teleportation circuit for the T gate with realistic components is involved, but now simulatable using the \texttt{bosonic} backend of \texttt{Strawberry Fields}.}
	\label{fig:Tgate_wigner}
\end{figure*} 

First, in \cref{fig:Tgate_wigner}, we present the result of a T gate applied to $\ket{+^\epsilon}_{\rm gkp}$ with $\epsilon=0.1$ (10 dB); in the ideal limit ($\epsilon\rightarrow0$), this should return the state $\ket{M}_{\rm gkp}$. For sake of clarity, in this figure we assume that the $\CZ$ gate and the phase gate are applied perfectly, but employ a finite-energy resource state 
\begin{equation}
    \ket{M^\epsilon}_{\rm gkp}=\hat{E}(\epsilon)[\frac{1}{\sqrt{2}}(e^{-i\pi/8}\ket{0}_{\rm gkp}+e^{i\pi/8}\ket{1}_{\rm gkp})],
\end{equation} 
also with $\epsilon=0.1$ (10 dB). In (a) we present the Wigner function for a finite energy $\ket{M^\epsilon}_{\rm gkp}$ to use for comparison, while in (b) we provide the Wigner function for the output of the T gate circuit given a measurement outcome on the ancillary GKP mode that does not require the feedforward phase. In (c), we present the same as (b) but for an instance of the T gate circuit that does require feedforward phase. We see that the gate works to some extent: the positive and negative peaks in (b) and (c) are in the same places as positive and negative peaks in (a), and we recall from Section \ref{subsec:GKP} that the logical information of the state is encoded in the weights. However, even with the $\CZ$ and feedforward phase gate implemented perfectly, we still see some differences between the Wigner functions. First, notice that the $p$ quadrature variance of each peak is now larger than the $q$ quadrature variance, even though they began as symmetric. This is because the $\CZ$ gate adds the $q$ variance of the resource magic state to the data state. Second, in (c), there is a shearing effect of both the overall envelope and the individual peaks due the phase gate, as also seen in Section \ref{subsubsec:Pgate}.

\begin{figure}
	\centering
        \includegraphics[width = \columnwidth]{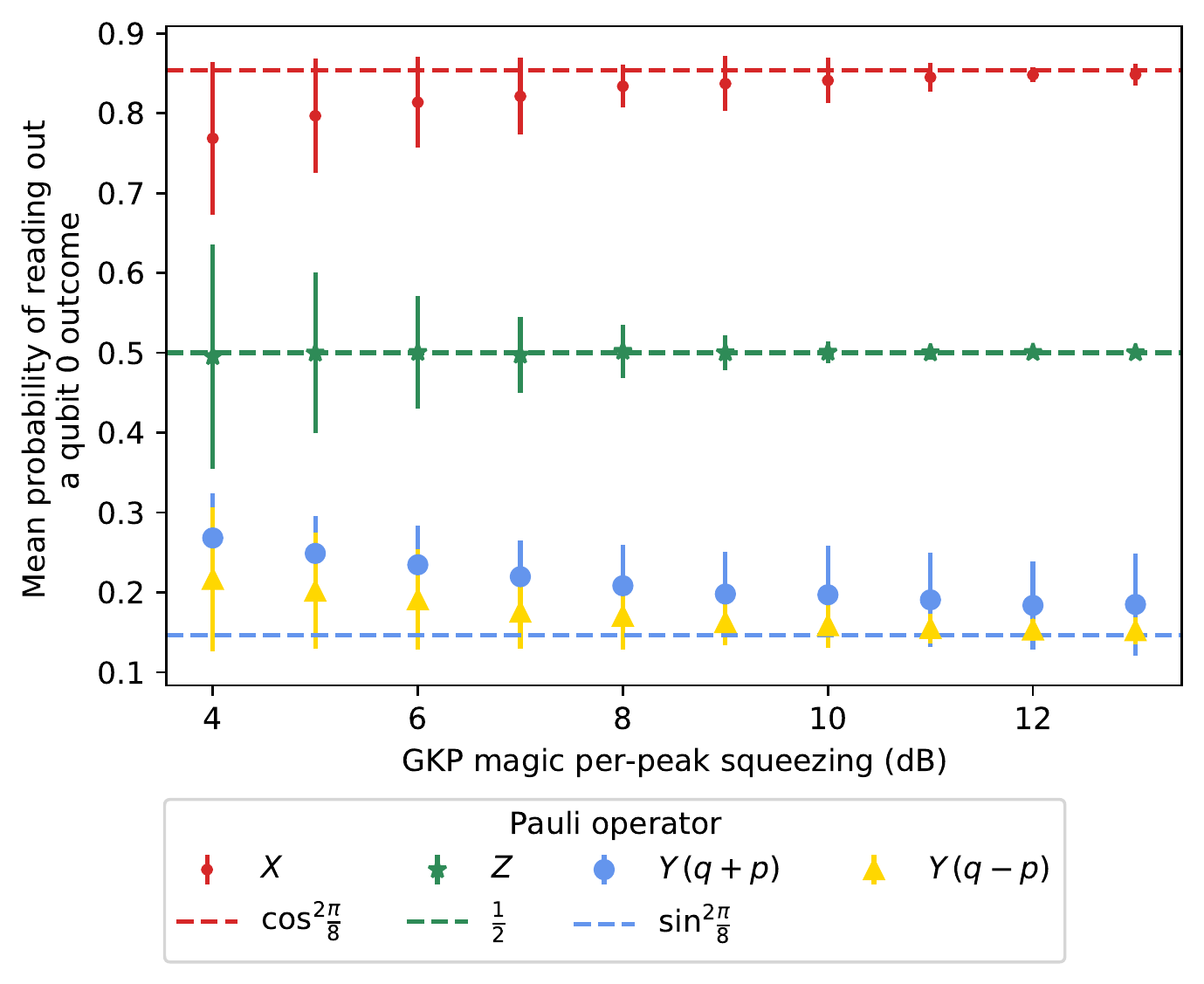}
	\caption{Mean probability of obtaining a qubit 0 outcome from different Pauli operator measurements after applying a T gate to $\ket{+^\epsilon}_{\rm gkp}$ with $\epsilon=0.1$ (10 dB). We vary the per-peak squeezing of the GKP magic resource state and see how well the qubit readout matches the ideal readout values for the Pauli operators (dashed lines). We see that a per-peak squeezing on the order of 11 dB (close to that of the data state) is sufficient to retrieve correct readout to within 1\%. The simulation provides another example of the usefulness of having two methods for performing Pauli Y measurements, as seen in \cref{fig:Pgate_marginal}; homodyne detection and binning along $q-p$ yields a more faithful outcome.}
	\label{fig:Tgate_readout}
\end{figure}

Next, we examine the effect of finite-energy magic states on the quality of the T gate teleportation, keeping the rest of the circuit ideal. We again simulate the T gate circuit applied to $\ket{+^\epsilon}_{\rm gkp}$ with $\epsilon=0.1$ (10 dB), this time varying the value of $\epsilon$ for $\ket{M^\epsilon}_{\rm gkp}$. For each $\epsilon$, we perform 500 simulations, each time producing a different output state based on the probabilistic measurement of the ancilla. For the output state of each simulation, we calculate marginal distributions in phase space to determine the probability of reading out a qubit 0 outcome for the four Pauli measurements (three Pauli operators, and two ways of measuring Pauli Y). In \cref{fig:Tgate_readout}, we plot how the mean probability of obtaining a qubit 0 readout value for the Pauli measurements varies as a function of the magic state's finite energy parameter $\epsilon$. We additionally provide what the ideal probabilities for readout should be for the Pauli measurements. We find that, with smaller epsilon (increased per-peak squeezing), the statistics become more closely aligned with the ideal values, along with less variation in the output state of each instance of the circuit, with $\epsilon$ on the order of $11$ dB seeming to be sufficient to achieve correct readout to within 1\%. It is good news that the quality of magic state required is not significantly higher than that of the data state. We also note that measuring the Pauli Y operator by performing homodyne along $q-p$ again yields more faithful readout of the qubit outcomes.

Finally, we perform the same simulation of a GKP T gate applied to $\ket{+^\epsilon}_{\rm gkp}$, this time implementing the $\CZ$ and phase gates using realistic components: lossy beam-splitters, inefficient homodyne measurements, and measurement-based squeezing. We fix the magic state to have $\epsilon = 0.1$ (10 dB); we set the efficiency for the four homodyne measurements (three for measurement-based squeezing in the $\CZ$ and feedforward phase gate, and one on the ancillary magic state) to be 99\%; and we choose a squeezing level of 12 dB for the three squeezed states used for measurement-based squeezing. We investigate how the Pauli operator readout varies with respect to a loss parameter $\eta$, which we apply in multiple places throughout the circuit: right after each state (GKP or squeezed) is initialized and after each beam-splitter. 

\begin{figure}
	\centering
        \includegraphics[width = \columnwidth]{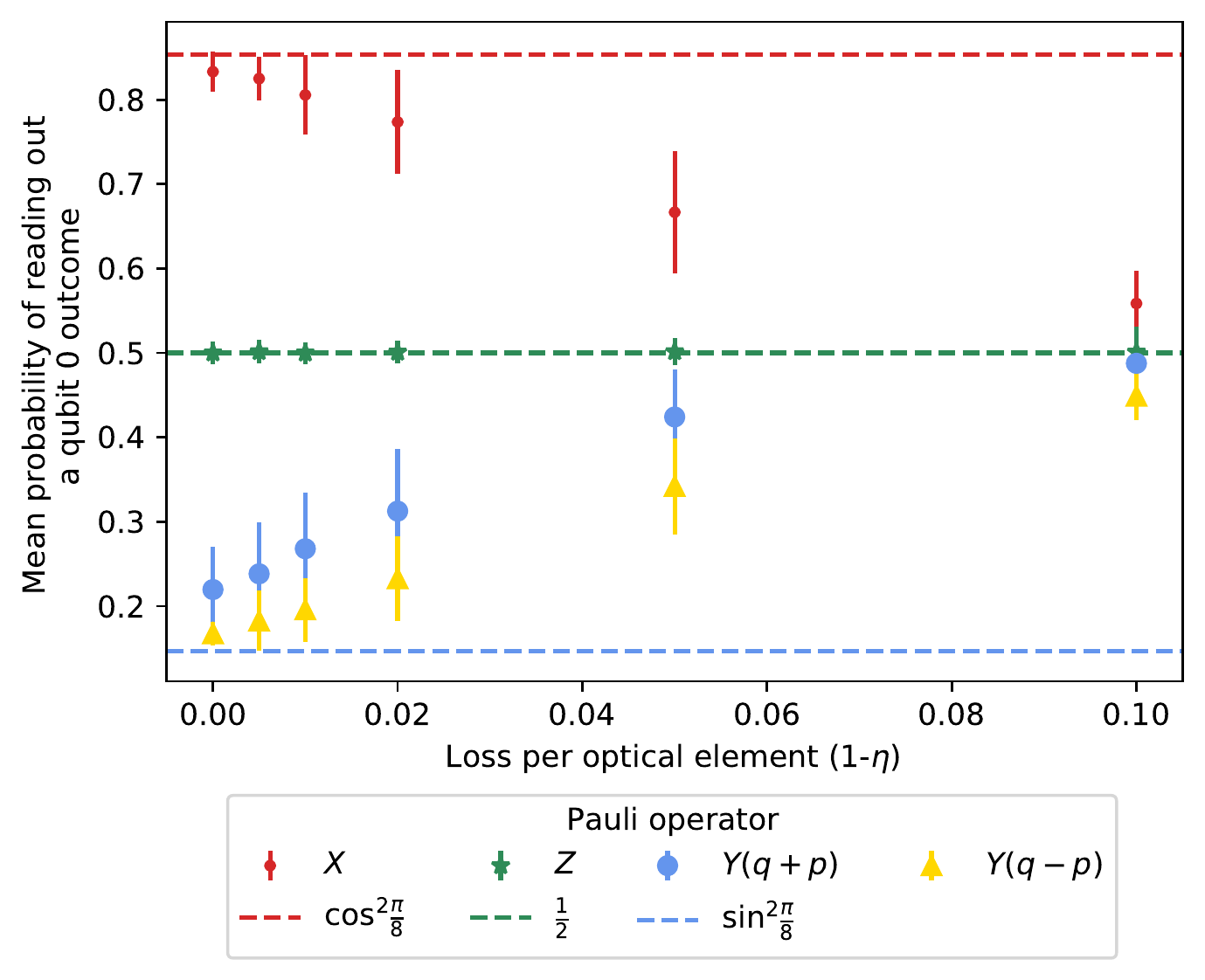}
	\caption{Mean probability of obtaining a qubit 0 outcome from different Pauli operator measurements after applying a T gate to $\ket{+^\epsilon}_{\rm gkp}$ with $\epsilon=0.1$ (10 dB). Here, the gate teleportation is effected using imperfect optical elements (see Figs. \ref{fig:Tgate} and \ref{fig:mbsq_circuits} for circuit breakdowns). We fix the ancillary magic state to have the same $\epsilon$ as the data state; we employ 12 dB squeezed states for measurement-based squeezing, and set the homodyne efficiency to 99\%. We see how readout changes as we vary $\eta$, which parametrizes the loss channel we apply after each of the five states are initialized and after each of the five beam-splitters.}
	\label{fig:Tgate_loss}
\end{figure} 

A few comments are in order. First, in the case of no loss, we already see that the readout is not as good as when the $\CZ$ and phase gates are applied perfectly; this is undoubtedly due to the change from idealized squeezing to measurement-based squeezing; even with squeezed resource states of 12 dB, the readout has room for improvement, motivating deployment of even more highly squeezed states. Second, as loss per optical element is increased, the readout quality quickly deteriorates. This is likely due to the number of loss channels the data state undergoes, since it must pass through four beam-splitters to perform the teleportation. Notably, only the Pauli $\Z$ readout is still close to the desired value, but this is coincidence: a broad, noisy distribution is essentially uniform over the bins applied to the quadrature outcomes, resulting in a 50\% chance of reading out a logical 0 from a Pauli measurement. The sensitivity of the optically-teleported T gate to loss, due to its many components, is additional motivation for pursuing passive optical implementations of computing with GKP states~\cite{walshe2020continuous}; by potentially eliminating in-line squeezing, these architectures reduce opportunities for loss and noise brought in by measurement-based squeezing.

\section{Summary and open problems}\label{sec:conclusion}
In this work, we have introduced a formalism for simulating continuous-variable quantum states.
At its heart is a representation of these states as linear combinations of Gaussian functions in phase space. 
This novel framework can be used to analyze and simulate valuable classes of CV states, namely bosonic qubits like GKP, cat, and Fock states, under realistic transformations. 
Our mathematical framework inherits the simplicity and convenience of the Gaussian CV formalism, requiring only to track how transformations and measurements update the weights, means, and covariance matrices of the Gaussian functions in the linear combination. 
In addition to the straightforward inclusion of Gaussian channels in our formalism, a salient class of non-Gaussian transformations and measurements---photon-number-resolving measurements and gate teleportation for GKP and cat qubits---can also be simulated using our framework. 

Enabled by our formalism, we provided simulation methods that outperformed state-of-the-art CV simulators that employ the Fock basis.
Using our new method, we were able to perform novel simulations of bosonic qubits in realistic settings, informing future design decisions for quantum computers based on bosonic qubits. 
We focused on GKP qubits, examining how they transform under gates that leverage measurement-based squeezing, how they interface with CV clusters, and how they transform under non-Clifford gates implemented via gate teleportation.
While we only investigated how the readout of Pauli operators was affected in these settings, we emphasize that our simulator outputs complete state information in the form of the elements required to construct the Wigner function, which can be used as input to other simulators or to more advanced analytical tools, such as the modular subsystem decomposition~\cite{pantaleoni2020modular,pantaleoni2021}.

Numerical simulations leveraging our formalism and methods were performed in the new \texttt{bosonic} backend of \texttt{Strawberry Fields}, an existing feature-rich Python library for simulating CV optical circuits. 
The backend, implemented by the authors, is accessible through a user-friendly frontend of the public repository \cite{bosonic_github}, along with some tutorials on its use \cite{tutorial1,tutorial2,tutorial3}; we hope this encourages exploration by those interested in CV quantum information.

We anticipate several useful applications of our work. For one, our new formalism will prove valuable in the design of circuit modules for quantum computers that employ bosonic qubits, as it can identify, for example, where losses need to be reduced in a circuit, or what level of ancillary squeezing is required to attain a target gate fidelity. While our focus in the simulations has been on GKP qubits, similar analysis of realistic gate application for cat states can be performed. Second, it is known that in optical settings where deterministic high-order nonlinearities are currently lacking, bosonic qubits will need to be produced probabilistically in the near term. There are promising heralded state generation techniques that involve photon-number-resolving measurements on some modes of a multimode Gaussian state~\cite{eaton2019non,Tzitrin2020}, a scenario which we have already discussed falls neatly within our formalism.
Previous investigations and optimizations of this technique were limited by the speed and memory limitations of the Fock basis; a straightforward extension would be to use our new formalism to develop these bosonic state preparation methods even further. Yet another state preparation proposal for GKP states relies on interacting or ``breeding" cat states with beam-splitters and homodyne measurements on some of the modes~\cite{vasconcelos2010all,weigand2018generating}. Since the circuits in those protocols consist of Gaussian transformations and measurements applied to cat states, our formalism applies here too.

Finally, there are several avenues for future research.
First, it is worth investigating the connection between other analytical representations of CV states (such as Riemann-Theta functions for GKP qubits~\cite{Baragiola2019,Matsuura2020}, or the shifted Fock basis representation of cat states~\cite{chamberland2020building}) and the ones we have presented here to identify any additional opportunities for faster or more accurate simulation.
Second, for the class of states and maps that we have considered, it would be of interest to determine the exact mathematical conditions for physicality in terms of the parameters of the states (weights, vectors of means and covariances matrices). Moreover, while we considered a wide class of transformations which fall under this formalism, it would be valuable to determine the most general class of transformations which can be reduced to manipulating the weights, means and covariances of the Gaussian functions in the linear combination.
Lastly, large-scale simulation of bosonic qubits under realistic noise models is challenging due to exponential scaling. Our formalism is a step towards numerically tractable models capturing some of those effects, either by developing suitable approximations of the states based on the mathematical framework provided, or by using the simulator to develop more detailed models for the evolution of qubit-level information or noise, then leveraging existing work in the field of quantum fault-tolerance for qubits.

\begin{acknowledgements}
We are grateful for discussions with Rafael Alexander, Giacomo Pantaleoni, Daiqin Su and Barbara Terhal. 
J.E.B. is supported through an Ontario Graduate Scholarship, and by Mitacs through the Mitacs Accelerate program grant. 
I.T. is supported by Mitacs through the Mitacs Accelerate program grant. 
\end{acknowledgements}

\onecolumngrid
\appendix{}
\section{Weyl Transform of the Outer Product of Two Gaussian States}\label{app:dyad}
In this appendix we obtain the Wigner function of the outer product $\hat{O} = \ket{\psi} \bra{\phi}$ where 
\begin{align}
\ket{\psi} &= \hat{D}(\gamma) \hat{S}(r) \ket{0}, \quad \bra{\phi} = \bra{0} \hat{S}^\dagger(r) \hat{D}^\dagger(\delta),
\end{align}
are pure Gaussian states.
To obtain the Wigner function recall that it can be calculated as (cf. Eq. 4.5.21 of Ref.~\cite{barnett2002methods})
\begin{align}\label{eq:wigner_general}
W_{\hat{O}}(\alpha) = \frac{2}{\pi^2} \int d^2 & \beta \braket{0|\hat{D}^\dagger(\alpha+\beta) \hat{O} \hat{D}(\alpha - \beta)|0} 
\exp\left(\alpha^* \beta - \alpha \beta^* \right),
\end{align}
where $\alpha,\beta \in \mathbb{C}$ and we parametrize $\alpha = \tfrac{1}{\sqrt{2\hbar}} (q+ i p) $.
We now recall (cf. Eq. 3.6.30 of Ref.~\cite{barnett2002methods}) the composition law for displacement operators
\begin{align}
\hat{D}(\alpha) \hat{D}(\alpha') = \hat{D}(\alpha+\alpha')  \exp\left(\tfrac{1}{2} \left[ \alpha \alpha'^{*} - \alpha^* \alpha' \right]  \right),
\end{align}
and the expression for the vacuum to vacuum amplitude of a Gaussian transformation (cf. Appendix 1 of Ref.~\cite{miatto2020fast})
\begin{align}
\braket{0|\hat{D}(\alpha)\hat{S}(r)|0} = \tfrac{1}{\sqrt{\cosh r}} \exp\left[-\tfrac{1}{2} \left\{|\alpha|^2+\alpha^{*2} \tanh r \right\}\right].
\end{align}
With these two expressions it is direct to write the integrand in Eq.~\eqref{eq:wigner_general} as a Gaussian in the real and imaginary parts of $\beta$ and perform the integral to find the Wigner function to be a normalized Gaussian in $\bm{\xi} = (q,p)$ with covariance matrix and (complex) means
\begin{align}
\bm{\Sigma} &= \tfrac{\hbar}{2} \text{diag}(e^{-2r},e^{2r}), \quad \bm{\mu} = \sqrt{\tfrac{\hbar}{2}} \begin{bmatrix}
\Re(\gamma)+\Re(\delta) + i e^{-2r}\{ \Im(\gamma) - \Im(\delta) \} \\
\Im(\gamma) + \Im(\delta) + i e^{2 r} \{ \Re(\delta) - \Re(\gamma) \}
\end{bmatrix}
\end{align}
multiplied by the prefactor
\begin{align}
c = \exp\Big[&-\tfrac{1}{2} e^{-2 r} (\Im(\gamma )-\Im(\delta ))^2 -\tfrac{1}{2} e^{2 r} (\Re(\gamma )-\Re(\delta ))^2 
-i\Im(\delta ) \Re(\gamma )+ i \Im(\gamma ) \Re(\delta) \Big].
\end{align}
In the limit where $\delta \to \gamma$ we find $c=1$ and $\bm{\mu} = \sqrt{2 \hbar} (\Re(\gamma),\Im(\gamma))$ as expected for a pure Gaussian state.
Finally, note that so far we only considered outer products $\hat O = \ket{\psi} \bra{\phi}$ where the squeezing magnitude and direction are the same. One can consider the more general situation in which they are not. 
The final expression for the Weyl transform of $\hat{D}(\gamma)\hat{S}(r e^{-i \theta}) \ket{0} \bra{0}S^{\dagger}(s e^{-i\phi}) \hat{D}^\dagger(\delta)$ is a Gaussian in the real and imaginary parts of $\alpha,\delta$ and $\gamma$ given by $\mathcal{R} \exp\left( -\tfrac{1}{2} \bm{\nu}^T \bm{A} \bm{\nu} \right) $ with
\begin{align}
\mathcal{R} &= \frac{2}
{\pi \sqrt{\cosh r \cosh s -e^{i (\theta-\phi) } \sinh r \sinh s }},\quad 
   \bm{\nu}^T = [\Re(\alpha),\Im(\alpha), \Re(\gamma), \Im(\gamma),\Re(\delta),\Im(\delta)], \\
   \bm{A} &= -\left(
\begin{array}{cccccc}
 c_+ & d_0 & -\frac{c_+}{2} & d_+ & -\frac{c_+}{2} & d_- \\
 d_0 & c_- & d_- & -\frac{c_-}{2} & d_+ & -\frac{c_-}{2} \\
 -\frac{c_+}{2} & d_- & \frac{c_+}{4} & \frac{d_0}{4} & \frac{c_+}{4} &
   -\frac{d_-}{2} \\
 d_+ & -\frac{c_-}{2} & \frac{d_0}{4} & \frac{c_-}{4} & -\frac{d_+}{2} &
   \frac{c_-}{4} \\
 -\frac{c_+}{2} & d_+ & \frac{c_+}{4} & -\frac{d_+}{2} & \frac{c_+}{4} &
   \frac{d_0}{4} \\
 d_- & -\frac{c_-}{2} & -\frac{d_-}{2} & \frac{c_-}{4} & \frac{d_0}{4} &
   \frac{c_-}{4} \\
\end{array}
\right),\\
c_\pm&= \frac{2}{\frac{1}{1\pm e^{i \phi } \coth s}-\frac{1}{1 \pm e^{i \theta } \tanh r}},\quad 
d_\pm=\frac{i \left(\mp 1+e^{i \theta } \tanh r\right) \left(\tanh s \pm e^{i \phi
   }\right)}{e^{i \theta } \tanh r \tanh s-e^{i \phi }},\quad 
d_0 = -d_+ - d_-.
\end{align}

\section{Coefficients of Ideal GKP} \label{anc:coeffs}
For ideal GKP qubits in the state $\vert \psi \rangle = \cos{\frac{\theta}{2}} \vert 0 \rangle_{\rm gkp} + e^{-i \phi} \sin{\frac{\theta}{2}} \vert 1 \rangle_{\rm gkp}$, the coefficients for the Wigner function are provided by~\cite{Garcia_2021}:
\begin{eqnarray}
        c_{k,\ell}(\theta,\phi)=
        \begin{cases}
            \frac{1}{4\sqrt{\pi}}  & \mbox{ for $k \bmod{2} = 0$, $\ell \bmod{2} = 0$,} \\
            \frac{1}{4\sqrt{\pi}} \cos \theta & \mbox{ for $k \bmod{4} = 0$,  $\ell \bmod{2} = 1$,} \\
            -\frac{1}{4\sqrt{\pi}} \cos \theta & \mbox{ for $k \bmod{4} = 2$, $\ell \bmod{2} = 1$,} \\
            \frac{1}{4\sqrt{\pi}} \sin \theta \cos \phi & \mbox{ for } \begin{cases} \mbox{$k\bmod{4} = 3$, $\ell\bmod{4} = 0$,} \\ \mbox{$k\bmod{4} = 1$, $\ell\bmod{4} = 0$,} \end{cases} \\
            \frac{-1}{4\sqrt{\pi}} \sin \theta \cos \phi & \mbox{ for } \begin{cases} \mbox{$k \bmod{4} = 3$, $\ell \bmod{4} = 2$,} \\ \mbox{$k \bmod{4} = 1$, $\ell \bmod{4} = 2$,}
            \end{cases} \\
            \frac{-1}{4 \sqrt{\pi}} \sin \theta \sin \phi & \mbox{ for } \begin{cases} \mbox{$k \bmod{4} = 3$, $\ell \bmod{4} = 3$,} \\ \mbox{$k \bmod{4} = 1$, $\ell \bmod{4} = 1$,}
            \end{cases} \\
            \frac{1}{4 \sqrt{\pi}} \sin \theta \sin \phi & \mbox{ for } \begin{cases} \mbox{$k \bmod{4} = 3$, $\ell \bmod{4} = 1$,} \\ \mbox{$k \bmod{4} = 1$, $\ell \bmod{4} = 3$.} \end{cases}
        \end{cases}
\end{eqnarray}

\section{Derivation of the Wigner function of a Finite-Energy GKP State} \label{app:wigner_GKP}

This Appendix explicitly computes the Wigner function shown in eq.~\eqref{eq:gkp_wigner_explicit} using two distinct mathematical methods: direct calculation using the Moyal product in Section~\ref{subann:GKP_Moyal}, and the physical application of the Fock damping operator using an optical circuit in Section~\ref{subann:GKP_optical}.

\subsection{Real Weights, Means and Covariances \emph{Via} the Moyal Product} \label{subann:GKP_Moyal}

Recall that the Wigner function of the \emph{physical} state $W_{\psi}^{\text{phys}} \left(\bm \xi \right)$ is given by~\cite{HILLERY1984121}:
\begin{equation} \label{eqn:def_Wigner_real}
    W_{\psi}^{\text{phys}} = W^{\epsilon} \star W_{\psi}^{\text{ideal}} \star W^{\epsilon},
\end{equation}
where $\star$ denotes the Moyal product, $\psi$ represents the encoded qubit state, and $W^{\epsilon}$ denotes the Weyl transform of the Fock damping operator $\hat{E}(\epsilon)=e^{-\epsilon\hat{n}}$. The $\star$-product for a single mode is defined by
\begin{equation} \label{eqn:def_star_prod}
    A \star B := A e^{\frac{i \hbar}{2} \left( \overleftarrow{\partial}_q \overrightarrow{\partial}_p - \overleftarrow{\partial}_p \overrightarrow{\partial}_q \right)} B,
\end{equation}
where the arrow above the derivative symbols indicate to take the derivative to the function to the left or right-hand side. The $\star$-product is linear and associative, but not commutative.

Since $W_{\psi}^{\text{ideal}}$ takes the form
\begin{equation} \label{eqn:Wigner_ideal}
    W_{\psi}^{\text{ideal}} \left( \bm \xi \right) = \sum_{\bm{\mu}_0} c_{\bm{\mu}_0} \left( \psi \right) \delta \left(\bm \xi - \bm{\mu}_0 \right),
\end{equation}
we find
\begin{align} \label{eq:W_phys}
    W^{\text{phys}}_{\psi} &= \sum_{\bm \mu_0} c_{\bm \mu_0} \left( \psi \right) W^{\text{phys}}_{\bm \mu_0}, \nonumber \\
    W_{\bm \mu_0}^{\text{phys}} &= W^{\epsilon} \star W_{\bm \mu_0}^{\text{ideal}} \star W^{\epsilon},
\end{align}
where we have defined $W_{\bm \mu_0}^{\text{ideal}} = \delta \left( \bm \xi - \bm \mu_0 \right)$.

Before calculating $W^{\text{phys}}_{\mu_0}$ we recall that the Fock damping operator is simply $e^{-\epsilon \hat{n}} = (1+\braket{\hat{n}}) \hat{\rho}_{\text{thermal}} $ with $\braket{\hat{n}}  = (e^{\epsilon}-1)^{-1}$ and ${\rho}^{\text{th}}$ is a thermal state (cf. Eq.~\eqref{eq:thermalstate}). We can then write the Weyl transform of the Fock damping operator as an unnormalized Gaussian
\begin{align}
W^{\epsilon}(\bm{\xi}) =  \frac{1}{1-e^{-\epsilon}} G_{\bm{0},\bm{\Sigma}(\epsilon)}(\bm{\xi}) \text{ with } \bm{\Sigma}(\epsilon) = \tfrac{\hbar}{2} \coth \epsilon \ \id . 
\end{align}
With this result in hand we again use that the $\star$-product is associative, and we have the identity~\cite{PhysRev.109.2198}:

\begin{equation}
    A \star B \left( \bm \xi \right) = \frac{1}{\pi^2}\int dy dk dy' dk' e^{2i \left(yk' - y'k \right)/\hbar} A\left(q+y, p+k \right) B \left(q+y', p+k' \right),
\end{equation}
where we implicitly use the fact that $A, B$ admit Fourier transforms.

We first evaluate
\begin{equation}
   W^{\epsilon} \star W^{\text{ideal}}_{\bm \mu_0} = N \int dy dk dy' dk' e^{2i\left(y k' - y' k \right)/\hbar} e^{- \frac{1}{2} \left(q+y, p+k \right)\Sigma^{-1}_{\epsilon} \left( \begin{smallmatrix} q+y \\ p+k \end{smallmatrix}\right) } \delta \left( q + y' - q_0 \right)\delta \left( p + k' - p_0 \right),
\end{equation}
where $N = \displaystyle{\frac{\hbar^{1/2}}{\pi^4\left(1 - e^{-\epsilon}\right) \vert \Sigma_{\epsilon} \vert^{\frac{1}{2}}}}$. Reorganizing the terms and performing the change of variable $k \rightarrow -k$, we get
\begin{equation}
    N \int dy' dk' \delta \left(q + y' -q_0 \right) \delta \left( p + k' - p_0 \right) \int dy dk e^{2i\left(yk' + ky' \right)/\hbar} e^{-\frac{1}{2} \left( y +q, k - p \right) \Sigma^{-1}_{\epsilon} \left(\begin{smallmatrix} y +q \\ k - p  \end{smallmatrix}\right) }.
\end{equation}
The second integral is readily recognised as the characteristic function of a multivariate normal probability distribution, so we find
\begin{equation}
    \frac{N \hbar^2}{4} \int dy' dk' \delta \left( q + \frac{y'\hbar}{2} - q_0 \right) \delta \left( p + \frac{k'\hbar}{2} - p_0 \right) e^{i \left( -q, p \right) \left(\begin{smallmatrix} k' \\ y' \end{smallmatrix}\right)} e^{- \frac{1}{2} \left(k', y' \right) \Sigma_{\epsilon} \left(\begin{smallmatrix} k' \\ y' \end{smallmatrix}\right)},
\end{equation}
Where we used the change of variable $y' \to 2y'/\hbar,\ k' \to 2k'/\hbar$. Performing the remaining integral, we find
\begin{equation}
    W^{\epsilon} \star W^{\text{ideal}}_{\bm \mu_0} = \frac{N\hbar^2}{4} e^{\frac{2i}{\hbar}\left[ q \left( p - p_0 \right) + p \left( q_0 - q \right) \right]} e^{-\frac{1}{2} \left(q - q_0, p - p_0 \right) (4\Sigma_{\epsilon}/\hbar^2 ) \left(\begin{smallmatrix} q - q_0 \\ p - p_0 \end{smallmatrix}\right)},
\end{equation}
where we used the fact that $\Sigma_{\epsilon}$ is diagonal.
Finally, we are set to calculate
\begin{align*}
    W^{\text{phys}}_{\bm \mu_0} = & \frac{N^2 \hbar^2}{4} \int dy dk dy' dk' e^{2i\left( yk' + y'k \right)/\hbar} e^{ \frac{2i}{\hbar}\left[ \left( q + y \right) \left(p + k - p_0 \right) - \left( p + k \right) \left( q + y - q_0 \right) \right]} \\ & \times e^{- \frac{1}{2} \left(q + y - q_0, p + k - p_0 \right) (4 \Sigma_{\epsilon}/\hbar^2) \left(\begin{smallmatrix} q + y - q_0 \\ p + k - p_0 \end{smallmatrix}\right)} e^{- \frac{1}{2} \left( q - y', p + k' \right) \Sigma^{-1}_{\epsilon} \left(\begin{smallmatrix} q - y' \\ p + k' \end{smallmatrix}\right)},
\end{align*}
where we made the change of variable $y' \to -y'$. Performing the integrals in $y'$ and $k'$, we find
\begin{equation}
    W^{\text{phys}}_{\bm \mu_0} = \frac{N^2 \hbar^4}{16} e^{ \frac{2i}{\hbar}\left(p q_0 - q p_0 \right)} \int dy dk e^{ \frac{2i}{\hbar}\left( k \left( q + q_0 \right) + y \left( p + p_0 \right) \right)}  e^{-\frac{1}{2} \left( y - q + q_0, k + p -p_0 \right) (4\Sigma_{\epsilon}/\hbar^2) \left(\begin{smallmatrix} y - q + q_0 \\ k + p - p_0 \end{smallmatrix}\right) } e^{- \frac{1}{2} \left( y , k \right) (4\Sigma_{\epsilon}/\hbar^2) \left(\begin{smallmatrix} y \\ k \end{smallmatrix}\right)}.
\end{equation}
Defining the variables $q' = q + q_0$ and $p' = p + p_0$ and rearranging the Gaussians, we find
\begin{align}
    W^{\text{phys}}_{\bm \mu_0} &= \frac{N^2 \hbar^4}{16} e^{ \frac{2i}{\hbar}\left( pq_0 - q p_0 \right)} e^{-\frac{1}{2} \left( q' - 2q_0, 2p_0 - p' \right) (2\Sigma_{\epsilon} /\hbar^2)\left(\begin{smallmatrix} q' - 2q_0 \\ 2p_0 - p' \end{smallmatrix}\right) }\nonumber\\ 
    &~~~~\times\int dy dk e^{\frac{2i}{\hbar} \left(kq' + yp' \right)} e^{\frac{1}{2} \left( y - \frac{q' - 2q_0}{2}, k - \frac{2p_0 - p'}{2} \right) (8 \Sigma_{\epsilon}/\hbar^2) \left(\begin{smallmatrix} y - \frac{q' - 2q_0}{2} \\ k - \frac{2p_0 - p'}{2} \end{smallmatrix}\right)}.
\end{align}
Performing the integral and replacing $x'$ and $p'$, we find
\begin{equation}
    W^{\text{phys}}_{\bm \mu_0} = \frac{N^2 \hbar^6}{64} e^{-\frac{1}{2} \left( q - q_0, p - p_0 \right) (2\Sigma_{\epsilon}/\hbar^2) \left(\begin{smallmatrix}
    q - q_0 \\ p - p_0 \end{smallmatrix}\right) } e^{- \frac{1}{2} \left( q + q_0, p + p_0 \right) \frac{\Sigma_{\epsilon}^{-1}}{2} \left(\begin{smallmatrix} q + q_0, p + p_0 \end{smallmatrix}\right)}.
\end{equation}
Finally regrouping the Gaussians together, we find
\begin{align*}
    W^{\text{phys}}_{\bm \mu_0} = & \ \frac{N^2 \hbar^6}{64} e^{- \frac{1}{2} \bm \mu_0^{\intercal} 4 \left(2\Sigma_{ \epsilon}/\hbar^2 + \left(2\Sigma_{\epsilon}\right)^{-1} \right)^{-1} \bm \mu_0} \\
    & \times e^{- \frac{1}{2} \left(\bm \xi - \left( 2\Sigma_{\epsilon}/\hbar^2 - \left( 2\Sigma_{\epsilon} \right)^{-1} \right) \left( 2\Sigma_{\epsilon}/\hbar^2 + \left( 2\Sigma_{\epsilon} \right)^{-1} \right)^{-1} \bm \mu_0 \right)^{\intercal} 2\Sigma_{\epsilon}/\hbar^2 + \left( 2\Sigma_{\epsilon}\right)^{-1} \left(\bm \xi - \left( 2\Sigma_{\epsilon}/\hbar^2 - \left( 2\Sigma_{\epsilon} \right)^{-1} \right) \left( 2\Sigma_{\epsilon}/\hbar^2 + \left(2\Sigma_{\epsilon}^{-1} \right)\right)^{-1} \bm \mu_0 \right)}.
\end{align*}
Replacing the values for $N$, noting that $2\Sigma_{\epsilon}/\hbar^2 + \left(2\Sigma_{\epsilon}\right)^{-1} = 4 \Sigma_{2\epsilon}/\hbar^2$, we find
\begin{equation} \label{eq:wigner_physical}
    W^{\text{phys}}_{\mu_0} = \frac{\hbar^4}{16\pi^8 \left(1 + e^{-\epsilon} \right)^2} e^{- \frac{1}{2} \bm \mu_0^{\intercal} \Sigma_{2 \epsilon}^{-1}  \bm \mu_0} e^{- \frac{1}{2} \left( \bm \xi - \bm \mu_{\epsilon} \right)^{\intercal} 4\Sigma_{2\epsilon}/\hbar^2 \left( \bm \xi - \bm \mu_{\epsilon} \right) },
\end{equation}
with $\bm \mu_{\epsilon} = \frac{2e^{-\epsilon}}{1 + e^{-2\epsilon}} \bm \mu_0$.

\subsection{Real Weights, Means and Covariances \emph{Via} Optical Circuits} \label{subann:GKP_optical}

The index set, weights, means and covariances matrices for the ideal GKP state are provided in \cref{eq:gkp_ideal}. Finite energy GKP states can be recovered by applying the Fock damping channel $\hat{E}(\epsilon) = e^{-\epsilon \hat{n}}$ to the state, a channel which neatly fits into the Gaussian-inspired transformation framework introduced in Section \ref{subsec:gauss_transf}. For the Fock damping channel with $e^{-\epsilon} = \cos\theta$, we derive the exact update rules for states in Appendix \ref{app:fock_damping}. Thus, using \cref{eq:fock_damp_update}, the covariances of the ideal GKP under Fock damping become:
\begin{equation}
    \begin{split}
        \Sigma_m =& \lim_{\delta \to 0+}\delta \id\\
        \overset{\hat{E}(\epsilon)}{\rightarrow} & \lim_{\delta \to 0+} \left[ \cos^2\theta\delta \id + \hbar\sin^2\theta\id/2 - \cos^2\theta\sin^2\theta(\hbar\id/2 - \delta \id)\left[\sin^2\theta\delta \id + (\cos^2\theta+1)\hbar\id/2\right]^{-1}(\hbar\id/2 - \delta \id)\right]\\
        =& \hbar\sin^2\theta\id/2 - \cos^2\theta\sin^2\theta(\hbar\id/2)\left[(\cos^2\theta+1)\hbar\id/2\right]^{-1}(\hbar\id/2)\\
        =& \frac{\hbar}{2}\frac{1-e^{-2\epsilon}}{1+e^{-2\epsilon}} \id = \bm{\Sigma}_m(\epsilon),
    \end{split}
\end{equation}
while the means become:
\begin{equation}
    \begin{split}
        \bm{\mu}_{m} &\overset{\hat{E}(\epsilon)}{\rightarrow}  \lim_{\delta \to 0+}\left[\cos\theta\bm{\mu}_{m} - \cos\theta\sin^2\theta(\hbar\id/2 - \delta \id)\left[\sin^2\theta\delta \id + (\cos^2\theta+1)\hbar\id/2\right]^{-1}\bm{\mu}_{m}\right]\\
        &=\frac{2e^{-\epsilon}}{1+e^{-2\epsilon}}\bm{\mu}_{m} = \bm{\mu}_{m}(\epsilon).
    \end{split}
\end{equation}
The re-weighting of the peak is provided by \cref{eq:fock_damp_update_weights}:
\begin{equation}
\begin{split}
    c_m (\epsilon;\theta,\phi) &= \lim_{\delta \to 0+} \frac{w(\bm{0}|\sin\theta\bm{\mu}_{m},\ \sin^2\theta\delta\id + \hbar\cos^2\theta\id/2,\ \hbar\id/2)}{p(\bm{0};\ket{\psi(\theta,\phi)}_{\rm gkp},\hbar\id/2)}\\
    &= \frac{c_m(\theta,\phi) G_{ \sin\theta\bm{\mu}_{m},\hbar(1+\cos^2\theta)\id/2}\left(\bm{0}\right)}{p(\bm{0};\ket{\psi(\theta,\phi)}_{\rm gkp},\hbar\id/2)}\\
    &= \frac{c_m(\theta,\phi)}{\mathcal{N}_\epsilon}\exp\left[-\frac{1-e^{-2\epsilon}}{\hbar(1+e^{-2\epsilon})}\bm{\mu}_m^T \bm{\mu}_m\right]
\end{split}
\end{equation}
where $\mathcal{N}_\epsilon$ is an overall normalization. This completes the derivation.

\subsection{Complex Weights and Means, and Real Covariances \emph{Via} the Wavefunction to Wigner Transformation} \label{subann:GKP_alt}

The ideal GKP state for a single qubit is:
\begin{equation}
    \ket{\psi(\bm{a})} = a_0 \ket{0}_{\rm gkp}+a_1\ket{1}_{\rm gkp} = \sum_{k=0}^{1}a_k\sum_{t=-\infty}^{\infty} \ket{\sqrt{\pi\hbar}(2t+k)}_{q} ,|a_0|^2+|a_1|^2=1
\end{equation}
Next consider application of $\hat{E}(\epsilon)=e^{-\epsilon\hat{n}}$:
\begin{equation}
    \hat{E}(\epsilon)\ket{\psi} =\sum_{k=0}^{1} a_k \sum_{t=-\infty}^{\infty}\int ds \left[  \bra{s}\hat{E}(\epsilon)\ket{\sqrt{\pi\hbar}(2t+k)}_{q} \right] \ket{s}_q\\
\end{equation}

Next we use Mehler's kernel~\cite{mehler1866ueber}:
\begin{equation}
\begin{split}
    \langle s_1| \hat{E}(\epsilon)|s_2\rangle_{q} &=  \sum_{n=0}^{\infty}e^{-\epsilon n}\braket{s_1|n}_{q}\braket{n|s_2}_{q}\\
    &=  \sum_{n=0}^{\infty}\frac{e^{-\epsilon n}}{2^n n! \sqrt{\pi\hbar}} e^{-(s_1^2 + s_2^2)/2\hbar}H_n(s_1)H_n(s_2)\\
    &=  \frac{e^{\epsilon/2}}{\sqrt{2\pi\sinh(\epsilon)}}\exp\left[-\coth(\epsilon)(s_1^2+s_2^2)/2\hbar + \csch(\epsilon)s_1s_2/\hbar \right]\\
    &= \frac{e^{\epsilon/2}}{\sqrt{2\pi\hbar\sinh(\epsilon)}}\exp[-\frac{1}{2\hbar}\bm{s}^T \bm{\sigma_\epsilon}^{-1} \bm{s}]
\end{split}
\end{equation}
where the second last line was obtained with Mehler's formula. In the last line:
\begin{equation}
\begin{split}
    \bm{s} &=  (s_1, s_2)^T\\
    \bm{\sigma_\epsilon}^{-1} &=  \begin{pmatrix}
                                       \alpha & -\beta\\
                                       -\beta & \alpha
                                        \end{pmatrix}=  \begin{pmatrix}
                                       \coth(\epsilon) & \csch(\epsilon)\\
                                       \csch(\epsilon) & \coth(\epsilon)
                                        \end{pmatrix}
\end{split}
\end{equation}

Therefore, we have that the (subnormalized) wavefunction in position quadrature is given by:
\begin{equation}
\begin{split}
    \psi_{\bm{a},\epsilon}(x) =& \sum_{k=0}^1 a_k\sum_{t=-\infty}^{\infty}\bra{x}\hat{E}(\epsilon)\ket{\sqrt{\pi\hbar}(2t+k)}_{q} \\
     =& \sum_{k=0}^1 a_k\sum_{t=-\infty}^{\infty} \exp[-\alpha\pi (2t+k)^2/2]\exp[-\alpha x^2/2\hbar - \beta\sqrt{\pi}x(2t+k)/\sqrt{\hbar}].
\end{split}
\end{equation}
Note that we can write the amplitude inside the sum as
\begin{align}
\exp[-\alpha x^2/2\hbar - \beta\sqrt{\pi}x(2t+k)/\sqrt{\hbar}] =  \sqrt[4]{\frac{\pi \hbar }{\alpha }} e^{\frac{\pi  \beta ^2 (k+2 t)^2}{2
   \alpha }} \bra{x} \hat{D}\left(-\sqrt{\tfrac{\pi }{2}} (2 t+k) \tfrac{\beta}{\alpha} \right)\hat{S}\left(\tfrac{1}{2} \log_{\text{e}} \alpha \right)\ket{0},
\end{align}
where $\hat{D}$ and $\hat{S}$ are the single mode displacement and squeezing operators (cf. Eq.~\eqref{eq:sq_and_disp}) and $\ket{0}$ is the single mode vacuum.
We can use these results to write the Hilbert space vector of a finite energy GKP state as
\begin{align}
\ket{\psi_{\bm{a},\epsilon}} = \sum_{k=0}^1 a_k\sum_{t=-\infty}^{\infty} \kappa_t \ \hat{D}\left(\sqrt{\tfrac{\pi }{2}} (2 t+k) \text{sech}( \epsilon) \right)\hat{S}\left(-\tfrac{1}{2} \log_{\text{e}} \tanh(\epsilon) \right)\ket{0}, \quad 
\kappa_t = \sqrt[4]{\pi \hbar  \tanh (\epsilon )} e^{-\frac{1}{2} \pi  (k+2 t)^2
   \tanh (\epsilon )},
\end{align}
showing that finite-energy GKP states can be written as linear combinations of single-mode pure Gaussian states like the ones discussed in Eq.~\eqref{eq:gauss_superposition}.

Now we move on to the (subnormalized) Wigner function calculation:
\begin{equation}
\begin{split}
        W(\bm{\xi};\ket{\psi(\bm{a},\epsilon)}) &= \frac{1}{2\pi\hbar}\int dy \exp(-ipy/\hbar)\psi_{\bm{a},\epsilon}^*(x-y/2)\psi_{\bm{a},\epsilon}(x+y/2)\\
            &= \frac{1}{2\pi\hbar}\sum_{\ell,k=0}^1a_\ell a^*_k \sum_{s,t=-\infty}^{\infty}\exp[-\alpha\pi (2t+\ell)^2/2]\exp[-\alpha\pi (2s+k)^2/2]I(x,p;s,t,\ell,k)
\end{split}
\end{equation}
where:
\begin{equation}
\begin{split}
    I(x,p;s,t,\ell,k) =&
  \int dy \exp(-ipy/\hbar)\exp[-\alpha(x+y/2)^2/2\hbar - \beta\sqrt{\pi}(x+y/2)(2t+\ell)/\sqrt{\hbar}]\\
  &\hspace{1cm}\times\exp[-\alpha(x-y/2)^2/2\hbar - \beta\sqrt{\pi}(x-y/2)(2s+k)/\sqrt{\hbar}]\\
  =& \sqrt{\frac{4\pi\hbar}{\alpha}} \exp\left[-\frac{\alpha}{\hbar} x^2 -x\frac{\beta \sqrt{\pi}}{\sqrt{\hbar}} (2t+2s+\ell+k) \right] \exp\left[\frac{1}{\alpha\hbar}\left(\frac{\beta\sqrt{\pi\hbar}}{2}(2s-2t+k-\ell)-ip\right)^2\right]\\
\end{split}
\end{equation}
Here, the result was obtained by performing the Gaussian integral.

Putting it all together, we find that the normalized Wigner function is a linear combination of Gaussian functions in phase space:
\begin{equation}
        W(\bm{\xi};\ket{\psi(\bm{a},\epsilon)}) = \sum_{m \in {\mathcal M} } c_m G_{ \bm{\mu}_m,\bm{\Sigma}_m} \left( \bm{\xi} \right)
\end{equation}
with:
\begin{equation}
    \begin{split}
        &\mathcal{M} = \{m\equiv (k,\ell,s,t) \, | \, k, \ell \in \{ 0,1\} \,\& \, s,t \in \mathbb{Z} \}\\
        &\bm{\mu}_{m}^T = \left(\frac{-\beta \sqrt{\pi\hbar} (2t+2s+\ell+k)}{2\alpha}, \frac{-i\beta \sqrt{\pi\hbar} (2s-2t+\ell-k)}{2}\right)\\
        &\bm{\Sigma}_m = \frac{\hbar}{2}\begin{pmatrix}
                                       1/\alpha & 0\\
                                       0 & \alpha
                                        \end{pmatrix}\\
        &(\alpha,\beta) = (\coth(\epsilon), -\csch(\epsilon))\\
        &c_m= \frac{1}{\mathcal{N}(\bm{a},\epsilon)}a_\ell a^*_k\exp[-\alpha\pi (2t+k)^2/2]\exp[-\alpha\pi (2s+\ell)^2/2]\exp\left[ \frac{\beta^2\pi(2t+2s+k+\ell)^2}{4\alpha}\right]
    \end{split}
\end{equation}
where $\mathcal{N}(\bm{a},\epsilon)$ is an overall normalization chosen so that $\sum_{m\in\mathcal{M}}c_m = 1$, since the Gaussian functions are already normalized over phase space.

\section{Fock States as Linear Combinations of Gaussians in Phase Space} \label{app:Fock}

In this appendix we provide the mathematical details related to the expressions derived for Fock states expressed as linear combinations of Gaussians. 

We generalize the scheme in Section~\ref{subsec:Fock} to $n$-photon addition to the vacuum to prepare and $n$-photon Fock state (cf. Fig.~\ref{additioncirc}(b)).
We can now write for the state in mode 0
\begin{align}
\pr{n} \approx \hat{\rho}_{n} &= \frac{1}{p_{n}} \text{tr}_{1,\ldots,n}  \left( \hat{Q}_{ n}  \right),\quad
p_{ n} = \text{tr}\left( \hat{Q}_{ n}  \right),\quad
\hat{Q}_{ n} = \left\{ \bigotimes_{i=1}^n\left( \mathbb{\hat{I}}_i - \pr{0_i}  \right)\right\}  \pr{\Psi}, \label{nfock}
\end{align}
where $\ket{\Psi} = \bigotimes_{i=1}^n \hat{S}_{0,i}^{(2)}(r) \ket{0_0,\ldots,0_n}$ and $p_{n}$ is a normalization constant. Now note that each of the $2^n$ terms in the product inside curly braces in Eq.~\eqref{nfock} is a product of local identity operations or projections into vacuum. Upon partial tracing, both of these operations give rise to a Gaussian map and we have just shown that one can write an $n$-photon Fock state as a linear combination of $2^n$ Gaussians. As it turns out, this number can be significantly reduced. Indeed, note that in the limit $r \ll 1$ all the squeezing operations can be permuted with impunity. This implies that the covariance matrix is invariant under any permutation that does not involve the first mode.
Thus, for example, the following two projectors are equivalent when acted on the state $\pr{\Psi}$
\begin{align}
&\text{tr}_{ij}\left\{ (\mathbb{\hat{I}}_i - \pr{0_i})  \pr{0_j}  \pr{\Psi} \right\}= \text{tr}_{ij}\left\{ \pr{0_i}  (\mathbb{\hat{I}}_j - \pr{0_j})   \pr{\Psi} \right\} . \nonumber
\end{align}
One can state this assumption mathematically by writing the adjacency matrix (cf. Eq. 24 of Ref.~\cite{jahangiri2020point}) of this state,
\begin{equation}
\bm{B} = \bm{B}^T = \begin{pmatrix}
0 & r & \ldots & r \\
r & 0 & \ldots & 0 \\
\vdots & \vdots & \ddots & \vdots  \\
r & 0 & \ldots & 0
\end{pmatrix}.
\end{equation}
from which the covariance matrix is~\cite{Quesada_2019}
\begin{align}
\bm{\Sigma} = \hbar \bm{S}^T \bm{R}^\dagger \left\{  \begin{bmatrix} 
\id & -\bm{B} \\
-\bm{B}^* & \id
\end{bmatrix}^{-1} - \frac{\id}{2}
\right\} \bm{R} \bm{S}, \quad  \bm{R} = \frac{1}{\sqrt{2}} \begin{bmatrix} \id &  i \id \\
\id & -i \id \end{bmatrix}
\end{align}
and $\bm{S}^T$ is the permutation matrix that maps the vector $(q_0,\ldots,q_n,p_0,\ldots,p_n)$ into the vector $(q_0,p_0,\ldots,q_n,p_n)$. Note that the vector of means of this state is $\bm{\mu} = \bm{0}$, and since we will be heralding using threshold detectors any heralded state will also have a zero mean vector.

The exact permutation symmetry of the state allows us to group the $2^n$ terms in Eq.~\eqref{nfock} into $n+1$ terms with binomial coefficients as
\begin{equation}
\hat{Q}_{ n} = \sum_{j=0}^n {\binom{n}{j}} (-1)^{(n-j)} \left\{ \bigotimes_{i=1}^j \pr{0}_i \right\} \pr{\Psi}.
\end{equation}
Now consider the term with $j$ vacuum detections. For this term we need to trace out $n-j$ modes.
One can write the covariance matrix of the $j+1$ modes after tracing out in the ordering $(q_0,p_0,q_1,p_1,\ldots,q_j,p_j)$ as
\begin{equation}
\bm{\Sigma} = \begin{bmatrix}
\bm{\Sigma}_A & \bm{\Sigma}_{AB} \\
\bm{\Sigma}_{AB}^T & \bm{\Sigma}_B
\end{bmatrix},
\end{equation}
where $A$ is used to indicate the $0^{\text{th}}$ mode and $B$ the remaining $1$ through $j$ modes and where the submatrices are
\begin{align}
\bm{\Sigma}_A &= \tfrac{\hbar}{2}\left(\frac{2}{1-n r^2}-1 \right)\id_2, \\
\bm{\Sigma}_{AB} &= \tfrac{\hbar}{2} \frac{2 r}{1- n r^2} \left[Z, \ldots, Z \right] \in \mathbb{R}^{2 \times 2j}, \\
\bm{\Sigma}_B &= \tfrac{\hbar}{2} \id_{2j}+ \tfrac{\hbar}{2} \frac{2r^2}{1-n r^2}\begin{bmatrix}
\id_2 &\id_2 & \ldots & \id_2 \\
\id_2 & \ddots  &        & \id_2 \\
\vdots & & & \vdots \\
\id_2 & \id_2 & \ldots & \id_2
\end{bmatrix} \in \mathbb{R}^{2j \times 2j}.
\end{align}
and for which physicality alone entails $r < 1/\sqrt{n}$.
In the last equation we used $\id_j$ for the $j \times j$ identity matrix and $Z=\text{diag}(1,-1)$.

We can now write the covariance matrix of the state after $n-j$ modes are traced and $j$ modes are projected onto vacuum. We use Eq.~5.142 from Ref.~\cite{book_serafini}
\begin{equation}
\bm{\Sigma}_A \to \bm{\Sigma}_{A j} =   \bm{\Sigma}_A - \bm{\Sigma}_{AB} \left( \bm{\Sigma}_B+\bm{\Sigma}_{\text{vac}} \right)^{-1} \bm{\Sigma}_{AB}^T .
\end{equation}
For our case we can write the answer analytically
\begin{equation}
\bm{\Sigma}_{A j } = \frac{\hbar}{2} \frac{1+(n-j)r^2}{1-(n-j)r^2} \id_{2}.
\end{equation}
Also note that the weight of this Gaussian function should be
\begin{equation}
p_{j} =  \binom{n}{j} (-1)^{(n-j)} \frac{1}{\sqrt{\det\left( \frac{1}{\hbar} \left\{\bm{\Sigma}_B+ \tfrac{\hbar}{2}\id_{2j} \right\}\right)}} = \frac{1-n r^2}{1-(n-j)r^2} \binom{n}{j} (-1)^{(n-j)}.
\end{equation}
up to normalization, which is given by
\begin{equation}
\mathcal{N}_n = \sum_{j=0}^n p_j = \frac{n! \left(n
   \left( \frac{-r^2-1}{r^2} \right)!+\left(-\frac{1}{r^2
   }\right)!\right)}{ \left( \frac{n r^2-1}{r^2} \right)!}.
\end{equation}
Thus we conclude that the coefficients are given by $c_j =  p_j/\mathcal{N}_n$.

\section{Cat States as Linear Combinations of Real-Valued Gaussian Functions}\label{app:realCat}

Equation~\eqref{eqn:Wigner_Cat} obviously contains complex terms. However, it is possible to approximate it by a real linear superposition with arbitrary precision. We give the explicit expression here.

The complex terms in Eq.~\eqref{eqn:Wigner_Cat} can be written as
\begin{align}
& e^{-2|\alpha|^2} \left( e^{- i \phi} \mathcal{G}_{\bm V_{\text{vac}}, \bm \mu_z}(\bm \xi) +\text{c.c.} \right) \nonumber \\
&= \frac{1}{\pi \vert \bm V_{\text{vac}} \vert^{\frac{1}{2}}} e^{-\frac{1}{2} \bm \xi^{\intercal} \bm V_{\text{vac}}^{-1} \bm \xi } \left[ \cos{\phi} \left( \cos{(2 \sqrt{\tfrac{2}{\hbar}} q \Im(\alpha) )} \cos{( 2 \sqrt{\tfrac{2}{\hbar}} p \Re(\alpha) )} + \cos{( 2 \sqrt{\tfrac{2}{\hbar}} q \Im(\alpha) - \tfrac{\pi}{2} )} \cos{ ( 2 \sqrt{\tfrac{2}{\hbar}} p \Re(\alpha) - \tfrac{\pi}{2} )} \right) \right. \nonumber  \\
&\quad + \left. \sin{\phi} \left( \cos{( 2 \sqrt{\tfrac{2}{\hbar}} q \Im(\alpha) - \tfrac{\pi}{2} )} \cos{( 2 \sqrt{\tfrac{2}{\hbar}} p \Re(\alpha) )} - \cos{( 2 \sqrt{\tfrac{2}{\hbar}} q \Im(\alpha) )} \cos{( 2 \sqrt{\tfrac{2}{\hbar}} p \Re(\alpha) - \tfrac{\pi}{2}  )}  \right) \right].
\end{align}

Recall that $\cos q$ can be expressed as~\cite{mazya}
\begin{equation} \label{eqn:cos_approx}
\cos q = \frac{e^{\frac{\pi^2 D}{4}}}{2 \sqrt{\pi D}} \sum_{m \in \mathbb{Z}} (-1)^m e^{-\frac{(\frac{q}{\pi} - m)^2}{D}} - R(q),
\end{equation}
with $R(q) = O(e^{-2\pi^2 D}) $, with $D$ a positive parameter controlling the accuracy of the approximation.

Considering the most complex case where $\Re(\alpha), \Im(\alpha) \neq 0$ , we find
\begin{align} \label{eqn:cat_cos_approx}
& e^{-2|\alpha|^2} \left( e^{- i \phi} \mathcal{G}_{\bm V_{\text{vac}}, \bm \mu_z}(\bm \xi) +\text{c.c.} \right) \nonumber \\
=& \frac{1}{4\pi^2 D | \bm V_{\text{vac}}|^{\frac{1}{2}}} e^{\frac{\pi^2 D}{2}} e^{-\frac{1}{2} \bm \xi^{\intercal} \bm V_{\text{vac}}^{-1} \bm \xi} \left[ \cos(\phi) \left( \sum_{m \in \mathbb{Z}} (-1)^m e^{- \frac{1}{2} \frac{16 \Im(\alpha)^2}{\pi^2 \hbar D} \left(q - \frac{m \pi \sqrt{\hbar}}{2 \sqrt{2} \Im(\alpha)} \right)^2} \right) \left( \sum_{m \in \mathbb{Z}} (-1)^m e^{- \frac{1}{2} \frac{16 \Re(\alpha)^2}{\pi^2 \hbar D} \left(p - \frac{m \pi \sqrt{\hbar}}{2 \sqrt{2}\Re(\alpha)} \right)^2} \right) \right. \nonumber \\
& \quad + \cos{(\phi)} \left( \sum_{m \in \mathbb{Z}} (-1)^m e^{-\frac{1}{2} \frac{16 \Im(\alpha)^2}{\pi^2 \hbar D} \left( q - \frac{(m + \frac{1}{2}) \pi \sqrt{\hbar}}{2 \sqrt{2} \Im(\alpha)} \right)^2} \right) \left( \sum_{m \in \mathbb{Z}} (-1)^m e^{-\frac{1}{2} \frac{16 \Re(\alpha)^2}{\pi^2 \hbar D} \left( p - \frac{(m+\frac{1}{2}) \pi \sqrt{\hbar}}{2 \sqrt{2} \Re(\alpha)} \right)^2} \right) \nonumber \\
& \quad + \sin{(\phi)} \left( \sum_{m \in \mathbb{Z}} (-1)^m e^{- \frac{1}{2} \frac{16 \Im(\alpha)^2}{\pi^2 \hbar D} \left(q - \frac{(m + \frac{1}{2}) \pi \sqrt{\hbar}}{2 \sqrt{2} \Im(\alpha)} \right)^2}\right) \left( \sum_{m \in \mathbb{Z}} (-1)^m e^{- \frac{1}{2} \frac{16 \Re(\alpha)^2}{\pi^2 \hbar D} \left(p - \frac{m \pi \sqrt{\hbar}}{2 \sqrt{2} \Re(\alpha)} \right)^2} \right) \nonumber \\
& \quad - \sin{(\phi)} \left. \left( \sum_{m \in \mathbb{Z}} (-1)^m e^{- \frac{1}{2} \frac{16 \Im(\alpha)^2}{\pi^2 \hbar D} \left(q - \frac{m \pi \sqrt{\hbar}}{2 \sqrt{2} \Im(\alpha)} \right)^2} \right) \left( \sum_{m \in \mathbb{Z}} (-1)^m e^{-\frac{1}{2} \frac{16 \Re(\alpha)^2}{\pi^2 \hbar D} \left( p - \frac{(m+\frac{1}{2}) \pi \sqrt{\hbar}}{2 \sqrt{2} \Re(\alpha)} \right)^2} \right) \right].
\end{align}

Defining
\begin{align}
\bm \mu^{\alpha}_{m, m'} & = \frac{\pi \sqrt{\hbar}}{2 \sqrt{2}}( m / \Im(\alpha), m' / \Re(\alpha) )^{\intercal}, \nonumber \\
\bm \Sigma_{\alpha} & = \frac{\pi^2 \hbar D}{16} \begin{pmatrix} \Im(\alpha)^{-2} & 0 \\ 0 & \Re(\alpha)^{-2} \end{pmatrix},
\end{align}
and inserting into Eq.~\ref{eqn:cat_cos_approx} gives

\begin{align}
& e^{-2|\alpha|^2} \left( e^{- i \phi} \mathcal{G}_{\bm V_{\text{vac}}, \bm \mu_z}(\bm \xi) +\text{c.c.} \right) \nonumber \\
=& \frac{1}{4 \pi^2 D |\bm V_{\text{vac}}|^{\frac{1}{2}}} e^{\frac{\pi^2 D}{2}} e^{-\frac{1}{2} \bm \xi^{\intercal} \bm V^{-1}_{\text{vac}} \bm \xi} \nonumber\\
&\times\left[ \cos{(\phi)} \sum_{m, m' \in \mathbb{Z}} (-1)^{m + m'} \left( e^{-\frac{1}{2} (\bm \xi - \bm \mu^{\alpha}_{m, m'})^{\intercal} \bm \Sigma_{\alpha}^{-1} (\bm \xi - \bm \mu^{\alpha}_{m, m'})} + e^{-\frac{1}{2} (\bm \xi - \bm \mu^{\alpha}_{m+1/2, m'+1/2})^{\intercal} \bm \Sigma_{\alpha}^{-1} (\bm \xi - \bm \mu^{\alpha}_{m+1/2, m'+1/2})} \right) \right. \nonumber \\
& \quad \left. + \sin{(\phi)} \sum_{m, m' \in \mathbb{Z}} (-1)^{m + m'} \left( e^{-\frac{1}{2} (\bm \xi - \bm \mu^{\alpha}_{m+1/2, m'})^{\intercal} \bm \Sigma_{\alpha}^{-1} (\bm \xi - \bm \mu^{\alpha}_{m+1/2, m'})} - e^{-\frac{1}{2} (\bm \xi - \bm {\mu}^{\alpha}_{m, m'+1/2})^{\intercal} \bm \Sigma_{\alpha}^{-1} (\bm \xi - \bm \mu^{\alpha}_{m, m'+1/2})}  \right) \right].
\end{align}

Combining product of Gaussians in $\bm \xi$ together results in

\begin{align}
& e^{-2|\alpha|^2} \left( e^{- i \phi} \mathcal{G}_{\bm V_{\text{vac}}, \bm \mu_z}(\bm \xi) +\text{c.c.} \right) \nonumber \\
=& \frac{| \bm \Sigma_{\alpha} |^{\frac{1}{2}}}{D} e ^{\frac{\pi^2 D}{2}} \left[ \cos{\phi} \sum_{m, m' \in \mathbb{Z}} (-1)^{m+m'} \left( G_{\bm \mu^{\alpha}_{m, m'}, \bm \Sigma_{\alpha} + \bm V_{\text{vac}}} (\bm 0) G_{\bm \mu^{\alpha}_{m, m'}, \bm V_{\text{vac}} \bm \Sigma_{\alpha} (\bm V_{\text{vac}} + \bm \Sigma_{\alpha})^{-1}} (\bm \xi) \right. \right. \nonumber \\
& \hspace{6cm} + \left. G_{\bm \mu^{\alpha}_{m+1/2, m'+1/2}, \bm \Sigma_{\alpha} + \bm V_{\text{vac}}} (\bm 0) G_{\bm \mu^{\alpha}_{m+1/2, m'+1/2}, \bm V_{\text{vac}} \bm \Sigma_{\alpha} (\bm V_{\text{vac}} + \bm \Sigma_{\alpha})^{-1}} (\bm \xi) \right) \nonumber \\
& \hspace{2cm} + \sin{\phi} \sum_{m, m' \in \mathbb{Z}} (-1)^{m+m'} \left( G_{\bm \mu^{\alpha}_{m+1/2, m'}, \bm \Sigma_{\alpha} + \bm V_{\text{vac}}} (\bm 0) G_{\bm \mu^{\alpha}_{m+1/2, m'}, \bm V_{\text{vac}} \bm \Sigma_{\alpha} (\bm V_{\text{vac}} + \bm \Sigma_{\alpha})^{-1}} (\bm \xi) \right. \nonumber \\
& \hspace{6cm} - \left. G_{\bm \mu^{\alpha}_{m, m'+1/2}, \bm \Sigma_{\alpha} + \bm V_{\text{vac}}} (\bm 0) G_{\bm \mu^{\alpha}_{m, m'+1/2}, \bm V_{\text{vac}} \bm \Sigma_{\alpha} (\bm V_{\text{vac}} + \bm \Sigma_{\alpha})^{-1}} (\bm \xi) \right) \Bigg].
\end{align}
where $\bm \mu_{m, m'}^{\alpha, \text{vac}} = \bm \Sigma_{\alpha}^{-1} (\bm \Sigma_{\alpha}^{-1} + \bm V_{\text{vac}}^{-1})^{-1} \bm \mu_{m, m'}^{\alpha}$.

While this general expression is quite cumbersome, it is straightforward to find the simpler form for the case where either $\Re(\alpha) = 0$ or $\Im(\alpha) = 0$ (the case of $\alpha = 0$ being trivial). Following the same procedure in case where $\alpha \in \mathbb{R}$, we find

\begin{align}
e^{-2|\alpha|^2} \left( e^{- i \phi} \mathcal{G}(\bm \xi; \bm V_{\text{vac}}, \bm \mu_z) +\text{c.c.} \right) \approx & \frac{\sqrt{\pi \hbar} e^{\frac{ \pi^2 D}{4}}}{4 \alpha \left(\Sigma_{\alpha} + v \right)^{\frac{1}{2}}}  \left[ \cos{(\phi)} \sum_{m \in \mathbb{Z}} (-1)^m e^{- \frac{ {b}^2_{m}}{2 (\Sigma_{\alpha} + v)}} G_{\bm \mu_m^{\alpha}, \bm \Sigma_m} (\bm \xi) \right. \nonumber \\
& \hspace{2.5cm}\left. - \sin{(\phi)} \sum_{m \in \mathbb{Z}} (-1)^m e^{- \frac{ {b}^2_{m+1/2}}{2 (\Sigma_{\alpha} + v)}} G_{\bm \mu_{m+1/2}^{\alpha}, \bm \Sigma_m} (\bm \xi) \right],
\end{align}
where we have
\begin{align}
\bm \mu_m &= \left( 0, \cfrac{4 \alpha \sqrt{\hbar} m}{\sqrt{2} \pi \left(16 \alpha^2 + 2 D \right)} \right), \nonumber \\
\bm \Sigma_m &= \begin{pmatrix} v & 0 \\ 0 & \frac{\Sigma_{\alpha} + v}{v \Sigma_{\alpha}} \end{pmatrix}, \quad v = \frac{\hbar}{2}, \nonumber \\
\Sigma_{\alpha} &= \frac{\pi^2 \hbar D}{16 \alpha^2}, \quad b_m = \frac{\pi \sqrt{\hbar} m}{2 \sqrt{2} \alpha}.
\end{align}

\section{Comb States as Linear Combinations of Real-Valued Gaussians} \label{app:comb}

The Wigner function of the $| 0 \rangle$ state of the squeezed comb state is given by~\cite{shukla2020squeezed}
\begin{align}
W_{\text{comb}} (q, p) & = \frac{\exp(-e^{-2r}p^2)}{\mathcal N \pi} \sum_{m, n = 1}^N \cos{\left[ p (q_n - q_m) \right]} \exp\left[-e^{2r} \left( q - \frac{q_n + q_m}{2} \right)^2 \right]\nonumber \\
& = \frac{\exp(-e^{-2r}p^2)}{\mathcal N \pi} \left( \sum_{m = 1}^N \exp\left[-e^{2r} \left( q - q_m \right)^2 \right] + \sum_{\substack{m, n = 1 \\ m \neq n}}^N \cos{\left[ p (q_n - q_m) \right]} \exp\left[-e^{2r} \left( q - \frac{q_n + q_m}{2} \right)^2 \right] \right),
\end{align}
where the terms in the sum have been split for convenience. Inserting Equation~\eqref{eqn:cos_approx}, we find

\begin{align}
W_{\text{comb}} (q, p) \approx & \frac{\exp(-e^{-2r}p^2)}{\mathcal N \pi} \left( \sum_{m = 1}^N \exp\left[-e^{2r} \left( q - q_m \right)^2 \right] \right. \nonumber \\
& \left. + \frac{e^{\frac{\pi^2 D}{4}}}{2 \sqrt{\pi D}} \sum_{\substack{m, n = 1 \\ m \neq n \\ k \in \mathbb Z}}^N (-1)^k e^{-\frac{(q_n - q_m)^2}{\pi^2 D}(p - \frac{\pi k}{q_n - q_m})^2} \exp\left[-e^{2r} \left( q - \frac{q_n + q_m}{2} \right)^2 \right] \right).
\end{align}
Regrouping exponential terms together, we find

\begin{align}
W_{\text{comb}} (q, p) \approx & \frac{1}{\mathcal N \pi} \sum_{m = 1}^N \exp\left[-e^{2r} \left( \left( q - q_m \right)^2 + p^2 \right) \right] \nonumber \\
& + \frac{e^{\frac{\pi^2 D}{4}}}{2 \mathcal N \sqrt{\pi^3 D}} \sum_{\substack{m, n = 1 \\ m \neq n \\ k \in \mathbb Z}}^N (-1)^k \exp\left(- \frac{\pi^2 k^2}{(q_n - q_m)^2 e^{2r} + \pi^2 D} \right) \exp\left( - \frac{(p - p_{k, m, n})^2}{2\sigma_{m, n}^2} \right) \exp\left[-e^{2r} \left( q - \frac{q_n + q_m}{2} \right)^2 \right],
\end{align}
with $p_{k, m, n} = \frac{\pi k (q_n - q_m)}{(q_n - q_m)^2 + \pi D e^{-2r}}$ and $\sigma_{m, n}^2 = \frac{1}{2} \frac{\pi^2 D}{(q_n - q_m)^2 + \pi^2 D e^{-2r}}$.

\section{Fock Damping}\label{app:fock_damping}

Here we show how a state with a Wigner function of the form in Eq.  \eqref{eq:lin_comb_gauss} is transformed by the Fock damping operator in \cref{eq:fock_damping}. Recall that the $\hat{E}(\epsilon)$ operator can be viewed as arising from passing a state through a beam-splitter of transmissivity $\cos\theta=e^{-\epsilon}$ with an ancillary mode in a vacuum state, then measuring vacuum on the ancillary mode~\cite{Noh2019capacity,Tzitrin2020}.

Let the state have means and covariance $(\bm{\mu}_{m,0},\bm{\Sigma}_{m,0})$. First we pass the state and an ancillary vacuum mode (covariance matrix of $\hbar\id/2$) through a beam-splitter represented by symplectic matrix $\bm{S}_\theta = \begin{pmatrix}
\cos\theta\id & \sin\theta\id \\ -\sin\theta\id & \cos\theta\id
\end{pmatrix}$ where we have chosen the mode-wise ordering $(q_1,p_1,q_2,p_2)$:
\begin{equation}\label{eq:fock_damp_update}
        \bm{\Sigma}_m = \bm{S}_\theta \begin{pmatrix}\bm{\Sigma}_{m,0} & \textbf{0} \\ \textbf{0} & \hbar\id/2\end{pmatrix}\bm{S}_\theta^T\\= \begin{pmatrix}\cos^2\theta\bm{\Sigma}_{m,0} + \hbar\sin^2\theta\id/2 & \cos\theta\sin\theta(\hbar\id/2 - \bm{\Sigma}_{m,0}) \\ \cos\theta\sin\theta(\hbar\id/2 - \bm{\Sigma}_{m,0}) & \sin^2\theta\bm{\Sigma}_{m,0} + \hbar\cos^2\theta\id/2\end{pmatrix}
\end{equation}
\begin{equation}\label{eq:fock_damp_update_weights}
    \bm{\mu}_m = \bm{S}_\theta \begin{pmatrix}\bm{\mu}_{m,0}\\ \textbf{0}\end{pmatrix}
    = \begin{pmatrix}\cos\theta\bm{\mu}_{m,0}\\ \sin\theta\bm{\mu}_{m,0}\end{pmatrix}
\end{equation}
where $\bm{\Sigma}_m$ and $\bm{\mu}_m$ now have the same form as \cref{eq:AB_mean_cov}.

Next, we project the second mode onto vacuum. Here we can use \cref{eq:update_covs_means}, setting $\bm{\Sigma}_j = \hbar\id/2$ and $\bm{r}_j = \bm{0}$, so that the full mapping becomes:
\begin{equation}
    \begin{split}
        \bm{\Sigma}_{m,0} &\rightarrow \cos^2\theta\bm{\Sigma}_{m,0} + \hbar\sin^2\theta\id/2 - \cos^2\theta\sin^2\theta(\hbar\id/2 - \bm{\Sigma}_{m,0})\left[\sin^2\theta\bm{\Sigma}_{m,0} + (\cos^2\theta+1)\hbar\id/2\right]^{-1}(\hbar\id/2 - \bm{\Sigma}_{m,0}), \\
        \bm{\mu}_{m,0} &\rightarrow  \cos\theta\bm{\mu}_{m,0} - \cos\theta\sin^2\theta(\hbar\id/2 - \bm{\Sigma}_{m,0})\left[\sin^2\theta\bm{\Sigma}_{m,0} + (\cos^2\theta+1)\hbar\id/2\right]^{-1}\bm{\mu}_{m,0}.
    \end{split}
\end{equation}
Finally, as in \cref{eq:update_weights} we find a per-peak re-weighting of:
\begin{equation}
    c_m \rightarrow \frac{ w(\bm{0}|\sin\theta\bm{\mu}_{m,0},\ \sin^2\theta\bm{\Sigma}_{m,A} + \hbar\cos^2\theta\id/2,\ \hbar\id/2)}{p(\bm{0};\hat{\rho},\hbar\id/2)},
\end{equation}
where $d_j = 1$ as the measurement is onto only a single Gaussian in phase space.

\section{Measurement-Based Gates that Employ Squeezed Ancillae}\label{app:mbgates}

In Table \ref{tab:mbsq_avg}, we provide a summary of the Gaussian CPTP maps corresponding to measurement-based squeezing, implemented using an ancillary squeezed state, a beam-splitter and homodyne measurement followed by a feedforward displacement.

\begin{table}[h]
\def\arraystretch{1.5}
\begin{centering}
\begin{tabular}{l|l}
\hline
{\bf Gate} & {\bf Average Map}\\
 \hline
\hline
$q$-squeezing &$\bm{X}^{(q)}_s = \begin{pmatrix}
    \cos{\theta} & 0 \\
    0 & \cos{\theta}^{-1}
    \end{pmatrix},~~ \cos{\theta}=e^{-s}, ~~ \bm{Y}^{(q)}_{s,r,\eta} = \frac{\hbar}{2}\begin{pmatrix}
    \sin^2{\theta} e^{-2r} & 0 \\
    0& \eta^{-1}\tan^2{\theta}(1-\eta)
    \end{pmatrix}$ \\
\hline
$p$-squeezing  & $\bm{X}^{(p)}_s = \begin{pmatrix}
    \cos{\theta}^{-1} & 0 \\
    0 & \cos{\theta}
    \end{pmatrix}, ~~~ \bm{Y}^{(p)}_{s,r,\eta} = \frac{\hbar}{2}\begin{pmatrix}
    \eta^{-1}\tan^2{\theta}(1-\eta)  & 0 \\
    0& \sin^2{\theta} e^{-2r}
    \end{pmatrix}$\\
\hline
\end{tabular}
\par\end{centering}
\caption{Average Gaussian CPTP maps for inline squeezing, as implemented with an ancillary squeezed vacuum state. Here, $\cos\theta=e^{-s}$ is the squeezing parameter, $r$ is the squeezing parameter of the ancillary state, and $\eta$ is the loss parameter of the inefficient homodyne measurement on the ancilla.}
\label{tab:mbsq_avg}
\end{table}

In Table \ref{tab:mbsq_single}, we provide a summary of the map for a single-shot run of ancilla-assisted squeezing.

\begin{table}[h]
\def\arraystretch{1.5}
\begin{centering}
\begin{tabular}{l|c|c}
\hline
&{\bf Initial} & {\bf Final}\\
 \hline
\hline
{\bf Covariance} & $\bm{\Sigma}_{m,A,0} = \frac{\hbar}{2}\begin{pmatrix} a_{m,0} & b_{m,0} \\ b_{m,0} & d_{m,0} \end{pmatrix}$ & $ \displaystyle{\cos^2{\theta}\bm{\Sigma}_{m,A,0} + \sin^2{\theta} \bm{\Sigma}_r - \frac{\hbar\eta( \sin{\theta} \cos{\theta})^2}{2F_m} \begin{pmatrix}
    b_{m,0}^2 &(d_{m,0}-e^{2r})b_{m,0}  \\
    (d_{m,0}-e^{2r})b_{m,0} &  (d_{m,0}-e^{2r})^2
    \end{pmatrix}} $\\
 &$\bm{\Sigma}_r = \frac{\hbar}{2}\begin{pmatrix} e^{-2r} & 0 \\ 0 & e^{2r}\end{pmatrix}$& $F_{m} = \left[\eta (s^2_{\theta} d_{m,0}+ c^2_{\theta} e^{2r}) + 1-\eta\right]$\\
\hline
{\bf Mean} &$\bm{\mu}_{m,A,0} = \begin{pmatrix}
    \bar{x}_{m,0}\\ \bar{p}_{m,0}\end{pmatrix}$ &  $\displaystyle{\cos{\theta} \bm{\mu}_{m,A,0} - \frac{\sqrt{\eta} \sin{\theta} \cos{\theta}(p_M-\sqrt{\eta} \sin{\theta} \bar{p}_{m,0} )}{F_m} \begin{pmatrix} b_{m,0} \\ d_{m,0} - e^{2r} \end{pmatrix} + \begin{pmatrix} f_x(p_M) \\ f_p(p_M) \end{pmatrix}}$\\
\hline
{\bf Peak weight} & $c_m$ &  $\displaystyle{ c_m\frac{\exp[-2(p_M-\sqrt{\eta} s_{\theta} \bar{p}_{m,0})^2/\hbar F_m]}{\sum_{n\in\mathcal{M}}c_n\exp[-2(p_M-\sqrt{\eta} s_{\theta} \bar{p}_{n,0})^2/\hbar F_n]}}$ \\
\hline
\end{tabular}
\par\end{centering}
\caption{Update rules for each Gaussian peak in a linear combination from a single-shot run of ancilla-assisted squeezing in $q$ quadrature. $r$ is the squeezing level of the ancilla state, $\theta$ is the angle of the beam-splitter with $\cos\theta=e^{-s}$ defining the squeezing parameter $s$, $\eta$ is the loss parameter of the homodyne detector, and $p_{M}$ is the $p$-homodyne outcome on the ancilla mode. To match the average case, one sets the feedforward operations to $f_x(p_M) = 0$ and $f_p(p_M) = p_M\tan\theta/\sqrt{\eta}$, and integrate over $p_M$.}
\label{tab:mbsq_single}
\end{table}

\subsection{Inline Squeezing Gate}

Here we derive the map effected by measurement-based squeezing, as described in~\cite{filip2005}, and shown in \cref{fig:mbsq_circuits} (a).

\paragraph*{Initial states}
Let the covariance matrix and mean of the initial state be of the form:
\begin{equation}
    \bm{\Sigma}_{m,A,0} = \frac{\hbar}{2}\begin{pmatrix} a_{m,0} & b_{m,0} \\ b_{m,0} & d_{m,0} \end{pmatrix},\ \bm{\mu}_{m,A,0} = \begin{pmatrix}
    \bar{x}_{m,0}\\ \bar{p}_{m,0}\end{pmatrix}.
\end{equation}
The covariance matrix and mean of the ancillary squeezed state is:
\begin{equation}
    \bm{\Sigma}_{B,0} = \frac{\hbar}{2}\begin{pmatrix} e^{-2r} & 0 \\ 0 & e^{2r}\end{pmatrix},\ \bm{\mu}_{B,0}= \bm{0}.
\end{equation}
The initial separable form of the joint covariance and mean is thus $\bm{\Sigma}_{m,0} = \bm{\Sigma}_{m,A,0} \oplus \bm{\Sigma}_{B,0}$ and $\bm{\mu}_{m,0} = \bm{\mu}_{m,A,0} \oplus \bm{\mu}_{B,0}$.

\paragraph*{Combine at a beam-splitter}
When combined on a beam-splitter we evolve the two-mode covariance matrix and means to
\begin{align}
\bm{\Sigma}_{m,1} = \begin{pmatrix}
\cos^2\theta \bm{\Sigma}_{m,A,0} + \sin^2\theta \bm{\Sigma}_{B,0} & -\sin\theta \cos\theta (\bm{\Sigma}_{m,A,0}-\bm{\Sigma}_{B,0})\\
-\sin\theta \cos\theta(\bm{\Sigma}_{m,A,0}-\bm{\Sigma}_{B,0}) & \sin^2\theta \bm{\Sigma}_{m,A,0} + \cos^2\theta \bm{\Sigma}_{B,0}
\end{pmatrix},
\bm{\mu}_{m,1} = \begin{pmatrix}
\cos\theta \bm{\mu}_{m,A,0}\\ \sin\theta \bm{\mu}_{m,A,0}
\end{pmatrix}
\end{align}
where symplectic matrix for a beam-splitter is given after Eq. \eqref{eq:fock_damping_update}. Note we are working with basis ordering $(q_1,p_1,q_2,p_2)$. For convenience we now move to $c_{\theta} \equiv \cos{\theta}, ~ s_{\theta} \equiv \sin{\theta}$.

\paragraph*{Inefficient homodyne measurement} We now apply loss to the second mode to model inefficient homodyne measurement. The Gaussian CPTP map for the loss channel is provided in Eq. \eqref{eq:loss}, so loss on the second mode only is simply $(\bm{X},\bm{Y}) = (\id \oplus \sqrt{\eta} \id, \bm{0}\oplus [1-\eta]\hbar\id/2)$. Applying this channel yields the updated covariance and means:
\begin{align}
    \bm{\Sigma}_{m} = \begin{pmatrix}
    c^2_{\theta} \bm{\Sigma}_{m,A,0} + s^2_{\theta} \bm{\Sigma}_{B,0} & -\sqrt{\eta} s_{\theta} c_{\theta} (\bm{\Sigma}_{m,A,0}-\bm{\Sigma}_{B,0})\\
-\sqrt{\eta} s_{\theta} c_{\theta} (\bm{\Sigma}_{m,A,0}-\bm{\Sigma}_{B,0}) & \eta (s^2_{\theta} \bm{\Sigma}_{m,A,0} + c^2_{\theta} \bm{\Sigma}_{B,0}) + (1-\eta)\hbar \id/2
    \end{pmatrix},
\bm{\mu}_{m} = \begin{pmatrix}
c_{\theta} \bm{\mu}_{m,A,0}\\ \sqrt{\eta} s_{\theta} \bm{\mu}_{m,A,0}
\end{pmatrix}
\end{align}
At this stage, $\bm{\Sigma}_m$ and $\bm{\mu}_m$ are in the form required for using \cref{eq:AB_mean_cov}.

\paragraph*{Measurement and conditional dynamics with feedforward} We now perform a ideal $p$-homodyne measurement on mode $B$ which yields outcome $\bm{r}_j^T = (0,p_M)$. Employing now \cref{eq:update_covs_means}, we find that the updated covariance and means for mode $A$ are provided by:
\begin{align}
    \bm{\Sigma}_{m,A}
&=    c^2_{\theta} \bm{\Sigma}_{m,A,0} + s^2_{\theta} \bm{\Sigma}_{B,0} \nonumber \\
&~~~~~ -  \eta(s_{\theta} c_{\theta})^2 (\bm{\Sigma}_{m,A,0}-\bm{\Sigma}_{B,0})^T[\eta (s^2_{\theta} \bm{\Sigma}_{m,A,0} + c^2_{\theta} \bm{\Sigma}_{B,0}) + (1-\eta)\hbar \id/2  + \bm{\Sigma}_{j}]^{-1} (\bm{\Sigma}_{m,A,0}-\bm{\Sigma}_{B,0}),
\end{align}
\begin{align}
\bm{\mu}_{m,A}&=c_{\theta}\bm{\mu}_{m,A,0}- \sqrt{\eta} s_{\theta} c_{\theta} (\bm{\Sigma}_{m,A,0}-\bm{\Sigma}_{B,0})^T[\eta (s^2_{\theta} \bm{\Sigma}_A + c^2_{\theta} \bm{\Sigma}_{B,0}) + (1-\eta)\hbar \id/2  + \bm{\Sigma}_{j}]^{-1} (\bm{r}_j - \sqrt{\eta} s_{\theta} \bm{\mu}_{m,A,0})
\end{align}
Here, $\bm{\Sigma}_{j}$ is the covariance matrix of the Gaussian state which is detected. In this case, we are considering ideal $p$-homodyne measurement, so it is given by $\bm{\Sigma}_{j}=\lim_{\epsilon\rightarrow 0}\frac{\hbar}{2}\begin{pmatrix}\epsilon^{-1} &0 \\ 0& \epsilon\end{pmatrix}$. We note that there is only one weight with $d_j = 1$, since the measurement is onto a Gaussian state.

Because of the form of $\bm{\Sigma}_{M}$, the inverse matrix in both the expressions for the covariance matrix and mean reduces to
\begin{align} \left[\eta (s^2_{\theta} \bm{\Sigma}_{m,A,0} + c^2_{\theta} \bm{\Sigma}_{B,0}) + (1-\eta)\hbar \id/2 +\bm{\Sigma}_{M}\right]^{-1} &= \frac{2}{\hbar}\begin{pmatrix} 0 & 0 \\  0 & 1/F_m \end{pmatrix}, \\  ~~F_{m} = \eta (s^2_{\theta} d_{m,0}+ c^2_{\theta} e^{2r}) + 1-\eta.
\end{align}
Let $\bm{T}^{(i,j)}$ denote the $i^{th}$ row and $j^{th}$ column of matrix $\bm{T}$. We find that after the feedforward displacement, the covariance and mean update as
\begin{align}
    \bm{\Sigma}_{m,A} &= c^2_{\theta} \bm{\Sigma}_{m,A,0} + s^2_{\theta} \bm{\Sigma}_{B,0}  \nonumber\\
    &~~~~~- \frac{\eta(s_{\theta} c_{\theta})^2}{F_m} \begin{pmatrix} [(\bm{\Sigma}_{m,A,0} -\bm{\Sigma}_{B,0})^{(1,2)}]^2 & (\bm{\Sigma}_{m,A,0} -\bm{\Sigma}_{B,0})^{(2,2)} (\bm{\Sigma}_{m,A,0} -\bm{\Sigma}_{B,0})^{(1,2)}\\ (\bm{\Sigma}_{m,A,0} -\bm{\Sigma}_{B,0})^{(2,2)} (\bm{\Sigma}_{m,A,0} -\bm{\Sigma}_{B,0})^{(1,2)} & [(\bm{\Sigma}_{m,A,0} -\bm{\Sigma}_{B,0})^{(2,2)}]^2\end{pmatrix} \\
    & = c^2_{\theta}\bm{\Sigma}_{m,A,0} + s^2_{\theta} \bm{\Sigma}_{B,0} - \frac{\hbar\eta(s_{\theta} c_{\theta})^2}{2F_m} \begin{pmatrix}
    b_{m,0}^2 &(d_{m,0}-e^{2r})b_{m,0}  \\
    (d_{m,0}-e^{2r})b_{m,0} &  (d_{m,0}-e^{2r})^2
    \end{pmatrix}
\end{align}
\begin{align}
    \bm{\mu}'_{m,A} &= c_{\theta} \bm{\mu}_{m,A,0} - \frac{\sqrt{\eta} s_{\theta} c_{\theta}(p_M-\sqrt{\eta} s_{\theta} \bar{p}_{m,0} )}{F_m} \begin{pmatrix} b_{m,0} \\ d_{m,0} - e^{2r} \end{pmatrix} + \begin{pmatrix} f_x(p_M) \\ f_p(p_M) \end{pmatrix},
\end{align}
Note that the outcome covariance matrix does not depend on the measurement outcome, while the mean does. Here, $f_x(p_M),f_p(p_M)$ correspond to the value of the feedforward displacement that is applied after the homodyne measurement.

\paragraph*{Single run peak reweighting} While we have found the covariance matrix and mean of mode $A$ after detecting outcome $p_M$ on mode B, we must also take into account the probability of detecting $p_M$, which is provided by \cref{eq:outcome_prob}:
\begin{align}
    p(p_M; \hat{\rho},\bm{\Sigma}_M) = \sum_{m\in\mathcal{M}}c_m\frac{\exp[-2(p_M-\sqrt{\eta} s_{\theta} \bar{p}_{m,0})^2/\hbar F_m]}{\sqrt{\pi\hbar F_m/2}}
\end{align}
Note that each term in the sum depends on $\bar{p}_{m,0}$ and $F_m$. This means that if mode $A$ is a linear combination of Gaussian functions, the coefficients update as in \cref{eq:update_weights}:
\begin{equation}
    c_m(p_M) =  c_m\frac{\exp[-2(p_M-\sqrt{\eta} s_{\theta} \bar{p}_{m,0})^2/\hbar F_m]}{p(p_M; \hat{\rho},\bm{\Sigma}_M)\sqrt{\pi\hbar F_m/2}}
\end{equation}
At this stage, we have all the information required to update a single-shot run of measurement-based squeezing.

\paragraph*{Average map} However, we can also use the probability to determine the average map applied to the state. Importantly, we find that the average map for specific $f_x(p_M),f_p(p_M)$ simply corresponds to a Gaussian CPTP map applied to mode $A$, which means we do not need to worry about per-peak reweighting, but can simply update the covariance and means of each peak as in \cref{eq:gauss_cptp}. Consider the characteristic function of the output state $\hat{\rho}_M$, which is a function of the input state and the measurement outcome $p_M$:
\begin{align}
    \chi(\hat{\rho}_M;\bm{\xi}) = \sum_{m\in\mathcal{M}} c_m(p_M)\exp[-\frac{1}{4} \bm{\xi}^T \bm{\Omega}^T \bm{\Sigma}_{m,A} \bm{\Omega} \bm{\xi} - i \bm{\xi}^T \bm{\Omega} \bm{\mu}'_{m,A} ], ~{\rm with} ~ \bm{\Omega} = \begin{pmatrix}0 & 1 \\
    -1 & 0 \end{pmatrix}
\end{align}

The measurement-averaged output state is then given by \begin{align}
    \hat{\rho} &= \int dp_M  p(p_M; \hat{\rho},\bm{\Sigma}_M) \hat{\rho}_M \nonumber
    \end{align}
The characteristic function of the output state is then given by
\begin{align}
    \chi(\hat{\rho};\bm{\xi}) = \int dp_M p(p_M; \hat{\rho},\bm{\Sigma}_M) \chi(\hat{\rho}_M;\bm{\xi})
\end{align}
Since the covariance matrix is independent of $p_M$:
\begin{align}
    \chi(\hat{\rho};\bm{\xi}) = \sum_{m\in\mathcal{M}}\frac{c_m}{ \sqrt{\pi\hbar F_m/2}} \exp[-\frac{1}{4} \bm{\xi}^T \bm{\Omega}^T \bm{\Sigma}_{m,A} \bm{\Omega} \bm{\xi}] \int dp_M \exp[-2(p_M-\sqrt{\eta} s_{\theta} \bar{p}_{m,0})^2/\hbar F_m] \exp[-i \bm{\xi}^T \bm{\Omega} \bm{\mu}'_{m,A}]
\end{align}
Let us rewrite $\bm{\mu}'_{m,A}$ as:
\begin{align}
    \bm{\mu}'_{m,A} &= \tilde{\bm{r}}(p_M) +  (p_M-\sqrt{\eta} s_{\theta} \bar{p}_{m,A})\bm{\nu}_m, ~~ \bm{\nu}_m =  -\frac{ \sqrt{\eta} s_{\theta} c_{\theta}}{F_m} \begin{pmatrix} b_{m,0} \\ d_{m,0} - e^{2r} \end{pmatrix},~~
    \tilde{\bm{r}}(p_M) = c_{\theta} \bm{\mu}_{m,0} + \begin{pmatrix} f_x[p_M] \\ f_p[p_M]\end{pmatrix}
\end{align}
Furthermore, we relabel $\tilde{p}_{M,m} = p_M-\sqrt{\eta} s_{\theta} \bar{p}_{m,A}$. The characteristic function then becomes:
\begin{align}
    \chi(\hat{\rho};\bm{\xi}) = \sum_{m\in\mathcal{M}}\frac{c_m}{ \sqrt{\pi\hbar F_m/2}} \exp[-\frac{1}{4} \bm{\xi}^T \bm{\Omega}^T \bm{\Sigma}_{m,A} \bm{\Omega} \bm{\xi}]   \int dp_M \exp[-i \bm{\xi}^T \bm{\Omega}^T \tilde{\bm{r}}(p_M) ] \exp[-2\tilde{p}_{M,m}^2/\hbar F_m] \exp[-i\tilde{p}_{M,m} \bm{\xi}^T \bm{\Omega} \bm{\nu}_m ]
\label{ffout}
\end{align}
Displacing mode $A$ by linear and quadratic functions of $p_M$, the above integral reduces to a Gaussian integral. Let us assume that \begin{align}\label{ffunc} f_x[p_M] = g_0 + g_1 p_M + g_2 p_M^2, ~~  f_p[p_M] = h_0 + h_1 p_M + h_2 p_M^2.
\end{align}

\paragraph*{Sanity check} As a sanity check, let us check what happens if we apply no feedforward. Performing the Gaussian integral, we get that:
\begin{align}
    \chi(\hat{\rho};\bm{\xi}) =  \sum_{m\in\mathcal{M}}c_m \exp[-\frac{1}{4} \bm{\xi}^T \bm{\Omega}^T \bm{\Sigma}_{m,A} \bm{\Omega} \bm{\xi}]    \exp[-i \bm{\xi}^T \bm{\Omega} c_{\theta} \bm{\mu}_{m,0} ] \exp[-\frac{\hbar F_m}{8} (\bm{\xi}^T \bm{\Omega} \bm{\nu}_m)^2]
\end{align}
This can be simplified into
\begin{align}
    \chi(\hat{\rho};\bm{\xi}) = \sum_{m\in\mathcal{M}}c_m \exp[-\frac{1}{4} \bm{\xi}^T \bm{\Omega}^T \tilde{\bm{\Sigma}}_{m,A} \bm{\Omega} \bm{\xi}] \exp[-i \bm{\xi}^T \bm{\Omega}c_{\theta} \bm{\mu}_{m,0} ], ~~\tilde{\bm{\Sigma}}_{m,A} = \bm{\Sigma}_{m,A} + \frac{\hbar F_m}{2} \bm{\nu}_m \bm{\nu}_m^T.
\end{align}
We see that we recover the partial trace condition, i.e. that $\bm{\Sigma}_{m,0} \to c_{\theta}^2\bm{\Sigma}_{m,0} + s_{\theta}^2 \bm{\Sigma}_{B,0}$ and $\bm{\mu}_{m,0} \to c_{\theta}\bm{\mu}_{m,0}$, which is exactly what happens if we pass through a beam-splitter and trace out mode B. \\

\paragraph*{Feedforward quadratic in $p_M$} Now we go back to a feedforward function quadratic in $m$. We then perform the necessary integrals in \cref{ffout} using \cref{ffunc}. We have that
\begin{align}
    \tilde{\bm{r}}(p_M) = c_{\theta} \bm{\mu}_{m,0} + \tilde{\bm{r}}_0 + p_M \tilde{\bm{r}}_1 + p_M^2 \tilde{\bm{r}}_2, ~~ \tilde{\bm{r}}_0 = \begin{pmatrix} g_0 \\ h_0 \end{pmatrix}, ~~ \tilde{\bm{r}}_1 = \begin{pmatrix} g_1 \\ h_1 \end{pmatrix}, ~~ \tilde{\bm{r}}_2 = \begin{pmatrix} g_2 \\ h_2 \end{pmatrix}.
\end{align}
Then \cref{ffout} reduces to
\begin{equation}
    \begin{split}
    \chi(\hat{\rho};\bm{\xi}) &= \sum_{m\in\mathcal{M}}\frac{c_m}{ \sqrt{\pi\hbar F_m/2}} \exp[-\frac{1}{4} \bm{\xi}^T \bm{\Omega}^T \bm{\Sigma}_{m,A} \bm{\Omega} \bm{\xi}]  \exp[-i \bm{\xi}^T \bm{\Omega} (c_{\theta} \bm{\mu}_{m,0} + \tilde{\bm{r}}_0)]\exp[-2\eta s^2_{\theta}\bar{p}_{m,A}^2/\hbar F_m] \\
    & ~~~\times\exp[i\sqrt{\eta} s_{\theta}\bar{p}_{m,A}\bm{\xi}^T\bm{\Omega}\bm{\nu}_m]   \int dp_M \exp\left[p_M\left[\frac{4\sqrt{\eta} s_{\theta}\bar{p}_{m,A}}{\hbar F_m}- i\bm{\xi}^T \bm{\Omega} (\tilde{\bm{r}}_1 + \bm{\nu}_m)\right] \exp[-p_M^2\left(\frac{2}{\hbar F_m} + i \bm{\xi}^T \bm{\Omega} \tilde{\bm{r}}_2\right)\right]
    \end{split}
\end{equation}
While the average map can be computed and is independent of $p_M$, we do not find that it provides a useful transformation of mode $A$.

\paragraph*{Feedforward linear in $m$} Consider instead linear feedforward, by setting $\tilde{\bm{r}}_2=\tilde{\bm{r}}_0=0$. Instead, we have that
\begin{equation}
    \begin{split}
    \chi(\hat{\rho};\bm{\xi}) &= \sum_{m\in\mathcal{M}}\frac{c_m}{ \sqrt{\pi\hbar F_m/2}} \exp[-\frac{1}{4} \bm{\xi}^T \bm{\Omega}^T \bm{\Sigma}_{m,A} \bm{\Omega} \bm{\xi}]  \exp[-i \bm{\xi}^T \bm{\Omega} (c_{\theta} \bm{\mu}_{m,0})]\exp[-2\eta s^2_{\theta}\bar{p}_{m,A}^2/\hbar F_m]\exp[i\sqrt{\eta} s_{\theta}\bar{p}_{m,A}\bm{\xi}^T\bm{\Omega}\bm{\nu}_m] \\
    & ~~~\times   \int dp_M \exp\left[p_M\left[\frac{4\sqrt{\eta} s_{\theta}\bar{p}_{m,A}}{\hbar F_m}- i\bm{\xi}^T \bm{\Omega} (\tilde{\bm{r}}_1 + \bm{\nu}_m)\right] \exp[-p_M^2\left(\frac{2}{\hbar F_m}\right)\right]
    \end{split}
\end{equation}
Let us further choose the linear feedforward to be $g_1=0$ and $h_1 = \sqrt{\eta^{-1}} \tan{\theta}$, which is based on the suggestion in~\cite{filip2005}. Then the output covariance matrix can be written as
\begin{align}
    \bm{\Sigma}_{A,out} &= \bm{\Sigma}_{m,A} + \frac{\hbar F_m}{2}(\bm{\nu}_m + \tilde{\bm{r}}_1 )(\bm{\nu}_m + \tilde{\bm{r}}_1 )^T = \bm{X} \bm{\Sigma}_{m,A,0} \bm{X}^T + \bm{Y}\nonumber, ~~ \bm{\mu}_{A,\rm out} =  \bm{X} \bm{\mu}_{m,A,0}
\end{align}
with
\begin{align}
    \bm{X} &= \begin{pmatrix}
    c_{\theta} & 0 \\
    0 & c_{\theta}^{-1}
    \end{pmatrix}, ~~~ \bm{Y} = \frac{\hbar}{2} \begin{pmatrix}
    s_{\theta}^2 e^{-2r} & 0 \\
    0& \eta^{-1}\tan^2{\theta}(1-\eta)
    \end{pmatrix}
\end{align}
This is a Gaussian CPTP map, with the $\bm{X}$ part of the map corresponding to squeezing mode $A$ in $q$ by a factor $\cos\theta$. An analogous derivation is possible for squeezing in $p$ by using a $p$-squeezed state, performing homodyne in $q$ and feedforward displacement in $p$. Moreover, we can achieve complex squeezing value $re^{i\phi}$ by placing the squeezing circuit between phase shifters.

\subsection{Gates that Employ Inline Squeezing}\label{subsec:mbsq_gates}
Given our examination of inline squeezing using squeezed vacuum ancillae, we now summarize valuable CV gates which employ inline squeezing.

\begin{figure}
\centering
\subfloat[Squeezing gate]{$$
\Qcircuit @C=1.0em @R=.7em {
\lstick{\Ket{0}_{\text{q}}}  & \multigate{1}{\hat{BS}(\theta)} & \measureD{p=p_0} & \cctrl{1}\\
\lstick{\Ket{\psi}} &   \ghost{\hat{BS}(\theta)} & \qw & \gate{\hat{X}\left[f(p_0)\right]} & \qw & &  {\hat{S}(r)\Ket{\psi}}
}
$$}
\hspace{2 cm}
\subfloat[Phase gate]{$$
\Qcircuit @C=1.0em @R=.7em {
\lstick{\Ket{\psi}}  & \gate{\hat{S}(re^{i\phi})} & \gate{\hat{R}(\theta)} &\qw & &  {\hat{P}(s)\Ket{\psi}}
}
$$}
\newline
\subfloat[CZ gate]{$$
\Qcircuit @C=1.0em @R=.7em {
\lstick{\Ket{\psi_1}}  & \multigate{1}{\hat{BS}(\theta+\frac{\pi}{2},\phi)} & \gate{\hat{S}(r)} & \multigate{1}{\hat{BS}(\theta,\phi)} &\qw& & \Ket{\psi_1}&& \multigate{1}{\hat{CZ}(s)} & \qw\\
\lstick{\Ket{\psi_2}} &   \ghost{\hat{BS}(\theta+\frac{\pi}{2},\phi)} & \gate{\hat{S}(r)} & \ghost{\hat{BS}(\theta,\phi)} &\qw&\push{\rule{.3em}{0em}=\rule{.5em}{0em}}& \Ket{\psi_2} &&\ghost{\hat{CZ}(s)}&\qw
}
$$}
\caption{Review of CV gates and their realistic optical implementations. (a) The circuit for measurement-based squeezing~\cite{filip2005}. An ancillary position eigenstate is combined at a beam-splitter of angle $\theta$ with the target mode. The ancilla is then measured in $p$ quadrature with efficiency $\eta$, and informing a feedforward $q$ displacement of $p_0 \tan\theta/\sqrt{\eta}$. The level of $q$-quadrature squeezing applied to the target mode is $r = \ln\cos\theta$. In practice, the quality of measurement-based squeezing is dependent on the level of squeezing of the ancillary state, and loss in the beam-splitter. Squeezing in $p$ can be achieved with a momentum eigenstate, a $q$ quadrature measurement, and a $p$ displacement. Complex squeezing parameters can be implemented by sandwiching the circuit between two phase shifters. (b) A CV phase gate applied using rotations and inline squeezing~\cite{Killoran2019}. Here, $r=\text{arccosh}\sqrt{1+\frac{s^2}{4}}$, $\theta=\arctan{\frac{s}{2}}$, and $\phi = -\frac{\pi}{2}\text{sign}(s)-\theta$. (c) A CV $\CZ$ gate~\cite{Killoran2019}, with $r=\text{arcsinh}{-\frac{s}{2}}$, $\phi=\frac{\pi}{2}$, $\sin{2\theta}=(\cosh{r})^{-1}$, and $\cos{2\theta}=\tanh{r}$. For the GKP encoding, the qubit phase and $\CZ$ gates corresponds to $s=1$. In practice, the inline squeezing in the phase and $\CZ$ gates can be implemented using measurement-based squeezing, and additional noise in the CZ gate can be modelled by lossy beam-splitters. 
}
\label{fig:mbsq_circuits}
\end{figure}
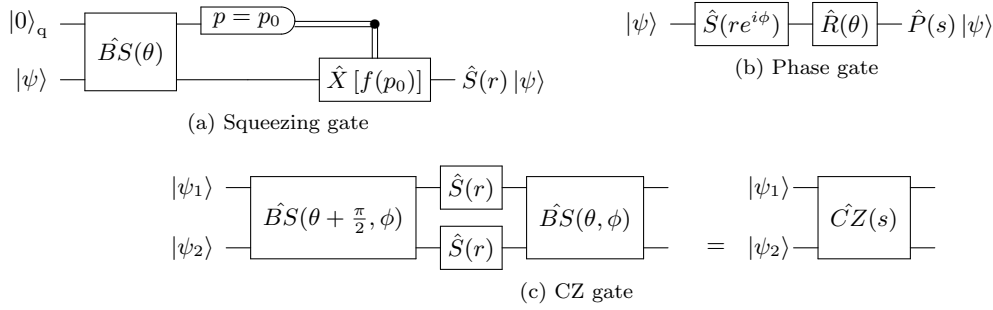

\paragraph*{Phase gate} The CV quadratic phase gate $\hat{P}(s)$ is just a squeezing gate $\hat{S}(re^{i\phi})$ composed with a rotation gate $\hat{R}(\theta)$. The ideal decomposition is given by~\cite{Killoran2019,Kalajdzievski2021exactapproximate}
\begin{align}
    \hat{P}(s) = \hat{R}(\theta)\hat{S}(re^{i\phi}), ~~ \theta = \tan^{-1}\left(\tfrac{s}{2}\right), ~ \phi = -\text{sign}(s)\frac{\pi}{2} - \theta, ~ r = \cosh^{-1}\left(\sqrt{1+\tfrac{s^2}{4}}\right).
\end{align}
as shown in \cref{fig:mbsq_circuits} (b), with values provided for the GKP qubit phase gate. Further noise could be assumed by using a lossy rotation gate ${\cal \hat{L}}(\eta) \hat{R}(\theta)$. If we assume perfect rotation operations, then the performance of the phase gate is limited by the squeezing gate.

\paragraph*{Entangling gates} The CV SUM gate can be decomposed as in terms of beam-splitters $\hat{BS}(\theta)$ and squeezing~\cite{Killoran2019}, where the control mode is the first mode:
\begin{align}
    \hat{CX}(s) = \hat{BS}(\theta + \pi/2) [\hat{S}(r) \otimes \hat{S}(-r)] \hat{BS}(\theta), ~ r = \sinh^{-1}\left(-\frac{s}{2}\right), ~ \theta = \frac{1}{2}\tan^{-1}\left(-\frac{2}{s}\right).
\end{align}
A simple noise model is to add single-mode losses to the beam-splitter outputs $\hat{\mathcal{L}}(\eta) \otimes \hat{\mathcal{L}}(\eta) \hat{BS}(\theta)$. A $\CZ$ gate can be achieved by applied Fourier rotations on mode $B$ both before and after the $\CX$ gate. The $\CZ$ gate is depicted in \cref{fig:mbsq_circuits} (c) with specific values provided for the GKP qubit CZ gate.

\paragraph*{Amplifier channel}
Amplification could be a useful in tackling photon loss effects~\cite{albert2018}. In the ideal average case, an amplifier modifies the covariance matrices and means in the following manner~\cite{book_serafini}:
\begin{align}
    \bm{\Sigma} \to \kappa^2 \bm{\Sigma} + (\kappa^2 -1) \hbar\id/2, ~~ \kappa >1, ~~ \bm{\mu} \to \kappa \bm{\mu}.
\end{align}
The ideal amplifier amplifies the input signal and adds a symmetric noise. To construct a realistic noise channel with $\kappa = \cosh{r}$, we use its Stinespring dilation to get~\cite{solomon2010}:
\begin{align}
    \hat{\cal A}(\kappa)[\hat{\rho}] &= \text{tr}_B[ \hat{S}_{0,1}(r)(\hat{\rho} \otimes |0\rangle \langle 0|_B)\hat{S}_{0,1}(r)^{\dagger}], \\
    \hat{S}_{0,1}(r) & = \hat{BS}(\pi/4)^{\dagger}[\hat{S}(r) \otimes \hat{S}(-r)] \hat{BS}(\pi/4),
\end{align}
where $\hat{S}_{0,1}(r)$ is a two-mode squeezing operation. Here, lossy beam-splitters and the noise we examined for inline squeezing form the dominant sources of error.

\clearpage
\twocolumngrid
\bibliography{boqes}

\end{document}